\documentclass[epjc3,twocolumn]{svjour3}

\pdfoutput=1
\journalname{Eur. Phys. J. C}

\usepackage{graphicx}
\usepackage[numbers,sort&compress]{natbib} 
\usepackage[english]{babel}
\usepackage{amsmath}
\usepackage{amssymb}
\usepackage{placeins}
\usepackage{mathtools}
\usepackage{tikz}
\usepackage{siunitx}
\usepackage[acronym,nohypertypes={acronym}]{glossaries}

\newcommand{\de}{\delta}
\newcommand{\eref}[1]{Eq.~(\ref{#1})}
\newcommand{\fref}[1]{Fig.~\ref{#1}}
\newcommand{\nnnl}{\nonumber\\}	
\newcommand{\strong}[1]{{\bf #1}\ }
\newcommand{\wt}[1]{\widetilde{#1}}
\newcommand\norm[1]{\left\lVert#1\right\rVert}
\newcommand{\specialcell}[2][c]{\def\arraystretch{1}\begin{tabular}[#1]{@{}c@{}}#2\end{tabular}}
\newcolumntype{L}[1]{>{\raggedright\arraybackslash}p{#1}} 
\newcolumntype{C}[1]{>{\centering\arraybackslash}p{#1}} 
\newcolumntype{R}[1]{>{\raggedleft\arraybackslash}p{#1}} 
\newcommand{\abs}[1]{|#1|}

\newcount\colveccount
\newcommand*\colvec[1]{
        \global\colveccount#1
        \begin{pmatrix}
        \colvecnext
}
\def\colvecnext#1{
        #1
        \global\advance\colveccount-1
        \ifnum\colveccount>0
                \\
                \expandafter\colvecnext
        \else
                \end{pmatrix}
        \fi
}

\newacronym{qcd}{QCD}{quantum chromodynamics}
\newacronym{ym}{YM}{Yang--Mills}
\newacronym{dse}{DSE}{Dyson--Schwinger equation}
\newacronym[longplural=Slavnov--Taylor identities]{sti}{STI}{Slavnov--Taylor identity}

\newacronym{proper}{1PI}{one-particle irreducible}

\newacronym{ir}{IR}{infrared}
\newacronym{uv}{UV}{ultraviolet}

\newacronym{rg}{RG}{renormalization group}
\newacronym{frg}{FRG}{functional renormalization group}

\newacronym{sf}{SF}{swordfish}
\newacronym{dt}{DT}{dynamic triangle}
\newacronym{glb}{GlB}{gluon box}
\newacronym{ghb}{GhB}{ghost box}
\newacronym{st}{ST}{static triangle} 

\begin{document}

\title{A Dyson--Schwinger study of the four-gluon vertex}
\date{\today}

\author{
Anton K. Cyrol\thanksref{addrDA,addrHD,eAC}
\and
Markus Q. Huber\thanksref{addrGRA,eMH}
\and
Lorenz von Smekal\thanksref{addrDA,addrGIE,eLS}
}

\thankstext{eAC}{e-mail: cyrol@thphys.uni-heidelberg.de}
\thankstext{eMH}{e-mail: markus.huber@uni-graz.at}
\thankstext{eLS}{e-mail: lorenz.smekal@physik.tu-darmstadt.de}

\institute{%
Technische Universit\"at Darmstadt, Institut f\"ur Kernphysik, Theoriezentrum, 64289 Darmstadt, Germany\label{addrDA}
\and
Ruprecht-Karls-Universit\"at Heidelberg, Institut f\"ur Theoretische Physik, 69120 Heidelberg, Germany\label{addrHD}
\and
University of Graz, Institute of Physics, NAWI Graz, 8010 Graz, Austria\label{addrGRA}
\and
Justus-Liebig-Universit\"at Gie\ss en, Institut f\"ur Theoretische Physik, 35392 Gie\ss en, Germany\label{addrGIE}
}

\maketitle

\begin{abstract}
We present a self-consistent calculation of the four-gluon vertex of Landau gauge Yang--Mills theory from a truncated Dyson--Schwinger equation. The equation contains the leading diagrams in the ultraviolet and is solved using as the only input results for lower Green functions from previous Dyson--Schwinger calculations that are in good agreement with lattice data. All quantities are therefore fixed and no higher Green functions enter within this truncation. Our self-consistent solution resolves the full momentum dependence of the vertex but is limited to the tree-level tensor structure at the moment. Calculations of selected dressing functions for other tensor structures from this solution are used to exemplify that they are suppressed compared to the tree-level structure except for possible logarithmic enhancements in the deep infrared. Our results furthermore allow one to extract a qualitative fit for the vertex and a running coupling.
\end{abstract}

\section{Introduction}

The non-perturbative analysis of quantum field theories is one of the great challenges in physics. One particular approach is to use functional equations for Green functions which are the basic building blocks of a quantum field theory. In \gls{qcd} they can be used as input for hadron phenomenology and strong-interaction matter studies; see, e.g., \cite{Alkofer:2000wg,Fischer:2006ub,Fischer:2009wc,Braun:2009gm,Braun:2010cy,Bashir:2012fs,Fischer:2012vc,Hopfer:2012qr,Eichmann:2013afa,Fischer:2013eca,Tripolt:2013jra}, but they also allow direct conclusions on non-perturbative aspects like confinement \cite{Braun:2007bx,Marhauser:2008fz,Fister:2013bh} or dynamical mass generation; see, e.g., \cite{Alkofer:2000wg,Fischer:2006ub,Bashir:2012fs} and the references therein.

The main challenge for functional methods is the necessity to truncate the originally infinite hierarchy of functional equations for Green functions and to quantify the resulting uncertainties. The most straightforward way to assess the quality of a particular truncation by going beyond it in a systematic way as provided, e.g., by derivative or vertex expansions is often rather difficult. Hence alternative possibilities for tests are welcome and widely used, such as comparisons with results from lattice simulations where they are available.

In this paper we focus on Yang--Mills theory in the Landau gauge. The good understanding of this particular covariant gauge that was established in the past provides the basis for many of the more phenomenological investigations of QCD. The Landau gauge propagators have been well studied with various methods, e.g., lattice simulations \cite{Cucchieri:2007md,Cucchieri:2008fc,Sternbeck:2007ug,Bogolubsky:2009dc,Oliveira:2012eh,Sternbeck:2012mf}, \glspl{dse} \cite{vonSmekal:1997is,vonSmekal:1997vx,Atkinson:1997tu,Zwanziger:2001kw,Lerche:2002ep,Zwanziger:2002ia,Fischer:2002hn,Zwanziger:2003cf,Boucaud:2008ji,Aguilar:2008xm,Alkofer:2008jy,Fischer:2008uz,Fischer:2009tn,Huber:2009tx,Pennington:2011xs,LlanesEstrada:2012my,Strauss:2012dg}, the \gls{frg} \cite{Pawlowski:2003hq,Fischer:2008uz}, a variational approach \cite{Quandt:2013wna}, a one-loop model calculation with gluon mass term \cite{Tissier:2010ts}, or the (refined) Gribov--Zwanziger framework  \cite{Gribov:1977wm,Zwanziger:1992qr,Zwanziger:1993dh,Dudal:2008sp,Dudal:2007cw,Dudal:2011gd}. Also three-point functions are by now better understood \cite{Schleifenbaum:2004id,Cucchieri:2008qm,Alkofer:2008dt,Ilgenfritz:2006he,Boucaud:2011eh,Fister:2011uw,Huber:2012kd,Pelaez:2013cpa,Aguilar:2013xqa,Aguilar:2013vaa,Blum:2014gna,Eichmann:2014xya} and their equations of motion can be solved self-consistently \cite{Huber:2012kd,Blum:2014gna,Eichmann:2014xya}. The qualitative behavior of propagators and vertices is well captured by standard truncations of functional equations, but their quantitative reliability still needs to be tested and improved. Based on our most recent results for the complete set of two- and three-point functions \cite{Blum:2014gna}, however, there is quite compelling evidence to expect that the system of DSEs truncated to the primitively divergent Green functions might yield a rather good approximation that does not rely on any further external input. The two pieces missing to confirm this are the four-gluon vertex, which was the only remaining model input in such calculations \cite{Blum:2014gna,Eichmann:2014xya}, and the two-loop diagrams in the gluon propagator DSE (of which one also contains the four-gluon vertex). Few direct calculations of the latter exist \cite{Bloch:2003yu,Mader:2013ru,Meyers:2014iwa}, but for the four-gluon vertex information is even more scarce. Here we provide further information from calculating the four-gluon vertex within a state-of-the-art truncation that takes into account its full momentum dependence. An interesting additional feature of this truncation for the four-gluon vertex DSE is that for the first time no model input is required here, since only primitively divergent lower $n$-point Green functions enter at this level which are all known sufficiently well. 

So far, little non-perturbative information on the four-gluon vertex is available to compare with, unfortunately. Even for the three-gluon vertex, available lattice results are limited to very restricted kinematical regions \cite{Cucchieri:2008qm}. Due to the existence of six kinematic variables in the four-gluon vertex (as compared to three for three-point functions) the situation is even much more difficult here. Thus, even if lattice data for the four-gluon vertex will become available in the future, a kinematically reasonably complete coverage will likely remain impossible for some time to come. A continuum method has a clear advantage in this respect, although the resolution of the full kinematic dependence is certainly a challenge here as well. Perturbative results at the symmetric point were presented in Refs.~\cite{Pascual:1980yu,Gracey:2014ola}. Studies beyond perturbation theory can be found in Refs.~\cite{Driesen:1998xc,Kellermann:2008iw,Binosi:2014kka}. In Ref.~\cite{Kellermann:2008iw} the box diagrams were studied in a certain momentum configuration which we will refer to as configuration $A$ below. As input non-perturbative propagators from the so-called scaling type were used \cite{vonSmekal:1997is,vonSmekal:1997vx,Fischer:2002hn}. In Ref.~\cite{Binosi:2014kka} this was extended to include all UV leading one-loop diagrams and propagators of the decoupling type were used as input \cite{Aguilar:2010gm}.

In this work we go beyond these previous studies in several ways. First of all, we take into account the full momentum dependence. This is useful when our results are used as input in future calculations, because  they provide a guideline to develop approximations that still capture the main features but are easier to handle than the full results. The full momentum dependence is also required to solve the equation self-consistently so that we can study the back coupling effects for the vertex. Finally, we use for the first time non-perturbative input for the three-gluon vertex that is in good agreement with lattice data.

In Sect.~\ref{sec:4g-DSE} we fix our notations and present a self-contained derivation of the four-gluon vertex \gls{dse}. This section also contains information on the truncation, the tensor basis, the kinematics and the renormalization. The input we employ is described in Sect.~\ref{sec:input} and our results are presented in Sect.~\ref{sec:results}. We conclude in Sect.~\ref{sec:summary}. Some technical details as regards color contractions, tensor bases, and the numerical calculations to solve the four-gluon vertex DSE can be found in the appendices.

\section{The four-gluon vertex DSE}
\label{sec:4g-DSE}

\subsection{Derivation of the four-gluon vertex DSE}

The Lagrangian density of Yang--Mills theory, fixed to the linear covariant gauge, is
\begin{align}
  \label{eq:LagrYM}
  \mathcal{L}_\text{YM}^\text{eff} &= \mathcal{L}_\text{G} + \mathcal{L}_\text{GF} + \mathcal{L}_\text{FP},\\
  \label{eq:LagrG}
  \mathcal{L}_\text{G} &= \frac{1}{4}F^a_{\mu\nu}F^{a}_{\mu\nu},\\
  \label{eq:LagrGF}
  \mathcal{L}_\text{GF} &= \frac{1}{2\xi} (\partial_\mu A_\mu^a)^2, \\
  \label{eq:LagrFP}
  \mathcal{L}_\text{FP} &= - i (\partial_\mu\bar{c}^a) D_\mu^{ab} c^b,
\end{align}
where $A$ is the gluon field and ($\bar{c}$) $c$ is the (anti-)ghost field. The gauge fixing parameter is denoted by $\xi$. For the Landau gauge it will be set to $0$ later.
The field strength tensor $F_{\mu\nu}^a$ and the covariant derivative $D_\mu^{ab}$ are given by
\begin{align}
   \label{eq:fieldStrengthTensor}
  F^a_{\mu\nu} &= \partial_\mu A_\nu^a - \partial_\nu A_\mu^a - g f^{abc}A_\mu^b A_\nu^c,\\
   D_\mu^{ab} &= \left(\delta^{ab}\partial_\mu + g f^{abc} A_\mu^c\right).
\end{align}
From the Lagrangian density the path integral is defined as
\begin{align}
\label{eq:Z}
\begin{split}
  & Z[J,\,\sigma,\,\bar{\sigma}]\\&\quad = \int \mathcal{D} [Ac\bar{c}]\,
  \exp \Big\{
  	 - \int \text{d}^4 x \, \mathcal{L}_\text{YM}^\text{eff} \\&\quad\quad
  	 + \int \text{d}^4 x \, \left( A_\mu^a J_\mu^a + \bar{\sigma}^a c^a + \bar{c}^a\sigma^a \right)
  \Big\},
  \end{split}
\end{align}
where $J$, $\sigma$ and $\bar{\sigma}$ are the sources for the gluon and ghost fields. For the derivation of Dyson--Schwinger equations the \gls{proper} action will be used, which is obtained from the path integral via a Legendre transformation:
\begin{align}
 \label{eq:effectiveAction}
  \Gamma[\Phi] = - \ln Z[J] + \Phi_i J_i.
\end{align}
Here $J_i$ represents the sources $J$, $\sigma$ and $\bar{\sigma}$ and $\Phi\in\{A_\text{cl},c_\text{cl},\bar{c}_\text{cl}\}$ denotes the classical fields determined by
\begin{align}
 A_\text{cl}=\frac{\delta \ln Z[J]}{\delta J}, \quad c_\text{cl}=\frac{\delta \ln Z[J]}{\delta \bar{\sigma}}, \quad \bar{c}_\text{cl}=\frac{\delta \ln Z[J]}{\delta \sigma}.
\end{align}
In the following we will drop the subscript $cl$ again. For convenience we use a multi index for the fields and sources that includes field species, Lorentz and color indices, and position (or alternatively momenta). Consequently, repeated indices entail summation over the discrete and integration over the continuous variables. For example, $\Phi_i J_i=\int \text{d}^4x (A_\mu^a(x) J_\mu^a(x) + \bar{\sigma}^a(x) c^a(x) + \bar{c}^a(x)\sigma^a(x))$.

The derivation of Dyson--Schwinger equations requires the existence of a well-defined generating functional. Although we use its path integral representation, this is not necessary and other possibilities exist; see, e.g., \cite{Rivers:1988pi}.
Before we derive the master equation, we change to renormalized quantities by introducing the standard renormalization constants for the gluon field ($Z_3$), the ghost field ($\widetilde{Z}_3$), the three-gluon vertex ($Z_1$), the ghost--gluon vertex ($\widetilde{Z}_1$), the four-gluon vertex ($Z_4$), the gauge fixing parameter ($Z_6$), and the coupling ($Z_g$) \cite{Rivers:1988pi}. Due to gauge invariance, these renormalization constants are related. For now we add a superscript R to all renormalized quantities. The renormalized fields $\phi^\text{R}$ are connected to the bare ones by $\phi=Z_\phi^{1/2} \phi^\text{R}$ and the renormalized coupling $g^\text{R}$ to the bare one by $g=Z_g g^\text{R}$. The action in terms of renormalized fields is denoted by $S[\phi^\text{R}]$. 
We start with the integral over the derivative of the path integral which --- assuming no boundary terms exist --- must vanish, viz.
\begin{align}\label{eq:DseStartingPoint}
  \int \mathcal{D}[\phi^\text{R}] \frac{\delta}{\delta \phi^\text{R}_j} \exp\left( -S[\phi^\text{R}] + \phi^\text{R}_i J^\text{R}_i \right) = 0,
\end{align}
where $S[\phi^\text{R}]=\int \text{d} x \mathcal{L}_\text{YM}^{\text{R,eff}}$ is the gauge-fixed, renormalized action and $\phi^\text{R} \in \{A^\text{R}, c^\text{R}, \bar{c}^\text{R} \}$. The sources were rescaled such that they correspond to the sources of the renormalized fields, $J^\text{R}_i=Z^{1/2}_\phi J_i$. Equation~\eqref{eq:DseStartingPoint} can be rewritten to
\begin{align}
  &\left(-\frac{\delta S[\phi^\text{R}]}{\delta \phi^\text{R}_j}\Bigg|_{\phi^\text{R}_j=\delta/\delta J^\text{R}_j}  + J^\text{R}_j \right)\nnnl
  &\quad\quad\int \mathcal{D}[\phi^\text{R}] \exp\left( -S[\phi^\text{R}] + \phi^\text{R}_i J^\text{R}_i \right)=\nnnl
  &\quad \quad \quad =\left(-\frac{\delta S[\phi^\text{R}]}{\delta \phi^\text{R}_j}\Bigg|_{\phi^\text{R}_j=\delta/\delta J^\text{R}_j} + J^\text{R}_j \right)Z[J^\text{R}] = 0.
\end{align}

With \eref{eq:effectiveAction} this can be reformulated to the master equation for the DSEs of \gls{proper} Green functions; see, e.g., \cite{Alkofer:2000wg,Roberts:1994dr,Huber:2011xc}:
\begin{align}
  \label{eq:BasicDseGamma}
  \frac{\delta\Gamma[\Phi^\text{R}]}{\delta \Phi^\text{R}_j} =\left. \frac{\delta S[\phi^\text{R}]}{\delta \phi^\text{R}_j} \right|_{\phi^\text{R}_j=\Phi^\text{R}_j+ D_{ji}^{\text{R},J}\delta/\delta \Phi^\text{R}_i}.
\end{align}
This equation relates a derivative of the renormalized action $S[\phi^\text{R}]$ to a derivative of the renormalized effective action $\Gamma[\Phi^\text{R}]$. Consequently, in addition to renormalized quantities also the renormalization constants appear. However, they are needed to guarantee a consistent renormalization and guarantee the disappearance of the perturbative UV divergences in the equations. Any truncation applied to the equations must not interfere with that. The renormalization scheme used here, discussed in Sect.~\ref{sec:renormalization}, fulfills this property.

The meaning of \eref{eq:BasicDseGamma} is that one should differentiate the action $S[\phi^\text{R}]$ with respect to a field $\phi^\text{R}_i$ and then replace every field $\phi^\text{R}_i$ by the classical field $\Phi^\text{R}_i$ plus a functional derivative multiplied by $D^{\text{R},J}_{ji}$. The \glspl{dse} resulting from this procedure can conveniently be represented by Feynman diagrams.
$D^{\text{R},J}_{ji}$ is the second derivative given by
\begin{align}
	\label{eq:inverseTwoPoint}
	D_{ji}^{\text{R},J}  = \frac{\delta^2 \ln Z[J]}{\delta J^\text{R}_j \delta J^\text{R}_i}.
\end{align}
If we set the sources to zero, this becomes the propagator $D_{ji}^{\text{R},J=0}$. In our case we have the ghost and the gluon propagators,
\begin{align}
 D^{\text{R},ab}(p) &= -\frac{G(p^2)}{p^2} \delta^{ab}, \\
 D^{\text{R},ab}_{\mu\nu}(p) &= P^\text{T}_{\mu\nu}(p)\frac{Z(p^2)}{p^2} \delta^{ab},
\end{align}
where $P^\text{T}$ is the transverse projector, $P^\text{T}_{\mu\nu}(p) = \delta_{\mu\nu} - p_\mu p_\nu/p^2$.

From the master equation \eqref{eq:BasicDseGamma} the \gls{dse} for any $n$-point function can be derived by $n-1$ field derivatives and subsequently setting the sources to zero. For $n=2$ one obtains the \glspl{dse} of the inverse propagators and for $n>2$ the equations for the vertices which we define as\footnote{The advantage of this choice of signs is that the signs of all diagrams are the same except for minus signs from Grassmann loops. However, it is completely arbitrary.}
\begin{align}
 \Gamma^\text{R}_{i_1\ldots i_n} = - \frac{\delta^n\Gamma[\Phi^\text{R}]}{\delta \Phi^\text{R}_{i_1} \ldots \delta \Phi^\text{R}_{i_n}}\Bigg|_{\Phi=0}.
\end{align}
As a characteristic feature every diagram in a \gls{dse} contains a bare vertex, which entails restrictions on the set of possible diagrams.
Due to the self-interaction of the gluons via the three- and the four-gluon vertices, the \gls{dse} of the four-gluon vertex is the four-point function with most terms, namely 60 terms of which there are 20 one-loop and 39 two-loop diagrams. In particular the tensorial structure of the four-gluon vertex is, because of its four color and four Lorentz indices, considerably more complicated than those of the pure ghost or the ghost--gluon four-point functions. So, in principle, the derivation of its \gls{dse} is straightforward but tedious, and we used the \emph{Mathematica} package \emph{DoFun} \cite{Alkofer:2008nt,Huber:2011qr} for this task.

Before we turn to the details of the four-gluon vertex DSE, we drop the superscript R again. In the following, all quantities are renormalized.

\begin{figure*}[tb]
  \includegraphics[width=\textwidth]{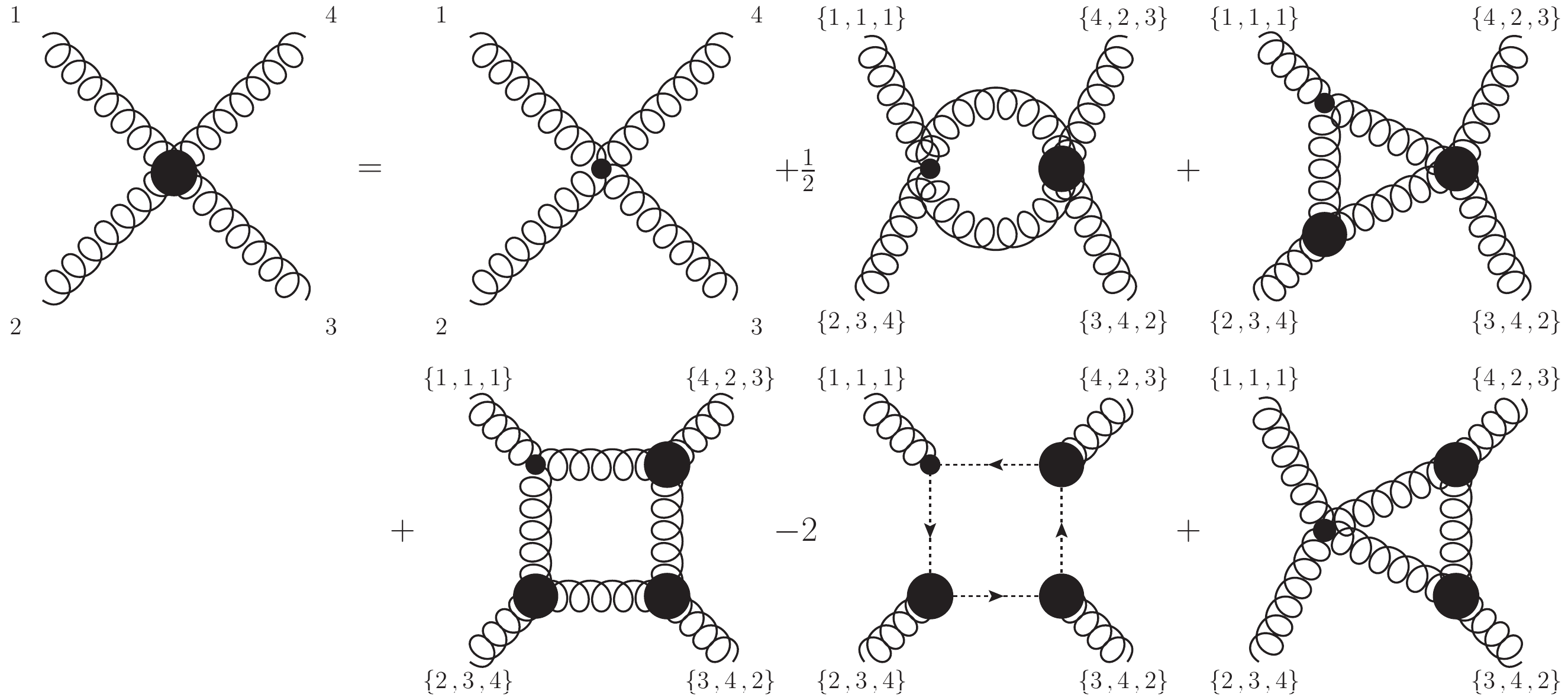}
  \caption{The truncated four-gluon vertex DSE.
  We use the following shorthand notation: \(\{i,\,j,\,k\}\) represents three diagrams where the indices at the first, second and third positions are chosen.
  We call the second and the third diagram on the right-hand side \acrfull{sf} and \acrfull{dt}, respectively.
  The diagrams in the second line are named \acrfull{glb}, \acrfull{ghb} and \acrfull{st} (from left to right).
  Further, we call the \acrlong{sf} and the \acrlong{dt} the dynamic diagrams since they depend on the four-gluon vertex.
  Accordingly, we call the diagrams of the second line static diagrams. Feynman diagrams were created with \textit{JaxoDraw} \cite{Binosi:2003yf}. }
  \label{fig:4g-DSE}
\end{figure*}

With the truncation discussed in Sect.~\ref{sec:truncation}, the DSE for the four-gluon vertex $\Gamma^{abcd}_{\mu\nu\rho\sigma}(p,\,q,\,r,\,s)$ is schematically written as
\begin{align}
\label{eq:four-gluon_DSE}
  \begin{split}
    &\Gamma^{abcd}_{\mu\nu\rho\sigma}(p,\,q,\,r,\,s)\\
    &\quad=\,\Gamma^{{(0)},abcd}_{\mu\nu\rho\sigma}
     +\, \Lambda^{abcd}_{\mu\nu\rho\sigma}(p,\,q,\,r,\,s)
     +\, \Lambda^{acdb}_{\mu\rho\sigma\nu}(p,\,r,\,s,\,q)\\
     &\quad\quad+\, \Lambda^{adbc}_{\mu\sigma\nu\rho}(p,\,s,\,q,\,r)
     +\ldots\,,
  \end{split}
\end{align}
where we used the fact that all one-loop diagrams appear in three permuted versions. Suppressing momentum arguments, the sum of all unpermuted one-loop diagrams, denoted by \(\Lambda^{abcd}_{\mu\nu\rho\sigma}\) in \eref{eq:four-gluon_DSE}, is (see also \fref{fig:4g-DSE}) 
\begin{align}
\label{eq:four-gluon_DSE_one-loop}
  \begin{split}
  &\Lambda^{abcd}_{\mu\nu\rho\sigma} \\&\quad=
      \int \frac{\text{d} k^4}{(2\pi)^4}
    	\Bigg(
       \frac{1}{2} \, ^\text{\acrshort{sf}}\Lambda^{abcd}_{\mu\nu\rho\sigma}
      + \, ^\text{\acrshort{dt}}\Lambda^{abcd}_{\mu\nu\rho\sigma}\\&\quad\quad
      + \, ^\text{\acrshort{glb}}\Lambda^{abcd}_{\mu\nu\rho\sigma}
      - 2 \, ^\text{\acrshort{ghb}}\Lambda^{abcd}_{\mu\nu\rho\sigma}
      + \, ^\text{\acrshort{st}}\Lambda^{abcd}_{\mu\nu\rho\sigma}
      \Bigg).
\end{split}
\end{align}
The subscripts denote the names of the diagrams as explained in \fref{fig:4g-DSE}. The five individual $^i\Lambda^{abcd}_{\mu\nu\rho\sigma}$ are called primitive diagrams from which other diagrams are constructed by permutation. They are given by
\begin{subequations}
\label{eq:Kernels}
\begin{align}
	\label{eq:SwordfishKernel}
  &^\text{\acrshort{sf}}\Lambda^{abcd}_{\mu\nu\rho\sigma}=\, Z_4 \,
  	\Gamma^{{(0)},b''aba'}_{\beta''\mu\nu\alpha'}\,
  	D^{a'a''}_{\alpha'\alpha''}\,
  	\Gamma^{a''cdb'}_{\alpha''\rho\sigma\beta'}\,
  	D^{b'b''}_{\beta'\beta''},\\
  \label{eq:DynamicTriangleKernel}
  &^\text{\acrshort{dt}}\Lambda^{abcd}_{\mu\nu\rho\sigma}= \, Z_1 \,
  	\Gamma_{\gamma''\mu\alpha'}^{{(0)},c''aa'}\,\nnnl&\quad
  	D^{a'a''}_{\alpha'\alpha''}\,
  	\Gamma_{\alpha''\nu\beta'}^{a''bb'}\,
  	D^{b'b''}_{\beta'\beta''}\,
  	\Gamma^{b''cdc'}_{\beta''\rho\sigma\gamma'}\,
  	D^{c'c''}_{\gamma'\gamma''}  , \\
  \label{eq:GluonBoxKernel}
  &^\text{\acrshort{glb}}\Lambda^{abcd}_{\mu\nu\rho\sigma}= \, Z_1 \,
  	\Gamma_{\delta''\mu\alpha'}^{{(0)},d''aa'}\,\nnnl&\quad
  	D^{a'a''}_{\alpha'\alpha''}\,
  	\Gamma_{\alpha''\nu\beta'}^{a''bb'}\,
  	D^{b'b''}_{\beta'\beta''}\,
  	\Gamma_{\beta''\rho\gamma'}^{b''cc'}\,
  	D^{c'c''}_{\gamma'\gamma''}\,
  	\Gamma_{\gamma''\sigma\delta'}^{c''dd'}
  	D^{d'd''}_{\delta'\delta''} ,\\
  \label{eq:GhostBoxKernel}
  &^\text{\acrshort{ghb}}\Lambda^{abcd}_{\mu\nu\rho\sigma}=\, \widetilde{Z}_1 \,
  	\Gamma_\mu^{{(0)},ad''a'}\,\nnnl&\quad
    D^{a'a''}_{}\,
    \Gamma_\nu^{ba''b'}\,
    D^{b'b''}_{}\,
    \Gamma_\rho^{cb''c'}\,
    D^{c'c''}_{}\,
    \Gamma_\sigma^{dc''d'}
    D^{d'd''}_{} ,\\
  \label{eq:GluonTriangleKernel}
  &^\text{\acrshort{st}}\Lambda^{abcd}_{\mu\nu\rho\sigma}=\, Z_4 \,
  	\Gamma^{{(0)},c''aba'}_{\gamma''\mu\nu\alpha'}\,\nnnl&\quad
    D^{a'a''}_{\alpha'\alpha''}\,
    \Gamma_{\alpha''\rho\beta'}^{a''cb'}\,
    D^{b'b''}_{\beta'\beta''}\,
    \Gamma_{\beta''\sigma\gamma'}^{b''dc'}\,
    D^{c'c''}_{\gamma'\gamma''}.
\end{align}
\end{subequations}
Purely gluonic vertices are denoted by $\Gamma$ with the corresponding number of Lorentz and color indices, while the ghost--gluon vertex has only one Lorentz index which is associated to the first color index.
Here the renormalization constants $Z_1$, $\widetilde{Z}_1$ and $Z_4$ of the three-gluon, the ghost--gluon and the four-gluon vertices appear.

\subsection{Asymptotic behavior and truncation}
\label{sec:truncation}

Our truncation scheme consists of two parts: As usual, we discard several diagrams based on their asymptotic behavior. As it happens, all remaining Green functions required for the calculation are already known and we need no model input. However, we further simplify the system by a restriction of the color and Lorentz bases for the four-gluon vertex. The latter aspect of the truncation is discussed in Sect.~\ref{sec:tensors}.

For the diagrammatic truncation of the four-gluon vertex DSE we follow the same arguments as employed for the three-gluon vertex \cite{Blum:2014gna,Eichmann:2014xya}. The main guidelines are the correct IR behavior and the inclusion of all diagrams contributing at one-loop order in the UV. Thus we retain only one-loop diagrams with primitively divergent Green functions. The presence of only one-loop diagrams means also that we do not have to deal with overlapping divergences since there are no sub-diagrams that have internal lines in common. To be precise, there are 20 one-loop diagrams. From them we discard one with a ghost--gluon five-point function, one with a gluonic five-point function, and three ghost triangles with a ghost--gluon four-point function leaving the 15 diagrams representing the $\Lambda$'s in  Eqs.~(\ref{eq:four-gluon_DSE}) and (\ref{eq:four-gluon_DSE_one-loop}), which are depicted in \fref{fig:4g-DSE}. The neglected one-loop diagrams do not contribute to the leading UV order, since they contain non-primitively divergent Green functions. However, they could contribute to the low- and mid-momentum behavior. For now, we adopt it as a working hypothesis that their contributions are smaller than those of the considered diagrams. This is motivated by the observation that, based on a comparison with lattice results, this is true for the three-gluon vertex \cite{Blum:2014gna}. Including such diagrams would require the inclusion of additional DSEs, which is beyond the scope of the present work.

For the discussion of the IR behavior of the vertex we have to elaborate shortly on the IR behavior of Landau gauge Yang--Mills theory in general. It is well known that the system of propagators allows two different types of solutions \cite{Boucaud:2008ji,Fischer:2008uz,Alkofer:2008jy}. One is called decoupling solution and actually consists of a family of solutions that have all in common that the ghost dressing function $G(p^2)$ and the gluon propagator $Z(p^2)/p^2$ become constant in the IR:
\begin{align}
 G(p^2)\rightarrow c, \quad \frac{Z(p^2)}{p^2}\rightarrow m_\text{gl}^2, \quad \text{for } p^2 \rightarrow 0.
 \end{align}
This solution is, besides by the DSE approach \cite{Aguilar:2008xm,Boucaud:2008ji,Fischer:2008uz,Alkofer:2008jy,LlanesEstrada:2012my}, also found by many other methods like lattice calculations, e.g., \cite{Cucchieri:2007md,Cucchieri:2008fc,Sternbeck:2007ug,Bogolubsky:2009dc,Oliveira:2012eh,Sternbeck:2012mf}, with the functional renormalization group \cite{Fischer:2008uz}, within the refined Gribov--Zwanziger framework \cite{Dudal:2007cw,Dudal:2008sp,Dudal:2011gd}, with an effective model \cite{Tissier:2010ts} or in a variational approach \cite{Quandt:2013wna}. From the functional perspective the different solutions are distinguished by the boundary condition imposed for the ghost propagator DSE \cite{Fischer:2008uz}. As a limiting case also the solution $c\rightarrow \infty$ exists. This is called the scaling solution and characterized by a power law behavior of all Green functions \cite{vonSmekal:1997is,Alkofer:2004it,Fischer:2009tn}. The propagator dressing functions can be written as
\begin{align}
 G(p^2)=c_\text{gh} (p^2)^{\delta_\text{gh}}, \quad Z(p^2)=c_\text{gl} (p^2)^{\delta_\text{gl}}.
\end{align}
The exponents $\delta_\text{gh}$ and $\delta_\text{gl}$ are related by $2\delta_\text{gh}+\delta_\text{gl}=0$. Typically they are given in terms of $\kappa:=-\de_\text{gh}$. Its value can be calculated analytically as $\kappa = (93 - \sqrt{1201})/98 \approx 0.6$ \cite{Lerche:2002ep,Zwanziger:2001kw}. This value can only change if the ghost--gluon vertex is not regular in the IR, viz. if its IR limit depends on the angle between two momenta \cite{Lerche:2002ep,Huber:2012zj}. Within modern truncations for the ghost--gluon vertex no such dependence was seen \cite{Huber:2012kd}.

A very convenient feature of the scaling solution is that the qualitative behavior of all Green functions can be derived without truncations by combining the two systems of functional equations given by the \gls{frg} and the \glspl{dse} \cite{Fischer:2009tn}. For a vertex with $2n$ ghost and $m$ gluon legs the dressing functions behave qualitatively as $(p^2)^{(n-m)\kappa}$ where $p$ is an IR momentum scale \cite{Alkofer:2004it,Fischer:2009tn}. Consequently we expect that for this type of solution the four-gluon vertex behaves as $(p^2)^{-4\kappa}$. This behavior is exhibited by all diagrams with a bare ghost--gluon vertex. Within our truncation these are the ghost boxes. However, other diagrams with ghost--ghost--gluon--gluon or ghost--ghost--gluon--gluon--gluon functions exist that have by power counting the same IR behavior but are discarded here. In fact, all diagrams can be classified in terms of their scaling behavior as determined by the bare vertex \cite{Huber:2007kc}.

For the decoupling type of solution such a straightforward classification is not possible. From the three-gluon vertex it is known that it diverges logarithmically in the IR \cite{Pelaez:2013cpa,Aguilar:2013vaa,Blum:2014gna,Eichmann:2014xya} and it was conjectured that this is also true for the four-gluon vertex \cite{Aguilar:2013vaa}. For both quantities these divergences stem from the ghost loops. We will come back to this point in Sect.~\ref{sec:results}.

\subsection{Tensor basis}
\label{sec:tensors}

The four-gluon vertex is undoubtedly the most complicated primitively divergent Green function of Landau gauge Yang--Mills theory. With four color indices it possesses a non-trivial color structure and the four Lorentz indices allow a multitude of Lorentz tensors. We start with a discussion of the former.

As specified in Sect.~\ref{sec:input}, we use only the totally anti-symmetric structure constant $f^{abc}$ for the three-point functions. To our knowledge no proof exists that the totally symmetric color part is non-zero in three-point functions. Thus we neglect the totally symmetric $d^{abc}$ from the beginning, but note that the color tensors we use can partly also be expressed via the $d$ symbols. The building blocks are then the Kronecker delta $\de^{ab}$ and the totally anti-symmetric structure constant $f^{abc}$. The basis constructed from them is
\begin{align}\label{eq:color_tensors_basis}
C^{abcd}_1 &= \delta^{ab}\delta^{cd}, \quad C^{abcd}_2 = \delta^{ac}\delta^{bd}, \quad C^{abcd}_3 = \delta^{ad}\delta^{bc},\nnnl
C^{abcd}_4 &= f^{abn'}f^{cdn'} \quad C^{abcd}_5 = f^{acn'}f^{dbn'}.
\end{align}
Another possible tensor, $C^{abcd}_6 = f^{adn'}f^{bcn'}$, is not included since it can be expressed via the Jacobi identity as $C^{abcd}_6=-C^{abcd}_4-C^{abcd}_5$. Furthermore, contractions of more anti-symmetric structure constants can be reduced to this set. In particular, in the four-gluon vertex DSE terms of the form 
\begin{align}\label{eq:C7}
C^{abcd}_7 & = f^{a'ab'}f^{b'bc'}f^{c'cd'}f^{d'da'}
\end{align}
appear. For $SU(3)$ it reduces to
\begin{align}\label{eq:C7_SU3identity}
 C^{abcd}_7 = \frac{3}{4}\left( C^{abcd}_1 + C^{abcd}_2 + C^{abcd}_3 \right) - C^{abcd}_4 - \frac{1}{2} C^{abcd}_5
\end{align}
and for $SU(2)$ to 
\begin{align}\label{eq:C7_SU2identity}
 C^{abcd}_7 = C^{abcd}_1 + C^{abcd}_3.
\end{align}
For $SU(N)$ with $N>3$ the tensor $C^{abcd}_7$ is linearly independent and must be considered as well. This is shown in Appendix~\ref{sec:app_color}.

The Lorentz space is even more complicated than the color space. If one constructs all possible Lorentz tensors from the metric tensor $\delta_{\mu\nu}$ and the three independent momenta, one arrives at $138$ tensors. They can be split into the following classes:
\begin{align*}
\begin{split}
  & 3 \text{ dimensionless tensors: }\\
  &\quad\delta_{\mu\nu}\delta_{\rho\sigma}\,,\; \delta_{\mu\rho}\delta_{\nu\sigma} \; \text{and} \; \delta_{\mu\sigma}\delta_{\nu\rho}\\
  & 54 \text{ tensors of dim. 2: }\\
  &\quad\delta_{\mu\nu} p^1_\rho p^2_\sigma\,,\; \delta_{\rho\sigma} p^1_\mu p^2_\nu\,,\; \delta_{\mu\rho} p^1_\nu p^2_\sigma\,,\; \delta_{\nu\sigma} p^1_\mu p^2_\rho\,,\; \delta_{\mu\sigma} p^1_\nu p^2_\rho\,,\;\\&\quad \delta_{\nu\rho} p^1_\mu p^2_\sigma\,,\quad p^i \in \{p,\,q,\,r\}\\
  & 81 \text{ tensors of dim. 4: }\\
  &\quad p^1_\mu p^2_\nu p^3_\rho p^4_\sigma\,, \quad p^i \in \{p,\,q,\,r\}
\end{split}
\end{align*}
However, from considerations along the lines of Refs. \cite{Carimalo:1992ia,Eichmann:2011vu,Eichmann:2014xya}, it turns out that there are only $136$ independent tensors \cite{Eichmann:2014pc}. In Landau gauge the completely transverse subspace is sufficient, which still contains 43 linearly independent tensors \cite{Driesen:1998xc}.

For a first study of the vertex within our truncation the full transverse basis is still by far too large. Thus we restrict ourselves here to the tree-level structure of the four-gluon vertex, which is given by
\begin{align}
  \label{eq:bare_four-gluon_vertex}
    \Gamma^{{(0)},abcd}_{\mu\nu\rho\sigma}&(p,q,r,s) =\nnnl
      -g^2 & \Big[\;\, 				\left(f^{acn'}f^{bdn'}-f^{adn'}f^{cbn'}\right)\delta_{\mu\nu}\delta_{\rho\sigma}\nnnl
           & +     \left(f^{abn'}f^{cdn'}-f^{adn'}f^{bcn'}\right)\delta_{\mu\rho}\delta_{\nu\sigma}\nnnl
           & +     \left(f^{acn'}f^{dbn'}-f^{abn'}f^{cdn'}\right)\delta_{\mu\sigma}\delta_{\rho\nu}\Big]
\end{align}
and we replace all full four-gluon vertices by
\begin{align}
 \Gamma^{abcd}_{\mu\nu\rho\sigma}(p,q,r,s)=\Gamma^{{(0)},abcd}_{\mu\nu\rho\sigma}D^\text{4g}(p,q,r,s).
\end{align}
The non-perturbative information is contained in the dressing function $D^\text{4g}(p,q,r,s)$. This approximation of the full vertex is motivated by two things: First, this strategy worked very well for the three-gluon vertex, where it was explicitly shown that other dressing functions are severely suppressed \cite{Eichmann:2014xya}. Second, as we will show in Sect.~\ref{sec:results} by considering also additional dressings, it turns out a posteriori that the dressing of the tree-level tensor provides the largest contribution of all calculated dressings. Thus we expect this to be a good first approximation of the full vertex, while a calculation with the full tensor basis is left to future studies.

\subsection{Bose symmetry}

A truncated DSE in general does no longer reflect all the symmetries of the full equation. For gluonic Green functions the Bose symmetry is of special importance. If we simply went ahead and calculated the truncated DSE \eref{eq:four-gluon_DSE}, the results would not be symmetric under the exchange of the leg attached to the bare vertices and another one. This would entail that the results depend on the way the four-gluon vertex is fed back into the DSE. Thus it is necessary to symmetrize the results. We want to stress that this effect comes from using dressed vertices in our setup and would be absent if all vertices were bare. The straightforward way for symmetrization is to average over the four DSEs with different momenta at the legs attached to the bare vertices. Within our truncation this corresponds to calculating all possible permutations of the primitive diagrams. The actual number of diagrams can be reduced using inherent symmetries of the equation. Nevertheless, this would lead to an increase in complexity and computing time, which we will avoid as explained below.

First of all, we consider how to extract the dressing function $D^\text{4g}(p,q,r,s)$ from the DSE given in \eref{eq:four-gluon_DSE}. For this the following projection is employed that explicitly gets rid of all longitudinal parts (momentum arguments are suppressed on the right-hand side):
\begin{equation}
	\label{eq:projected_one-loop}
		L(p,\,q,\,r,\,s) := 
		\frac{
			\Lambda^{abcd}_{\mu\nu\rho\sigma} \; P^\text{T}_{\mu\mu'} \, P^\text{T}_{\nu\nu'} \, P^\text{T}_{\rho\rho'} \, P^\text{T}_{\sigma\sigma'} \, \Gamma^{{(0)},abcd}_{\mu'\nu'\rho'\sigma'}
		}{
			\Gamma^{{(0)},efgh}_{\alpha\beta\gamma\delta} \; P^\text{T}_{\alpha\alpha'} \, P^\text{T}_{\beta\beta'} \, P^\text{T}_{\gamma\gamma'} \, P^\text{T}_{\delta\delta'} \, \Gamma^{{(0)},efgh}_{\alpha'\beta'\gamma'\delta'}
		}.
\end{equation}
Note that \(\Lambda^{adbc}_{\mu\sigma\nu\rho}\) does not represent the sum of all one-loop terms of the four-gluon vertex \gls{dse} but appears in three permuted versions in the \gls{dse}; see \eref{eq:four-gluon_DSE}.
From \eref{eq:projected_one-loop} one can obtain the symmetrized,  transversely projected tree-level dressing function by
\begin{align}
  \label{eq:effective_one-loop_DSE}
  \begin{split}
    & D^{\text{4g}}(p,\,q,\,r,\,s)\\&\quad = Z_4 + \frac{1}{4} \Big[
      L(p,\,q,\,r,\,s) +
      L(q,\,r,\,s,\,p) +\\&\quad\quad
      L(r,\,s,\,p,\,q) +
      L(s,\,p,\,q,\,r) +\\&\quad\quad
      L(p,\,r,\,s,\,q) +
      L(q,\,s,\,p,\,r) +\\&\quad\quad
      L(r,\,p,\,q,\,s) +
      L(s,\,q,\,r,\,p) +\\&\quad\quad
      L(p,\,s,\,q,\,r) +
      L(q,\,p,\,r,\,s) +\\&\quad\quad
      L(r,\,q,\,s,\,p) +
      L(s,\,r,\,p,\,q)
    \Big].
  \end{split}
\end{align}
Equation (\ref{eq:effective_one-loop_DSE}) is obtained from \eref{eq:four-gluon_DSE} by projecting it onto the transverse tree-level structure and symmetrizing it. Naively, one would expect \(4!=24\) terms. However, this number reduces to 12 since some diagrams turn out to be identical. The reasons are the Bose symmetry of the four-gluon vertex itself and the irrelevance of the direction of the loop momentum; see Fig.~\ref{fig:4g-DSE}.

To reduce the computational effort we did not calculate all $L$ explicitly. Rather, we computed the normalized one-loop expression, given by \eref{eq:projected_one-loop}, and from that the dressing function with \eref{eq:effective_one-loop_DSE}. In other words, calculating $L(p,\,q,\,r,\,s)$ with full momentum dependence gives us access to all other variants of $L$ appearing in \eref{eq:effective_one-loop_DSE} so that $D^{\text{4g}}(p,\,q,\,r,\,s)$ can be computed from the calculation of only one $L$.

We explicitly tested what happens when no symmetrization is employed and found that there is a considerable impact on the results. Furthermore, we want to mention that no transverse projection was employed in Ref.~\cite{Kellermann:2008iw}. Thus our results for the ghost and gluon boxes cannot directly be compared to theirs. Unfortunately, the transverse projection also increases the complexity of the four-gluon \gls{dse} significantly, by about an order of magnitude.

\subsection{Kinematics}

\begin{table*}
\def\arraystretch{2.75}
\centering
\begin{tabular}{lC{2.6cm}C{2.6cm}C{2.6cm}}
	Configuration & $\boldsymbol{A}$ & $\boldsymbol{B}$ & $\boldsymbol{C}$
\\ 
	Definition
	&
	\specialcell{
  	\(\;\,S^2 =R^2 =Q^2 =p^2\) \\
  	\(\theta_r = \theta_q = \psi_q = 0\)
 }
	&
	\specialcell{
  	\(\;S^2 =R^2 =Q^2 =p^2\) \\
  	\(\theta_r = \theta_q = \psi_q =  \frac{\pi}{2}\)
 	}
  &
	\specialcell{
  	\(\;\;\;S^2 = R^2 =p^2,\quad Q^2=2p^2\) \\
  	\(\theta_r = \frac{\pi}{2},\quad\theta_q=\frac{\pi}{4}, \quad \psi_q = 0\)
	} 
\\ 
  Visualization
  &  
  \specialcell{ \\
	  \begin{tikzpicture}
	 	  \draw[-stealth] (0,0) -- (0,1);
	 	  \node[align=left] at (0.,-0.35) {\(s\)};
	 	  \draw[-stealth] (0.5,0) -- (0.5,1);
	   	\node[align=left] at (0.5,-0.35) {\(r\)};
	   	\draw[-stealth] (1,0) -- (1,1);
	   	\node[align=left] at (1,-0.35) {\(q\)};
	   	\draw[-stealth,dashed] (1.5,1.5) -- (1.5,-1.5);
	   	\node[align=left] at (1.75,-0.35) {};
	  \end{tikzpicture}
  }
  &  
  \specialcell{ \\
  	\begin{tikzpicture}
	    \draw[-stealth] (0,0) -- (0,2);
			\node[align=left] at (0.25,1) {\(q\)};
			\draw (0,0.5) arc (90:0:0.5);
			\draw[fill] (45:0.25) circle[radius=0.5pt];
			\draw[-stealth] (0,0) -- (2,0);
			\node[align=left] at (1,-0.35) {\(r\)};
			\draw (0.6,0) arc (0:-135:0.6);
			\draw[fill] (-67.5:0.25) circle[radius=0.5pt];
			\draw[-stealth] (0,0) -- (-0.7,-0.7);
			\node[align=left] at (-1,-1) {\(s\)};
			\draw (0,0.4) arc (90:225:0.4);
			\draw[fill] (157.5:0.25) circle[radius=0.5pt];
	  \end{tikzpicture}
	}
  &
  \specialcell{ \\
	  \begin{tikzpicture}
	  	\draw[-stealth] (0,0) -- (0,1);
	  	\node[align=left] at (-0.25,0.5) {\(s\)};
	   	\draw[-stealth] (0,0) -- (1,0);
	   	\node[align=left] at (0.5,-0.3) {\(r\)};
	   	\draw[-stealth] (0,0) -- (1,1);
	   	\node[align=left] at (1.2,1.2) {\(q\)};
	   	\draw[-stealth,dashed] (0,0) -- (-2,-2);
	   	\node[align=left] at (-1,-0.7) {};
	  \end{tikzpicture}
	}
\\
 
\end{tabular}
\caption[Special configurations]{Definitions of the kinematic configurations $A$, $B$ and $C$ used in plots. The dashed lines represent the fourth momentum vector given by momentum conservation.}
\label{tab:SpecialConfigs}
\end{table*}

Another new aspect of our investigation is that we take into account the full momentum dependence of the vertex. This is due to the large number of kinematic invariants a considerably complex task. The vertex depends on three independent momenta, say, $s$, $r$ and $q$, from which six kinematic invariants can be formed, e.g., $s^2$, $r^2$, $q^2$, $s\cdot r$, $s\cdot q$ and $r\cdot q$. However, this choice has the disadvantage that the domains of the latter three are not independent. 
Hence, we directly use spherical coordinates to describe the three momenta at the cost of having some sets of coordinates that describe the same momentum vectors.\footnote{E.g., the vectors defined in \eref{eq:coordinates} do not depend on \(\psi_q\) if \(\theta_q=0\).} Exploiting the $O(4)$ symmetry we define
\begin{align}
  \label{eq:coordinates}
\begin{split}
  & s=S\colvec{4}{1}{0}{0}{0}, \quad
  r=R\colvec{4}{\cos(\theta_r)}{\sin(\theta_r)}{0}{0},\\
  & q=Q\colvec{4}{\cos(\theta_q)\phantom{\cos(\psi_q)}}{\sin(\theta_q)\cos(\psi_q)}{\sin(\theta_q)\sin(\psi_q)}{0},
\end{split}
\end{align}
where \(S,\,R,\,Q \in \mathbb{R}^+\) and \(\theta_r,\,\theta_q,\,\psi_q \in [0,\,\pi]\,\). From \eref{eq:coordinates} it is easy to see that the domain of $q\cdot r$ is not $[-Q R, Q R]$ but depends on the other angles. For example, for $\theta_r=\pi/2$ we have $q\cdot r=Q R \sin(\theta_q) \cos(\psi_q)$. This in general cannot be rewritten into a form $Q R \cos(\alpha_r)$ with $\alpha_r\in [0,\pi]$.
As a second example consider $\theta_r=0$. Then $q\cdot r=Q R\,\cos(\theta_q)$, but $\theta_q$ is already the free angle from $s\cdot q$. Thus we prefer to work with the angle $\psi_q$ instead of $r\cdot q$.

The dressing function itself is defined on a six-dimensional grid of the variables $s^2$, $r^2$, $q^2$, $\theta_r$, $\theta_q$ and $\psi_q$. Typically we use $15$ points for the radial and $7$ points for the angular variables. When we have a converged solution also points on a finer grid are calculated.

For visualization of our results we have to fix some variables. In Table~\ref{tab:SpecialConfigs} three kinematic configurations are shown, which will be used later. Configuration $A$ is the one also used in Refs. \cite{Kellermann:2008iw,Binosi:2014kka}.

\subsection{Renormalization}
\label{sec:renormalization}

The integrals of the four-gluon vertex DSE are logarithmically divergent. However, since we are using input that was obtained within the \textit{MiniMOM} scheme \cite{vonSmekal:1997is,vonSmekal:2009ae}, we are not free to subtract these divergences via a momentum subtraction. Within that scheme the renormalization constant of the ghost--gluon vertex $\widetilde{Z}_1$ is fixed to $1$ for the Landau gauge. The ghost and gluon propagators, on the other hand, were obtained from a self-consistent calculation which also entails certain values for their renormalization constants $\widetilde{Z}_3$ and $Z_3$, respectively. The \glspl{sti} then fix the renormalization constants of the three- and four-gluon vertices as
\begin{align}
Z_1&=\widetilde{Z}_1 Z_3/\widetilde{Z}_3=Z_3/\widetilde{Z}_3, \\
Z_4&=\widetilde{Z}_1^2 Z_3/\widetilde{Z}_3^2=Z_3/\widetilde{Z}_3^2.
\end{align}
The corresponding values can be found in Table~\ref{tab:paras}.

We use the given value for $Z_4$ in the tree-level expression of the vertex. However, the renormalization constants in front of the integrals in \eref{eq:Kernels} are treated differently. Motivated by the gluon propagator equation, where this is necessary to obtain the correct anomalous dimension, we replace the renormalization constants $Z_1$ and $Z_4$ by momentum dependent functions \cite{vonSmekal:1997vx,Huber:2012kd}. The purpose of these functions is to effectively add a UV dressing to the bare vertices which brings the equation in line with the \gls{rg}. The main requirement for these functions is to possess the correct UV behavior. This can be achieved by the following ansatz in case of $Z_4$:
\begin{align}
\label{eq:RGI_four-gluon}
 Z_4 \rightarrow D^{\text{4g}}_\text{\acrshort{rg}}(p,\,q,\,r,\,s) & =
G\left( \bar{p}^2 \right)^{\alpha_{4\text{g}}}
Z\left( \bar{p}^2 \right)^{\beta_{4\text{g}}}
\end{align}
where the momentum $\bar{p}$ is defined as $\bar{p}^2 = (p^2+q^2+r^2+s^2)/2$. The factor $2$ appears, because in the integral $\bar{p}$ approaches for high loop momenta the loop momentum itself. The exponents $\alpha_{4g}$ and $\beta_{4g}$ are determined from the anomalous dimensions of the ghost propagator ($\delta=-9/44$), the gluon propagator ($\gamma=-13/22$) and the four-gluon vertex ($\gamma_{4g}=2\delta-\gamma=2/11$) from the requirement $\alpha_{4g} \delta+\beta_{4g}\gamma=\gamma_{4g}$. As a second condition serves the IR finiteness of $D^{\text{4g}}_\text{\acrshort{rg}}(p,\,q,\,r,\,s)$ \cite{Huber:2012kd}. Solving for the exponents yields
\begin{subequations}\label{eq:Exponents}
\begin{align}
\alpha_{4\text{g}} & = -2-8\delta, &
\beta_{4\text{g}} & = -1-4\delta, &
	\, \text{(scaling);}\\
\alpha_{4\text{g}} & = 4 +1/\delta, &
\beta_{4\text{g}} & = 0, &
	\, \text{(decoupling).}
\end{align}
\end{subequations}
For $Z_1$ the replacement is given by \cite{Huber:2012kd}
\begin{align}
 Z_1  \rightarrow D^{3\text{g}}_\text{\acrshort{rg}}(p,\,q,\,r)  =
			G\left( \bar{p}^2 \right)^{\alpha_{3\text{g}}}
			Z\left( \bar{p}^2 \right)^{\beta_{3\text{g}}}
\end{align}
where $\bar{p}^2$ is $(p^2+q^2+r^2)/2$ and 
\begin{subequations}
\begin{align}
\alpha_{3\text{g}} & = -2-6\delta, &
\beta_{3\text{g}} & = -1-3\delta , &
	\, \text{(scaling);}\\
\alpha_{3\text{g}} & = 3 + 1/\delta, &
\beta_{3\text{g}} & = 0 , &
	\, \text{(decoupling).}
\end{align}
\end{subequations}

\section{Input}
\label{sec:input}

\begin{figure*}[tb]
\centering
  \includegraphics[width=0.38\textwidth]{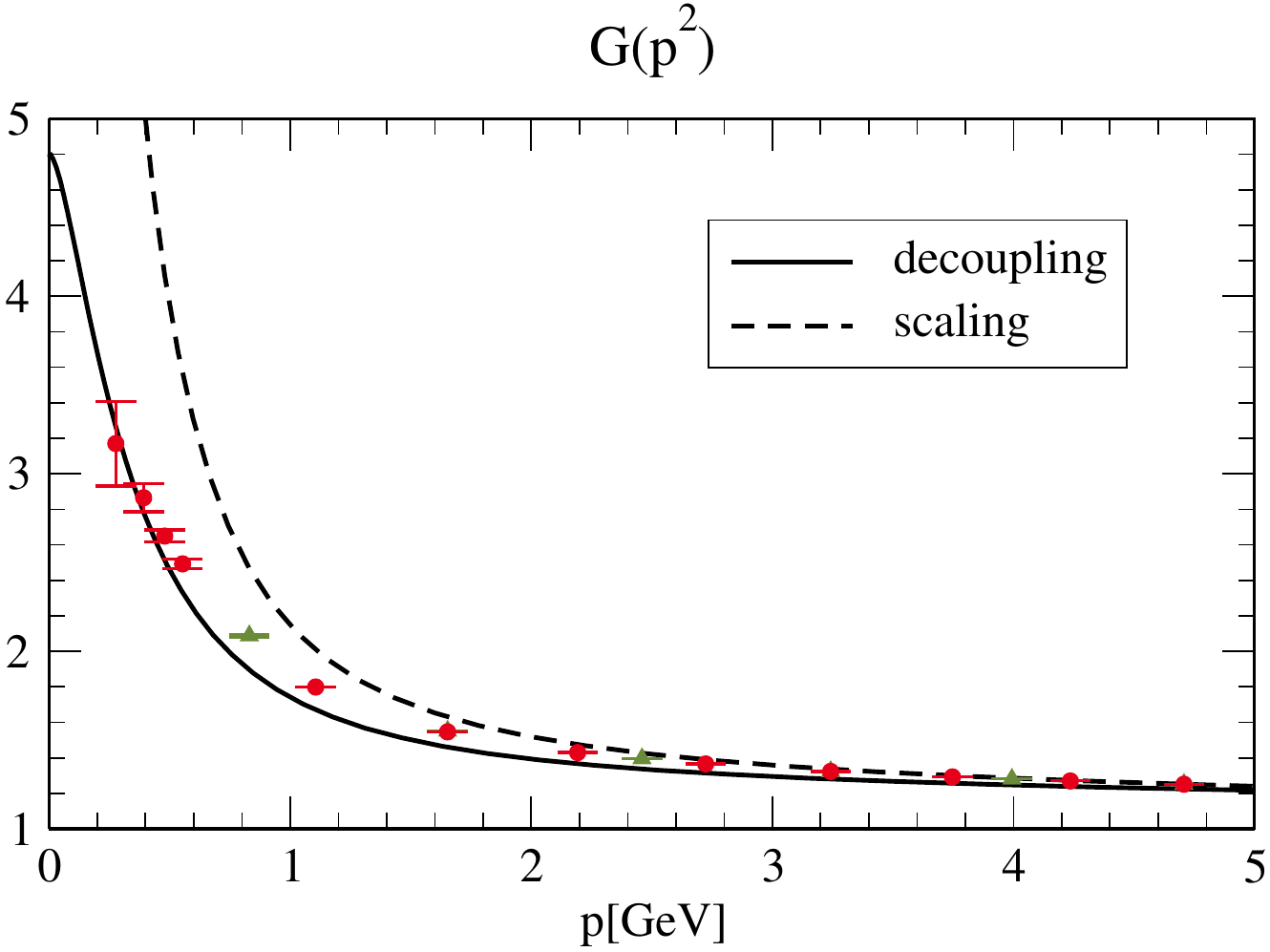}
  \hspace{0.06\textwidth}
  \includegraphics[width=0.38\textwidth]{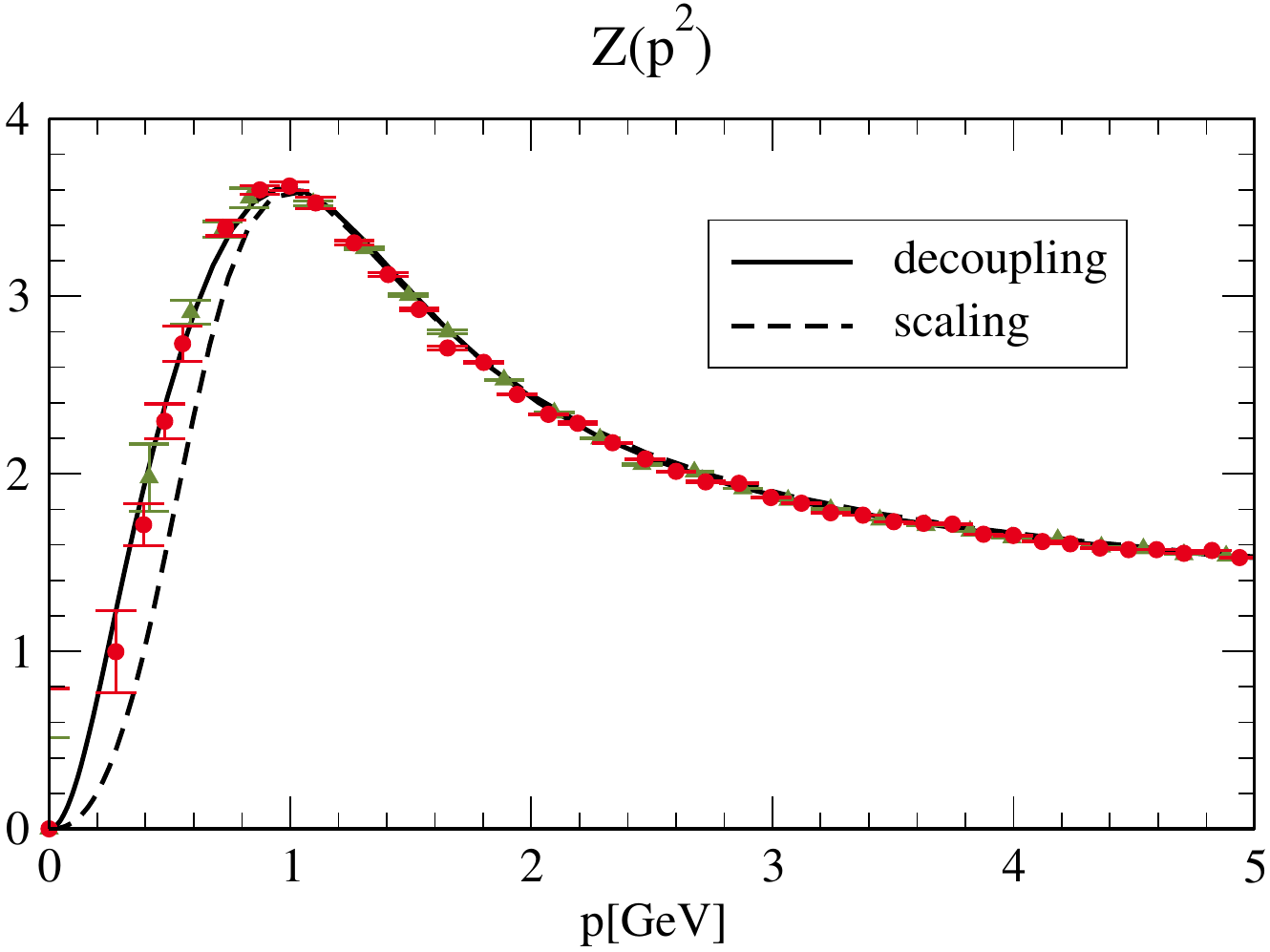}
  \caption{Propagator dressing functions for decoupling (solid line) \cite{Huber:2012kd} and scaling (dashed line) \cite{Huber:2014tva} in comparison to lattice data \cite{Sternbeck:2006rd} with $\beta=6$ and lattice sizes of $L=32$ (green) and $L=48$ (red).}
  \label{fig:input_props}
\end{figure*}

The four-gluon vertex DSE is calculated with input from other calculations whose results are in good agreement with lattice data. However, this was achieved by the dependence of the corresponding calculations on higher Green functions which were modeled based on the existing information from several sources. One important feature of our input is that we consider all important tensor structures. For the propagators and the ghost--gluon vertex this is trivially satisfied, because they each possess only one relevant tensor structure in the Landau gauge. For the three-gluon vertex there are four transverse tensors. However, we consider here only the one derived from the tree-level tensor, as it was shown in Ref.~\cite{Eichmann:2014xya} that the other three are severely suppressed in all momentum regimes.

\begin{figure*}[tb]
\includegraphics[width=0.32\textwidth]{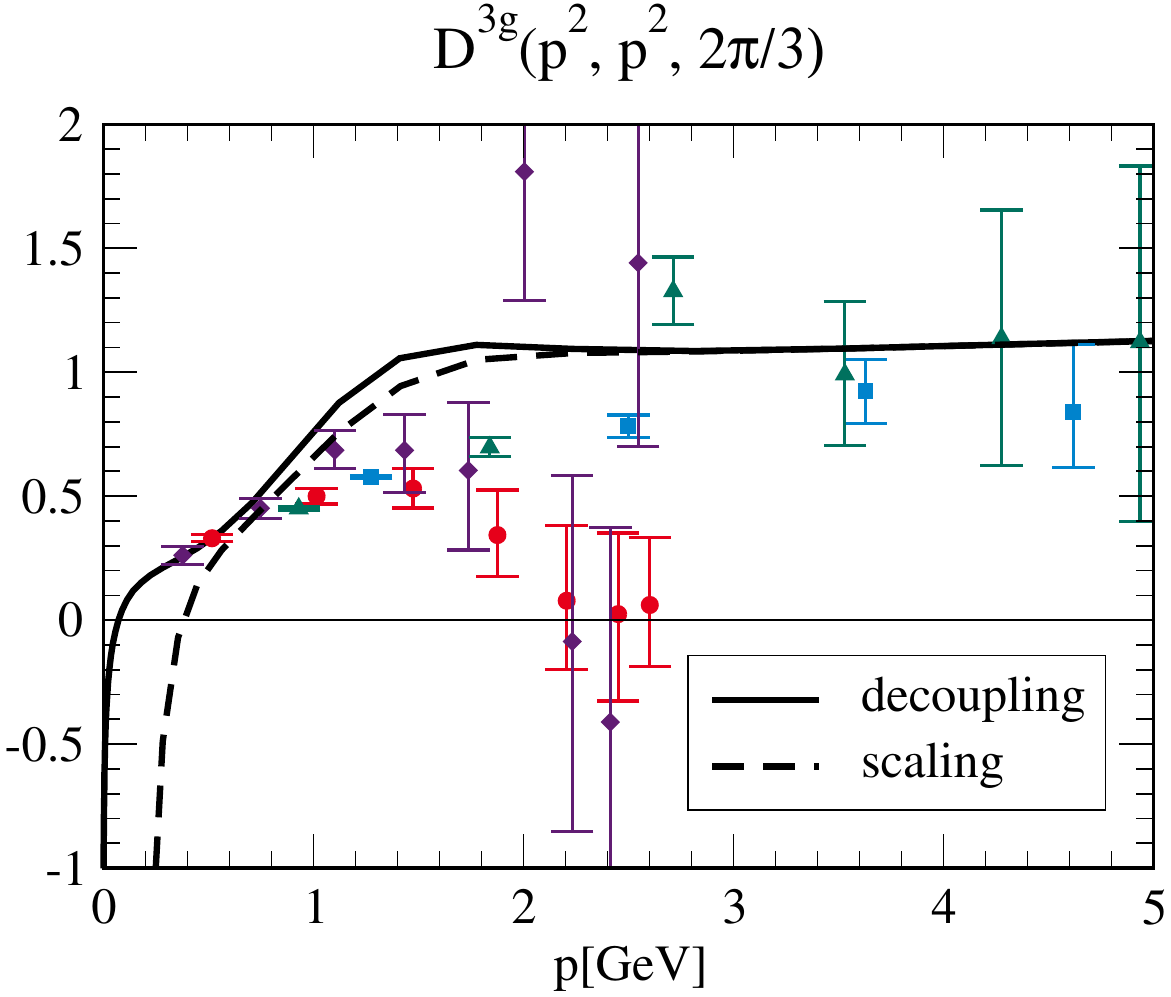}
 \includegraphics[width=0.32\textwidth]{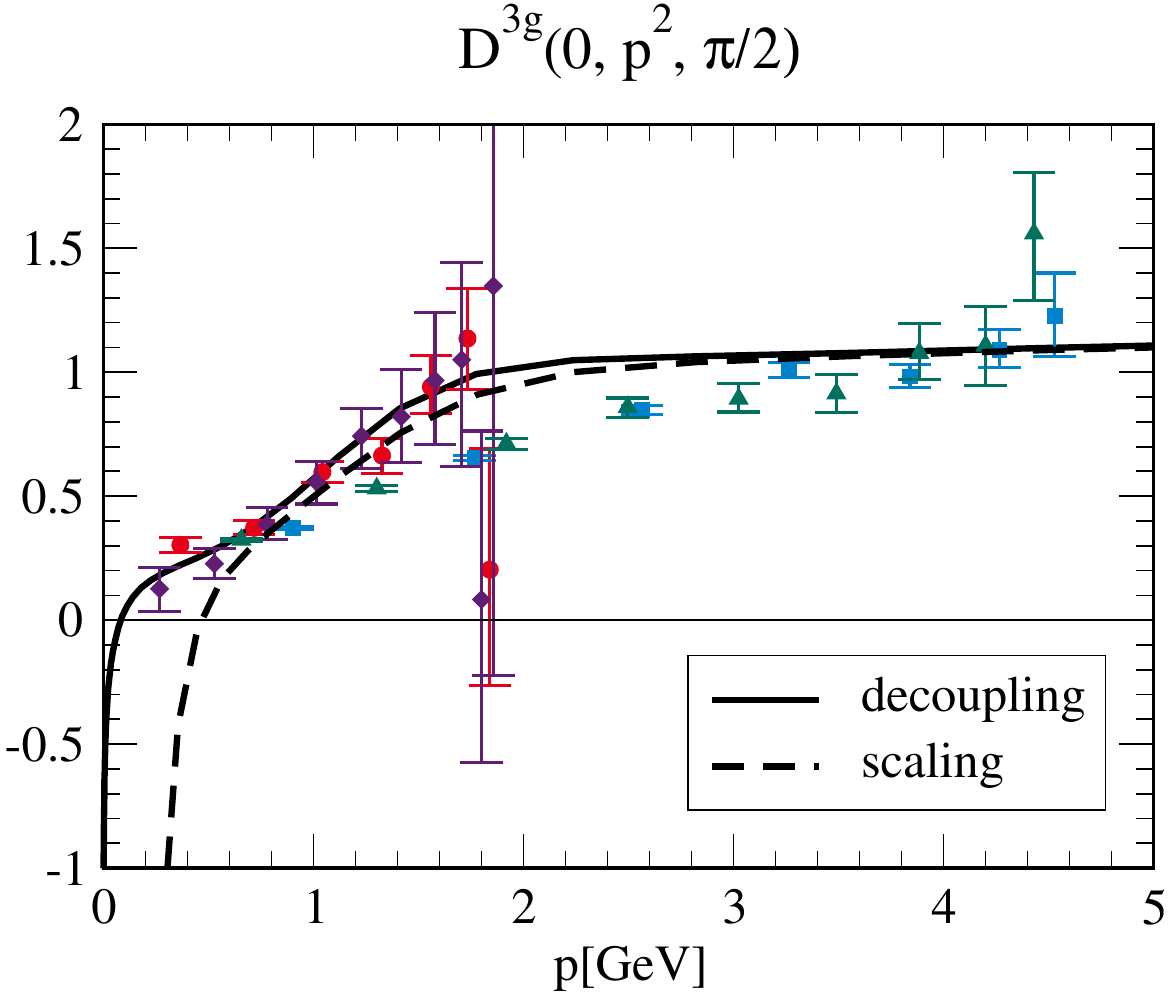}
 \includegraphics[width=0.32\textwidth]{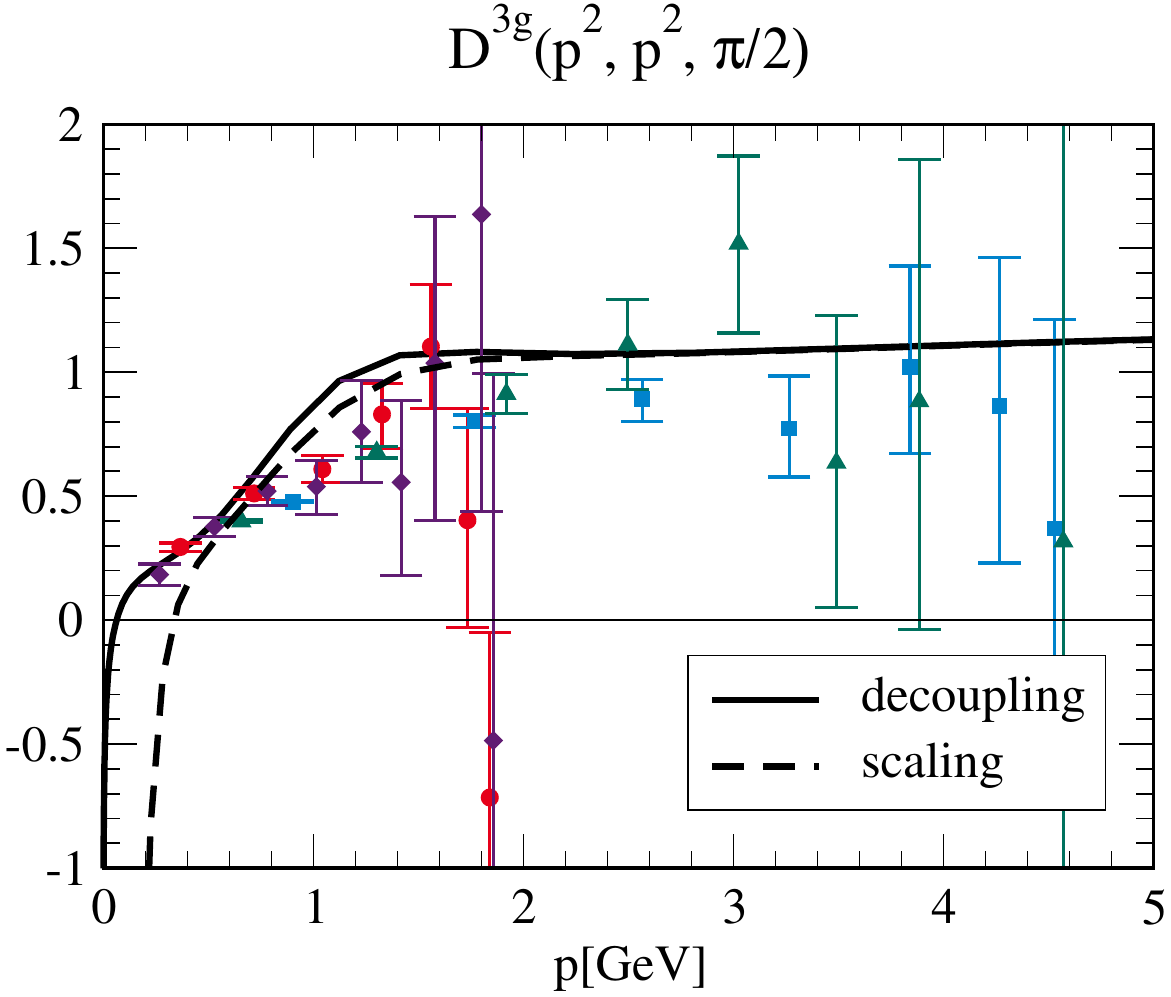}
 \caption{\label{fig:input_3g}Three-gluon vertex dressing function from Ref.~\cite{Blum:2014gna} (decoupling) and calculated with the scaling solution propagators obtained along the lines of Ref.~\cite{Huber:2014tva} in comparison with lattice data \cite{Cucchieri:2008qm} where different colors/symbols refer to different values of $\beta\in \{2.2, 2.5\}$ and different lattice sizes $1.4$ fm $<L<4.7$ fm. The solid lines correspond to the decoupling and the dashed lines to the scaling solution.}
\end{figure*}

The propagators we use for the decoupling case stem from Ref.~\cite{Huber:2012kd}, where the ghost--gluon vertex was dynamically included and an optimized effective three-gluon vertex was used. The results for the propagators, obtained for $\alpha(\mu)=g^2/4\pi=1$, are shown in \fref{fig:input_props}. Although the ghost--gluon vertex was also calculated there, we employ the bare vertex here. The reason is that it enters only in a static diagram. From the three-gluon vertex, where this is also the case, we know that the influence on the results is minor. We illustrate in \fref{fig:GhostBox} that this also holds true for the four-gluon vertex. On the other hand, for the propagators this is different and the mid-momentum regime of the gluon propagator is affected by this choice \cite{Blum:2014mt}.
For scaling we use as input data obtained along the lines of \cite{Huber:2014tva}. The corresponding dressing functions are shown in \fref{fig:input_props}.

The ghost--gluon vertex is described completely by one dressing function alone due to the transversality of the Landau gauge:
\begin{align}
 \Gamma_\mu^{abc}(k;p,q):=i\,g\,f^{abc} P^\text{T}_{\mu\nu}(k)p_\nu D^{A\bar{c}c}(k;p,q).
\end{align}
The gluon momentum is denoted by $k$ and the \mbox{(anti-)}ghost momentum by ($q$) $p$ and $P^\text{T}_{\mu\nu}(k)$ is the transverse projector. Except where mentioned in Sect.~\ref{sec:ghost_box} we use $D^{A\bar{c}c}(k;p,q)=1$. We note that there is also a longitudinal dressing function for the ghost--gluon vertex which is constrained by the Slavnov--Taylor identities. However, it decouples from the transverse part, as discussed in detail in Ref.~\cite{Fischer:2008uz}, and it is not required here.

The three-gluon vertex has four transverse tensors, but here we consider only the tree-level tensor and denote the full three-gluon vertex by
\begin{align}
 \Gamma_{\mu\nu\rho}^{abc}(p,q,k)=\Gamma_{\mu\nu\rho}^{{(0)},abc}(p,q,k) D^\text{3g}(p,q,k),
\end{align}
where the tree-level tensor is given by
\begin{align}
\begin{split}
  &\Gamma_{\mu\nu\rho}^{(0),abc}(p,\,q,\,r)\\&\quad =
   -i g f^{abc} \left[ (p-q)_\rho \delta_{\mu\nu} + (q-r)_\mu \delta_{\nu\rho} + (r-p)_\nu \delta_{\mu\rho} \right].
   \end{split}
\end{align}
The dressing function $D^\text{3g}(p,q,k)$ contains the non-perturbative information. Note that the restriction to the tree-level tensor is a very good approximation as demonstrated by an explicit calculation of the other dressing functions \cite{Eichmann:2014xya} which were found to be very small. For the longitudinal part the same argument applies as for the ghost--gluon vertex and we do not consider it here.

The data we use for the decoupling three-gluon vertex was calculated in Ref.~\cite{Blum:2014gna} with the propagators discussed before. Again, good agreement with lattice results was found as shown in \fref{fig:input_3g}. For that calculation a model for the four-gluon vertex was required; see \eref{eq:FourGluonVertexModelDec} below. However, we want to emphasize that coupling this vertex back into the gluon propagator reduces the agreement with lattice results again. Thus we expect that two-loop diagrams are important in the gluon propagator DSE, see also \cite{Bloch:2003yu,Mader:2013ru,Meyers:2014iwa}. The input we have here, on the other hand, can be interpreted essentially as equivalent to existing lattice results. For the scaling solution we calculated the three-gluon vertex from the propagators using the same four-gluon vertex model. Note that although the IR behavior of the corresponding diagram is not correct then, the IR behavior of the three-gluon vertex itself is because it is determined by the ghost triangle diagram. The scaling results for the three-gluon vertex are shown in \fref{fig:input_3g}.

All input quantities are rescaled in the plots, because the shown lattice data is not renormalized. In our calculations, on the other hand, we used the renormalized input data. The importance of consistently renormalized input data is discussed in Sect.~\ref{sec:results}.

\begin{table}
\begin{center}
\begin{tabular}{l||c|c|c|c|c}
 & $Z_3$ & $\widetilde{Z}_3$ & $Z_1$ & $Z_4$ & $\Lambda^2 [GeV^2]$\\
 \hline\hline
 decoupling & 4.528 & 1.714 & 2.642 & 1.541 & 48530\\
 scaling & 3.930 & 1.552 & 2.532 & 1.632 & 18814
\end{tabular}
\end{center}
\caption{\label{tab:paras} Renormalization constants and cutoffs for the used decoupling \cite{Huber:2012kd} and scaling input \cite{Huber:2014tva}.}
\end{table}

\section{Results}
\label{sec:results}

\subsection{Ghost box}
\label{sec:ghost_box}

\begin{figure*}
\centering
  \includegraphics[width=0.48\textwidth]{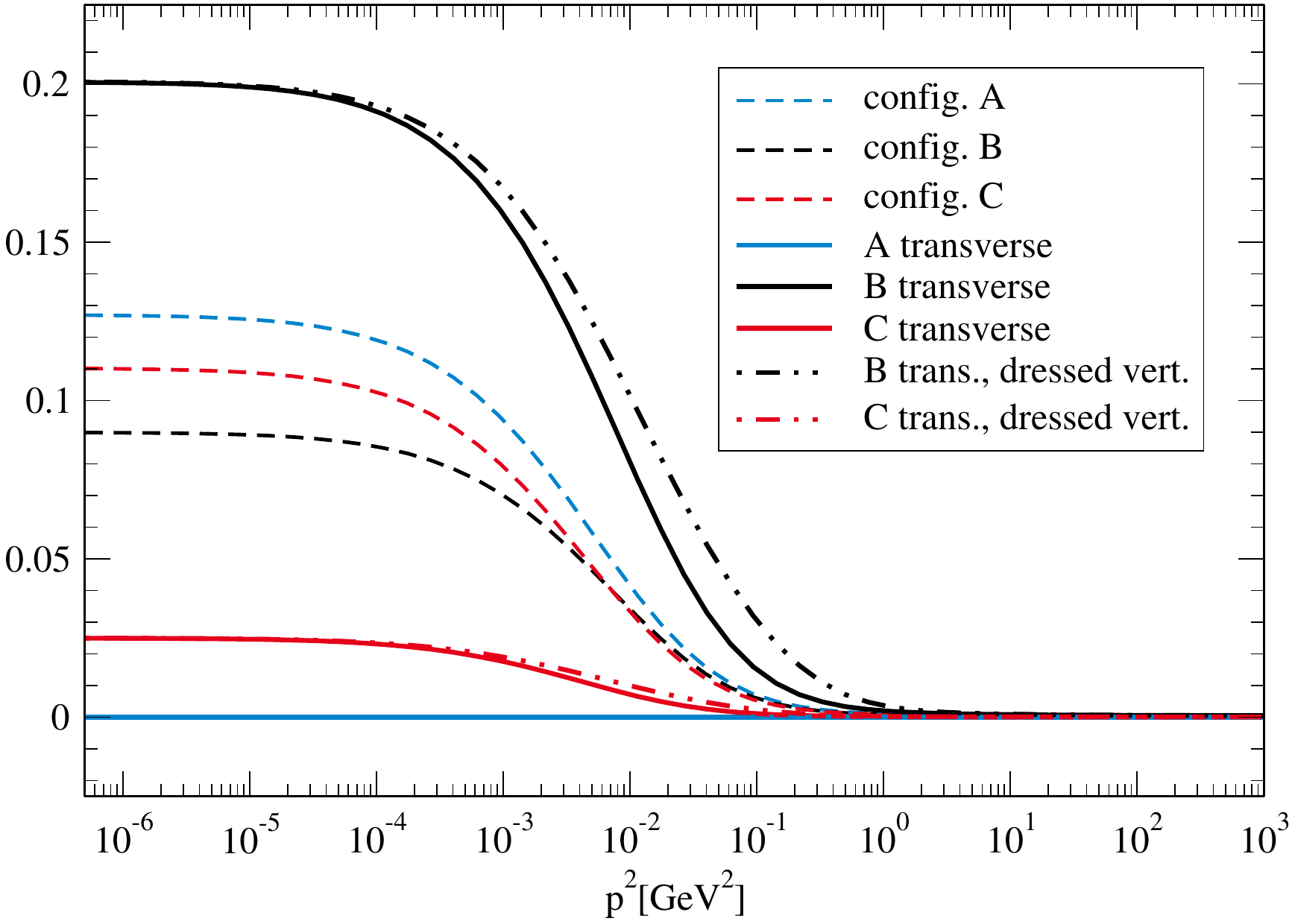}
  \hfill
  \includegraphics[width=0.48\textwidth]{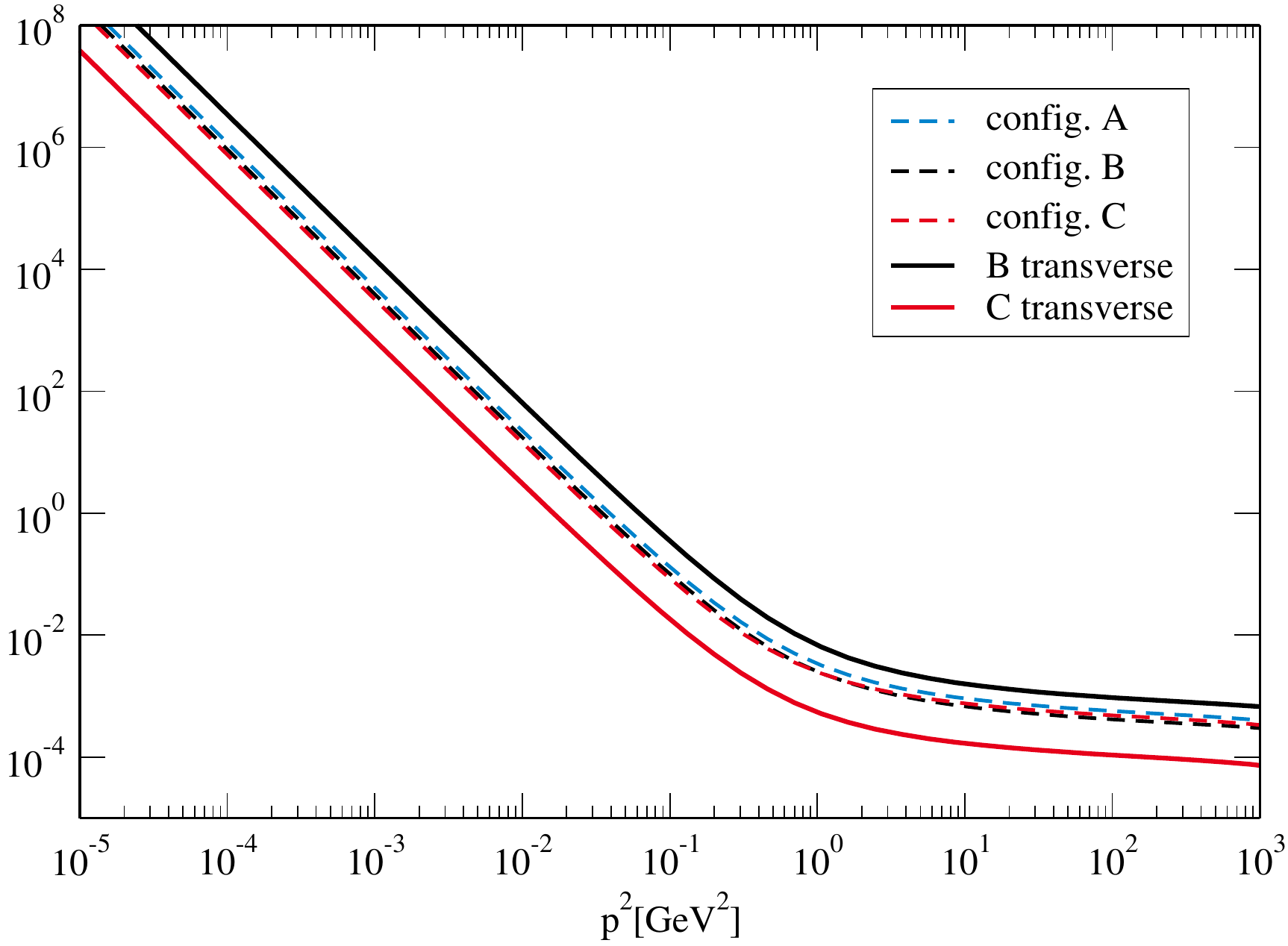}
  \caption[Ghost box contribution]{Symmetrized ghost box contributions for the configurations defined in Table~\ref{tab:SpecialConfigs} for decoupling and scaling (left/right). The continuous lines corresponds to the projector in \eref{eq:projected_one-loop} that projects onto the transverse part only. The dashed lines were obtained by dropping the transverse projectors in \eref{eq:projected_one-loop}. The dot-dashed lines were obtained with a full ghost--gluon vertex.}
  \label{fig:GhostBox}
\end{figure*}

Since it is expected that the ghost box yields the IR leading contribution to the four-gluon vertex both for scaling and decoupling solutions \cite{Alkofer:2004it,Fischer:2009tn,Aguilar:2013vaa} we start by a dedicated analysis of this diagram. No iteration is necessary and we can calculate specific kinematic configurations with a very high precision.

The ghost box contribution for the three special configurations defined in Table~\ref{tab:SpecialConfigs} is shown in Fig.~\ref{fig:GhostBox}.
To obtain the symmetrized results for configuration $C$, we calculate the (two) distinct permutations and then take the (weighted) average.
The other configurations need not be symmetrized since no distinct permutations exist.
The symmetrization for configuration $C$ is especially important if the transversely projected decoupling ghost box is calculated. In this case, contributions of different permutations diverge logarithmically in the \gls{ir} but the (weighted) sum approaches a finite value. Thus we confirm the finiteness of the tree-level dressing function beyond configuration $A$, for which this was already found in Ref.~\cite{Binosi:2014kka}. Indeed our calculations show that the ghost box is IR finite for all momentum configurations.

In \fref{fig:GhostBox} we also illustrate the effect of the transverse projection by showing results obtained from the projector given in \eref{eq:projected_one-loop} without transverse projections. As it turns out, this can have a sizable quantitative but not qualitative effect: The form of the curves stays the same, but the transverse projection can increase (config. B) as well as decrease (config. A and C) the contribution of the ghost box.

The influence of a dressed ghost--gluon vertex was also studied and is shown in \fref{fig:GhostBox}. For this we used the ghost--gluon vertex from Ref.~\cite{Huber:2012kd}. Note that with a dressed ghost--gluon vertex the symmetrization requires the calculation of more diagrams than in the case of a bare vertex. A dressed ghost--gluon vertex always leads to an increase in the region below $\SI{1}{GeV}$, but compared to the contributions of the other diagrams, discussed below, this is only a minor effect.

It can be shown analytically that the ghost box vanishes completely under transverse projection for configuration $A$.
It was this configuration that was used in Ref.~\cite{Kellermann:2008iw} to determine the \gls{ir} scaling fixed point of the running coupling. We will come back to this in Sect.~\ref{sec:runningCoupling}.
It is noteworthy that the angular dependence, viz., the dependence on the configuration, is stronger if the ghost box is transversely projected.

\subsection{Full calculation}

We will now turn to the results from the self-consistent solution of the truncated four-gluon vertex DSE. For this we take into account the complete momentum dependence and solve the equation by a fixed point iteration as explained in Appendix~\ref{sec:tech_details}.

\begin{figure*}[tb]
	\centering
 \includegraphics[width=0.48\textwidth]{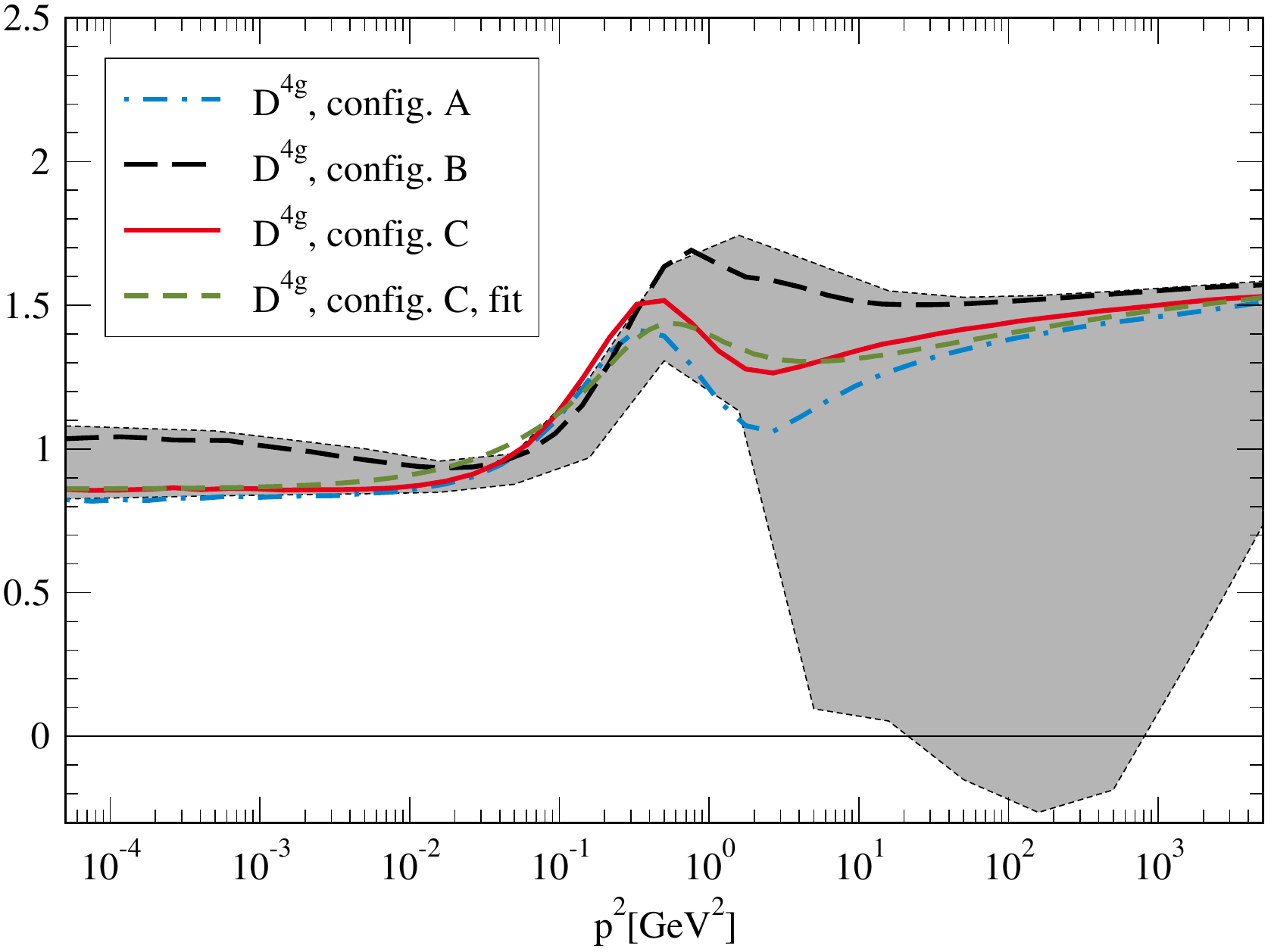}
 \includegraphics[width=0.48\textwidth]{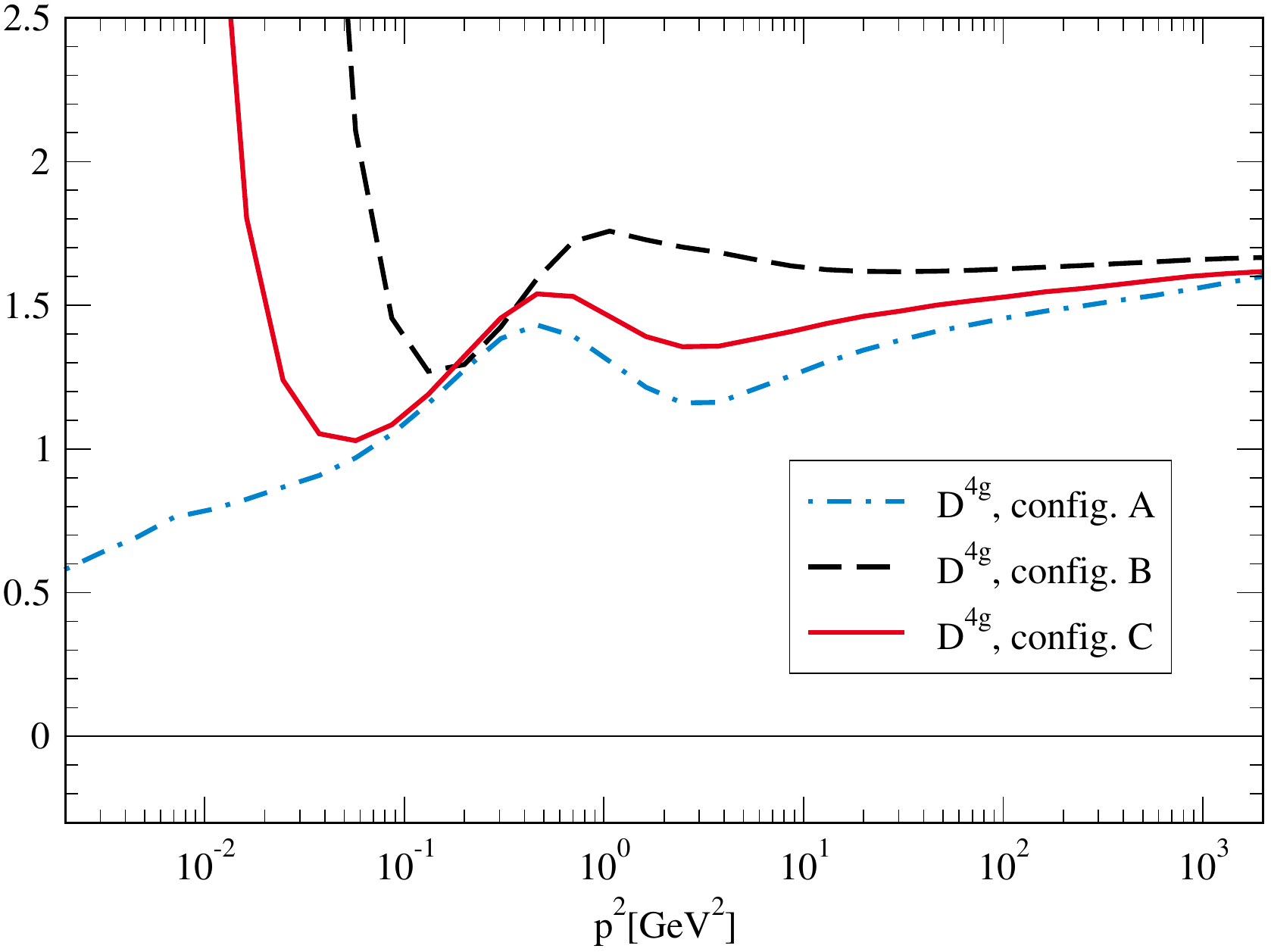} 
  \caption{Four-gluon vertex dressing function for three momentum configurations and its angle dependence indicated by the colored gray area. For the angle dependence a coarser grid was used than for the configurations and the points were only connected to guide the eye.
  The particular configurations can lie slightly outside the gray area. For example, in the case of configuration $C$ this is due to the strong rise in the mid-momentum regime and the fact that one of the squared momenta of configuration $C$ (see Table~\ref{tab:SpecialConfigs}) is indeed higher than \(p^2\).
  The dashed line corresponds to the fit to \eref{eq:FourGluonVertexModelDec}.
  Left: decoupling solution. Right: scaling solution.
  }
  \label{fig:Comparison}
\end{figure*}

\begin{figure*}[tb]
	\centering
	\begin{tabular}{cc}
		\hspace{1cm}\strong{Decoupling}
	&
		\hspace{1cm}\strong{Scaling}
		\\
		\vspace{0.1cm}
		\\
		\includegraphics[width=0.45\textwidth]{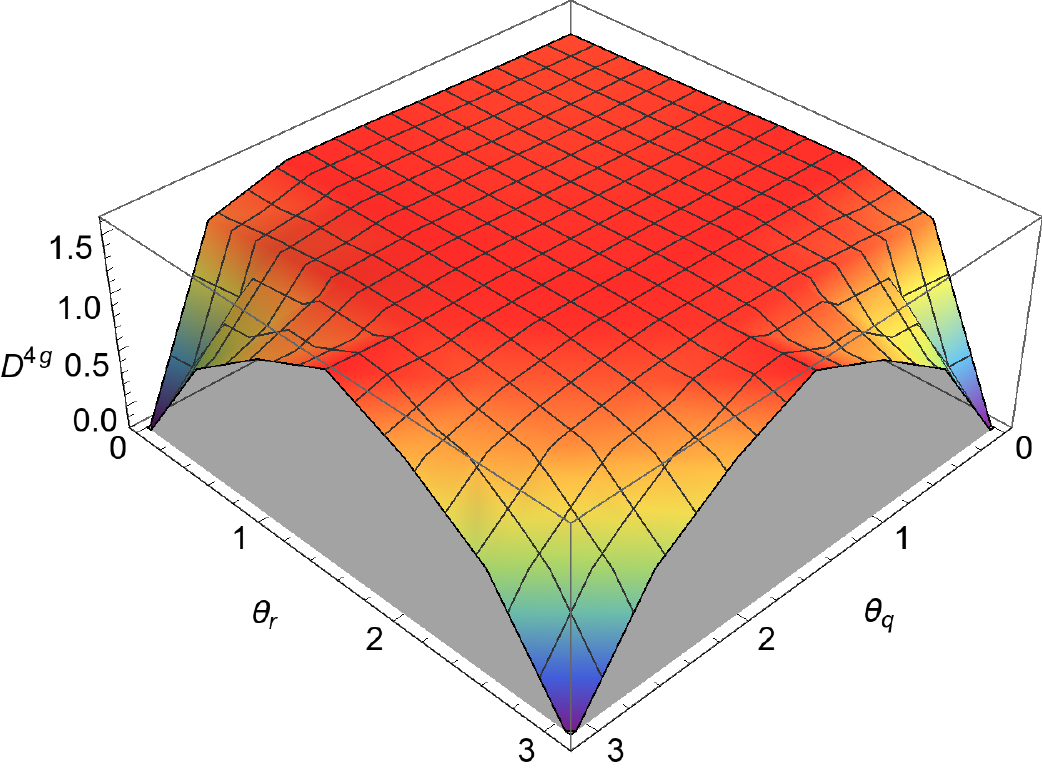}
	&  
  	\includegraphics[width=0.45\textwidth]{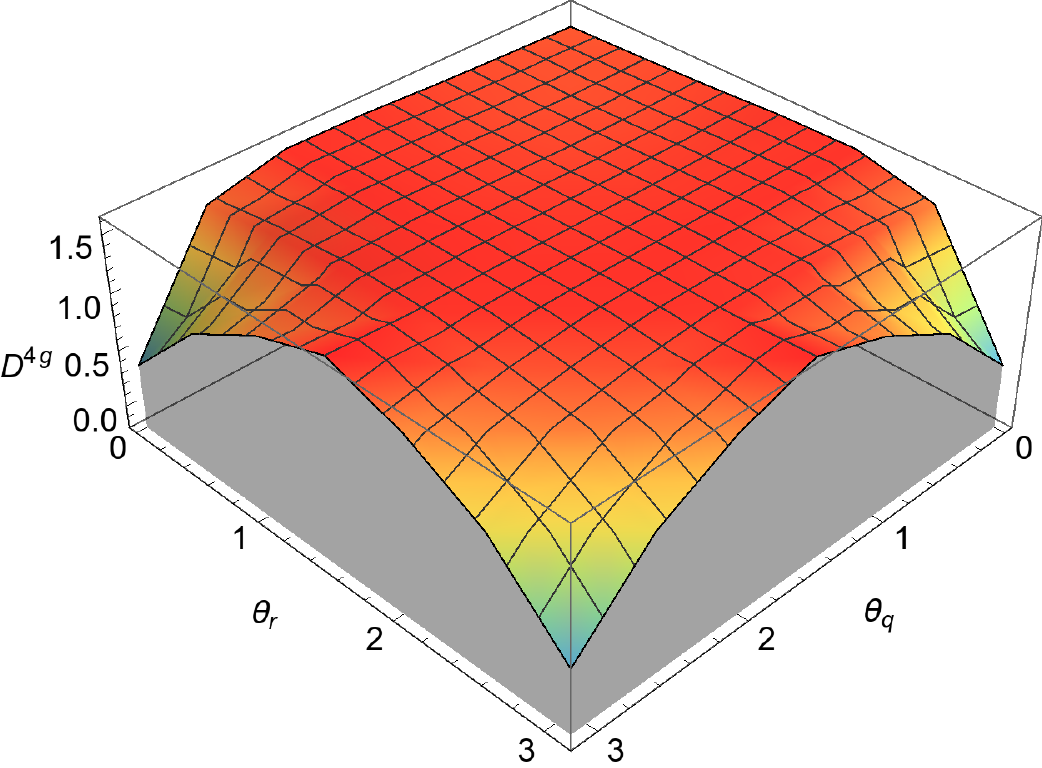} 
  	\\
  	\vspace{0.25cm}
		\\  	
  	\includegraphics[width=0.45\textwidth]{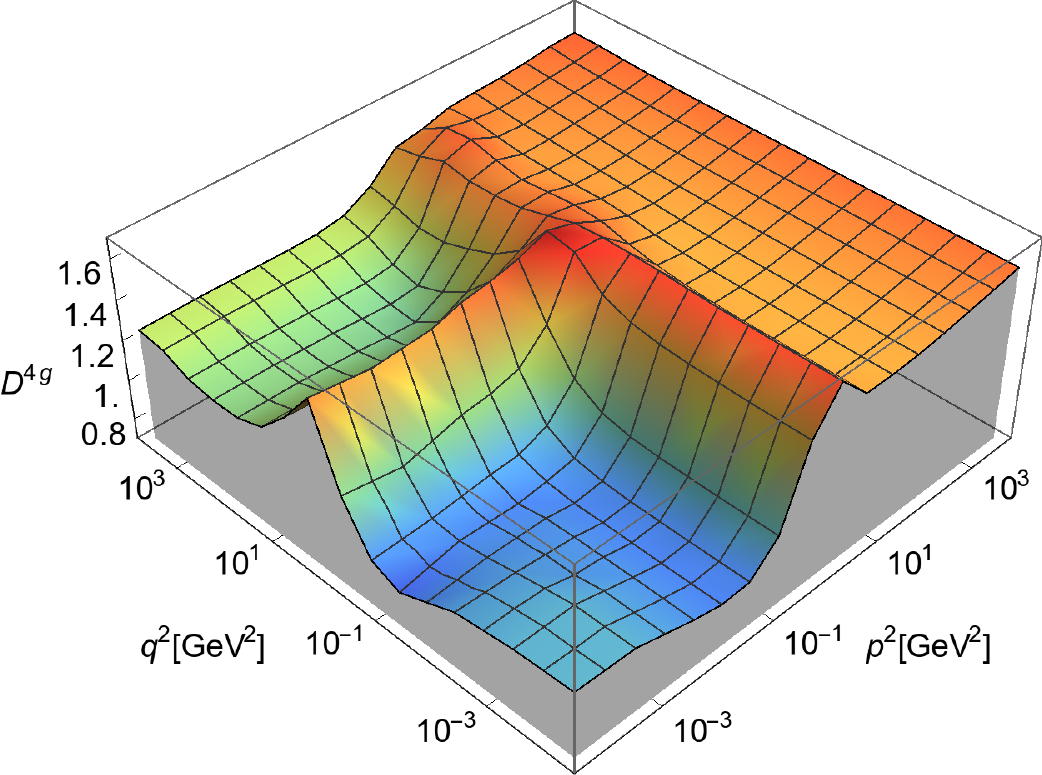}
	& 
		\includegraphics[width=0.45\textwidth]{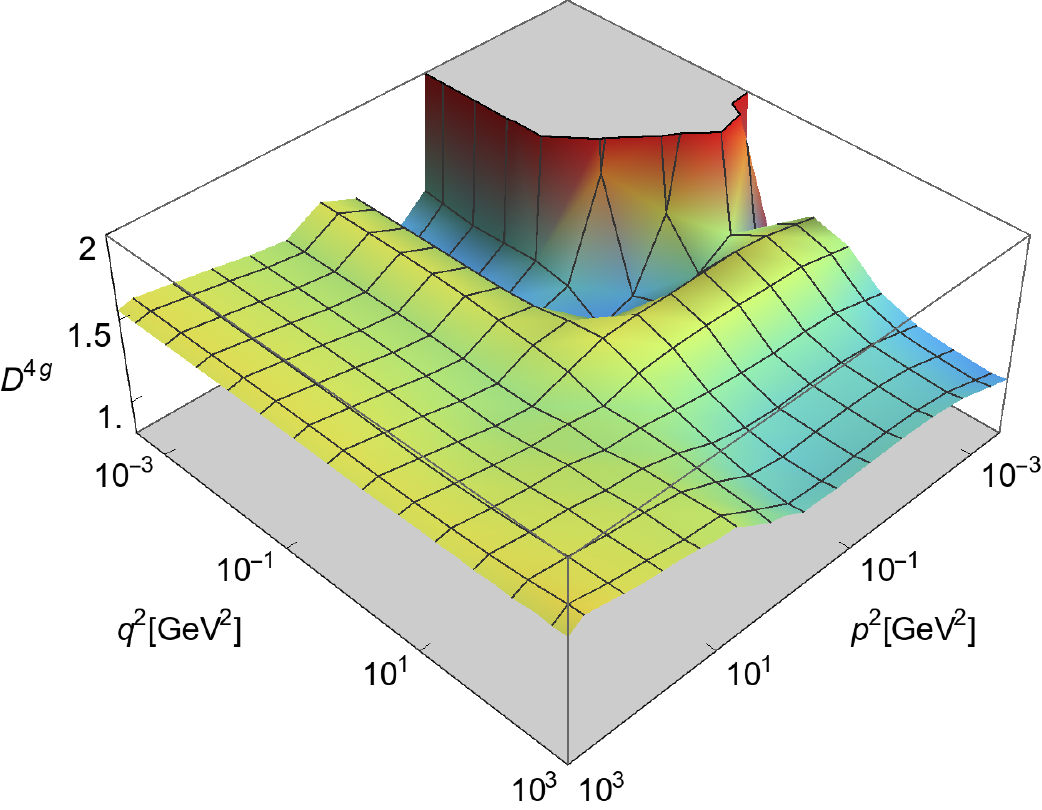}
	\end{tabular} 
  	
  \caption[3D plots]{Top: angular dependence of the dressing function. The squared momenta and the angle \(\psi_q\) are kept constant: \(S^2=R^2=Q^2=\SI{160}{\GeV\squared}\), \(\psi_q=0\). The shown configuration corresponds to the point with the largest angle dependence we found; see also \fref{fig:Comparison}.
  Bottom: momentum dependence of the dressing function: \(S^2=R^2=p^2\) and \(Q^2=q^2\).
  For the angles we chose \(\theta_r=\theta_q=\psi_q=\pi/2\) (decoupling) and \(\theta_r=\pi/4\), \(\theta_q=\pi/2\) and \(\psi_q=0\) (scaling).
  Note that the lower plots are shown from different viewpoints.}
  \label{fig:3DPlots}
\end{figure*}

\begin{figure}[tb]
	\centering
  \includegraphics[width=0.48\textwidth]{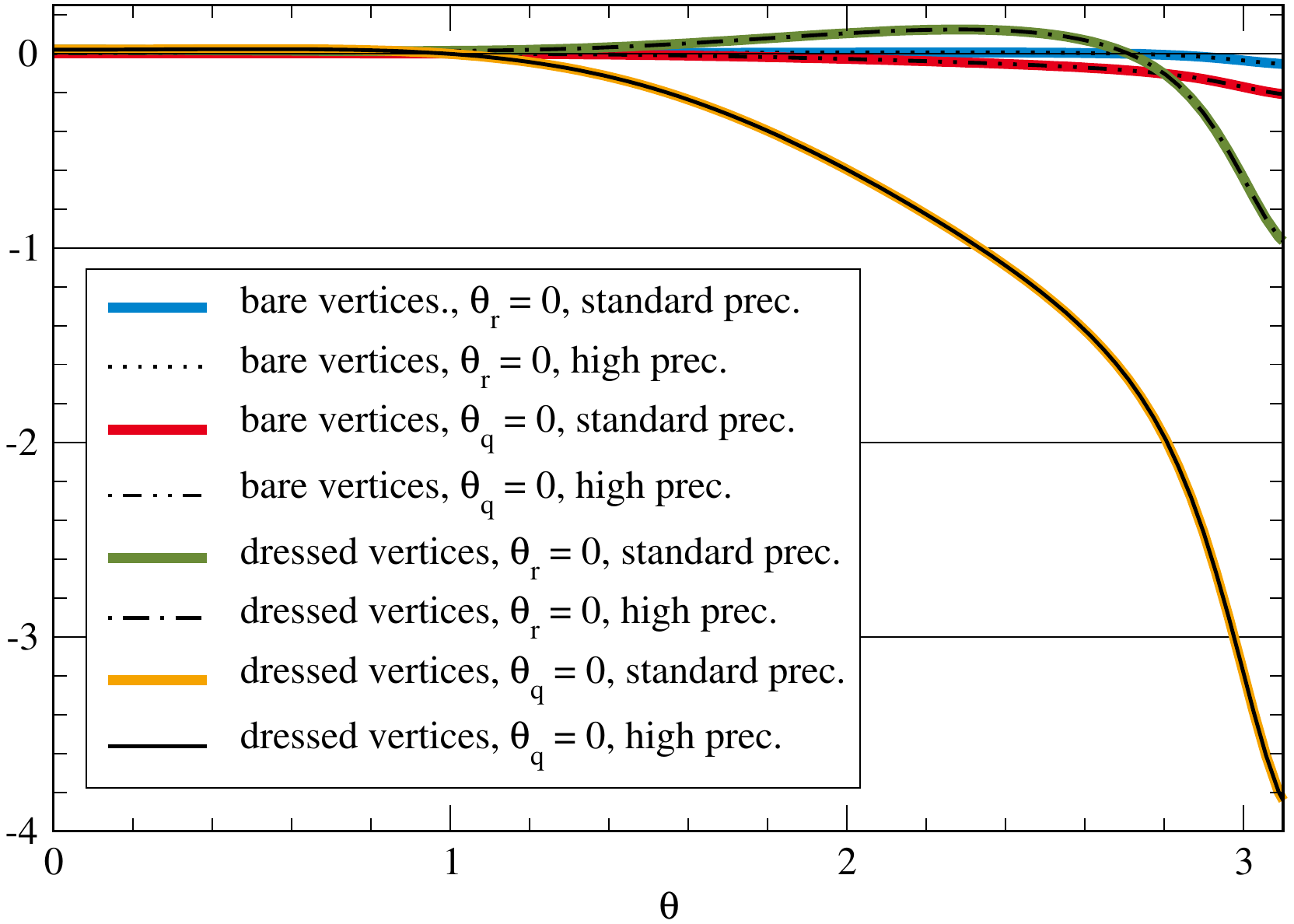}
  \caption[Unsymmetrized Gluon box]{Angle dependence of a single gluon box (no symmetrization employed) where $S^2=R^2=Q^2=\SI{50}{\GeV\squared}$ and $\psi_q=0$ are fixed. $\theta_q$ is varied while $\theta_r$ is kept fixed at $0$ and vice versa as indicated in the legend. Also shown is the effect a dressed three-gluon vertex has compared to a bare one: it enhances the angle dependence. For standard/high precision, 30/50 (12/25) nodes per radial (angular) integration region were used. Clearly the standard precision is sufficient as the lines lie on top of each other.}
  \label{fig:GluonBoxCriticalConfig}
\end{figure}

The results for the four-gluon vertex dressing function are shown in Figs.~\ref{fig:Comparison} and \ref{fig:3DPlots}. The former shows the kinematic configurations $A$, $B$ and $C$. The shaded area indicates the angle dependence. To determine it we used the configuration
\begin{equation*}
	S^2=R^2=Q^2 = p^2
\end{equation*}
and took the minimal and maximal values for the dressing when varying the angles.
However, this area has to be interpreted with a grain of salt, because what we plot as the boundaries of this area is determined by a few extreme points whereas the majority of the points lie around the solid lines. To illustrate this we show three-dimensional plots in \fref{fig:3DPlots} where configurations with the largest angle dependence are shown. Note that the typical angle dependence is much smaller. The main origin of the angle dependence is the gluon box. To check that this is not a numeric artifact, we calculated the gluon box for momenta and two angles fixed while varying the third angle. This calculation was repeated with increased numeric precision. The results, shown in \fref{fig:GluonBoxCriticalConfig}, clearly illustrate that increasing the numeric precision has no impact. Figure~\ref{fig:GluonBoxCriticalConfig} also shows the effect of a dressed three-gluon vertex compared to a bare one: It strongly enhances the angle dependence. Furthermore, the importance of the symmetrization can be seen. The plot shows the contribution of a single gluon box. If it were already symmetric, the results would be the same independent of varying $\theta_q$ or $\theta_r$. Note that even in the case of bare three-gluon vertices there is a difference, because a single gluon box is not symmetric under the exchange of $q$ and $r$, see \fref{fig:4g-DSE_Routing}, but only the sum of the three diagrams appearing in the DSE.

\begin{figure*}[tb]
	\centering
	\includegraphics[width=0.48\textwidth]{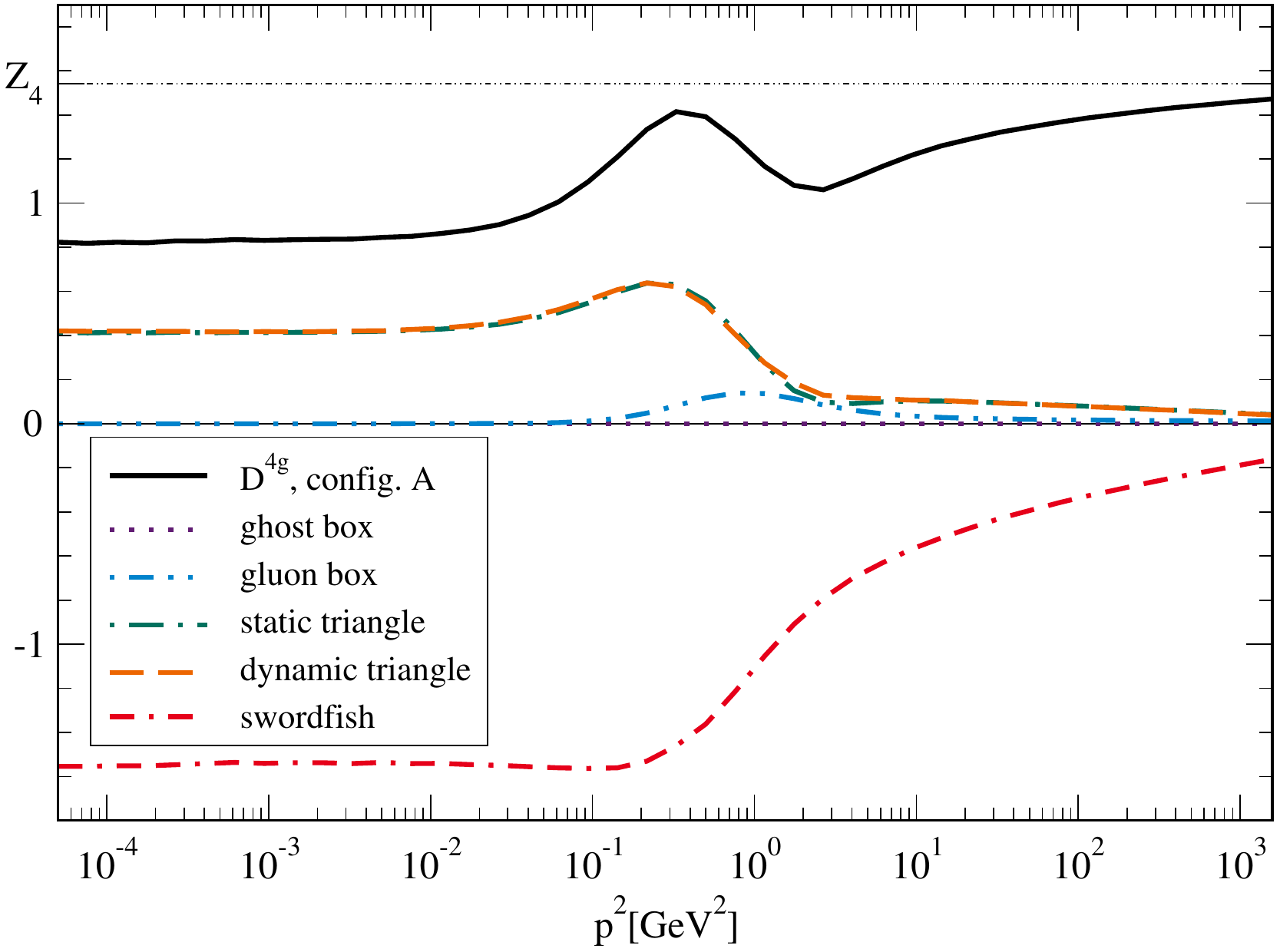}
  \includegraphics[width=0.48\textwidth]{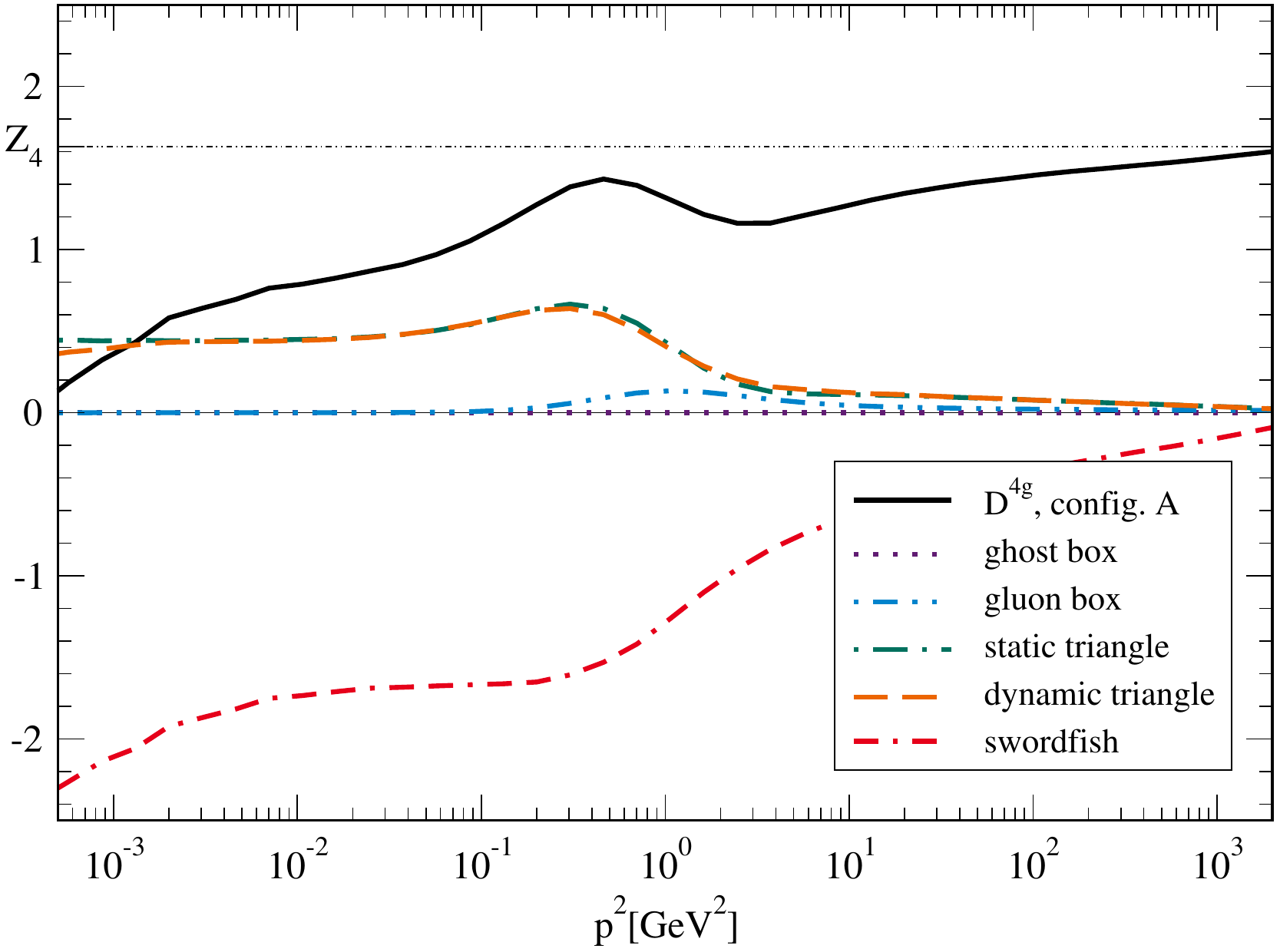}
  \\
	\includegraphics[width=0.48\textwidth]{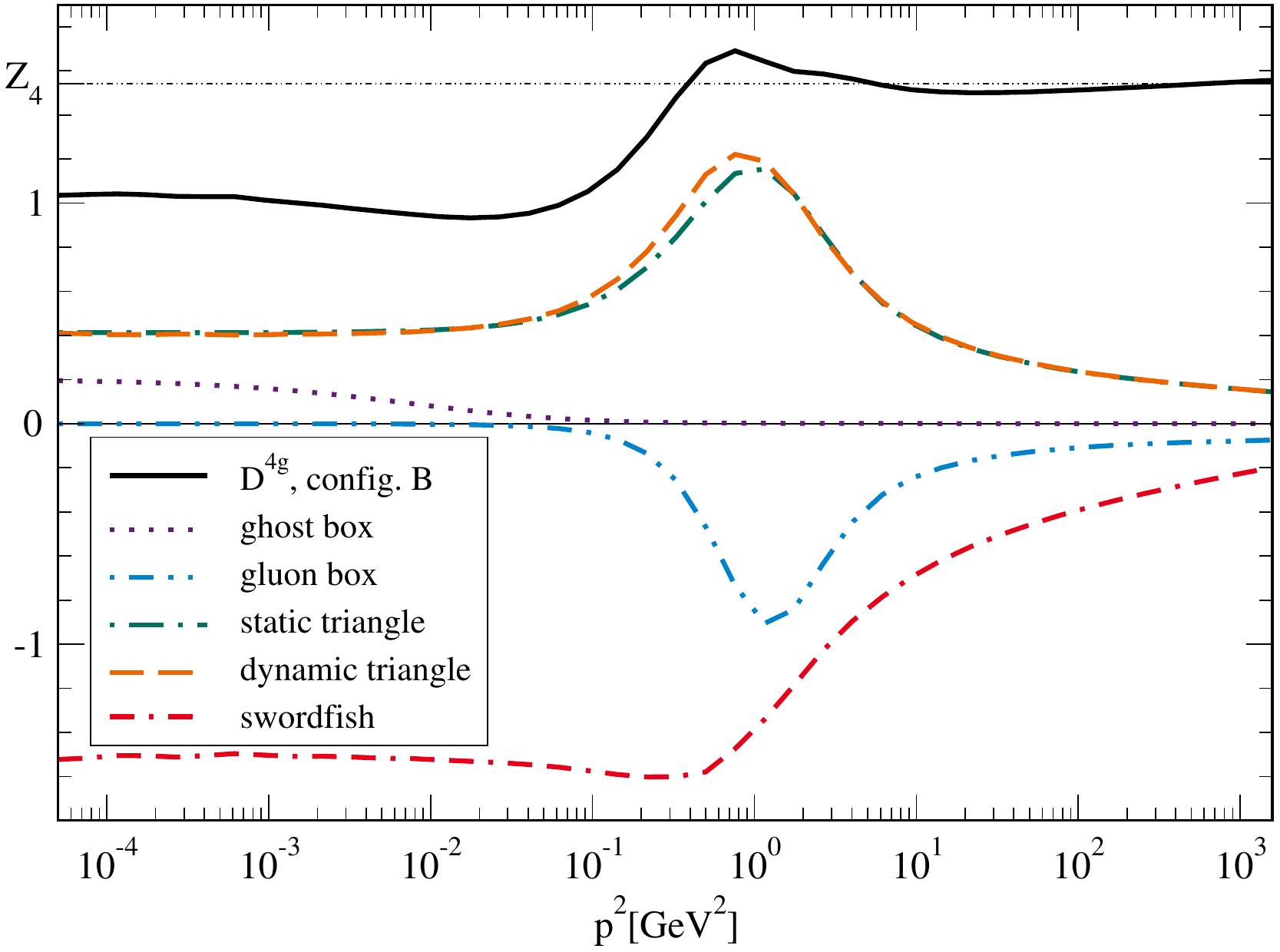}
  \includegraphics[width=0.48\textwidth]{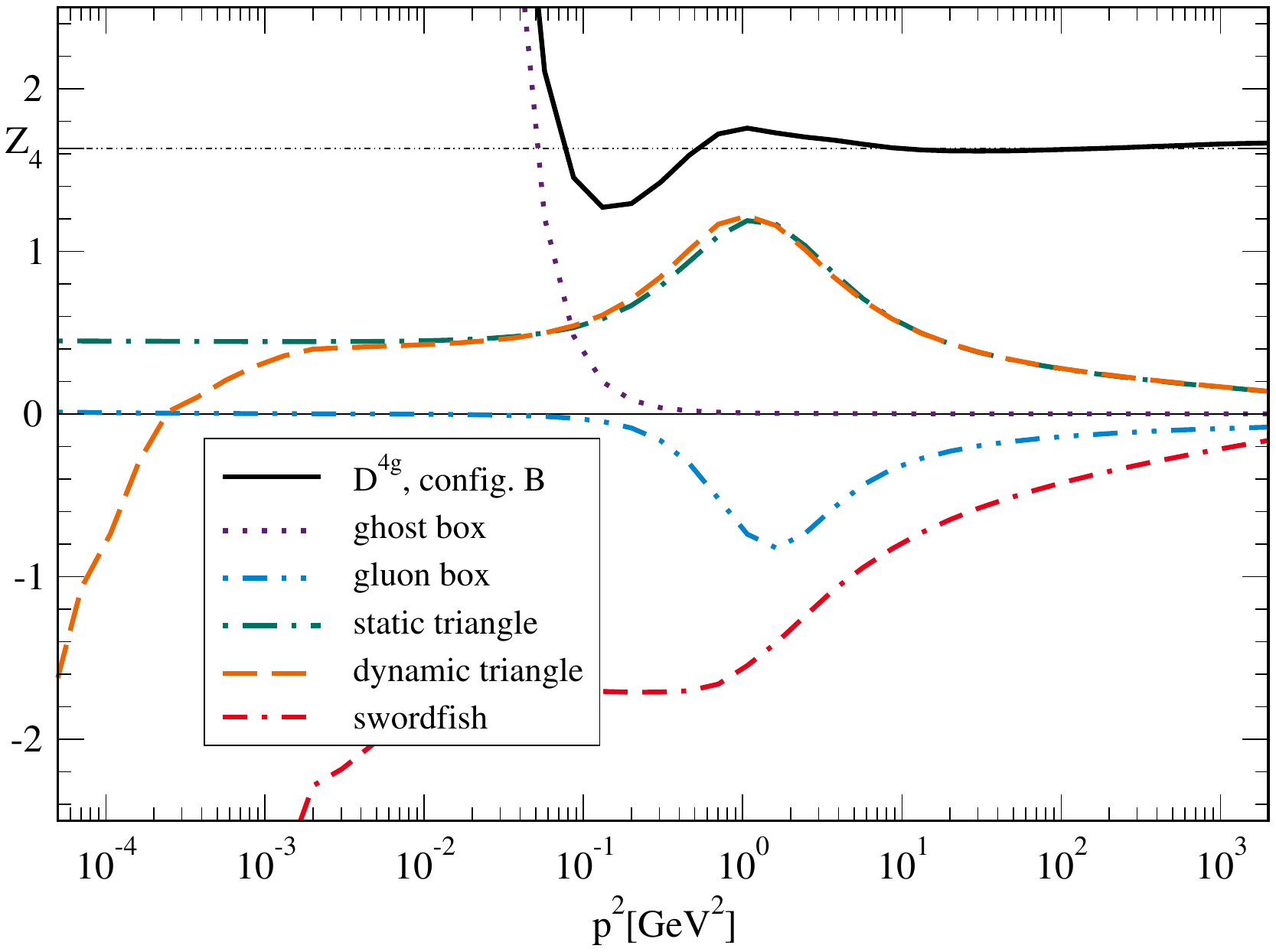}
  \\
	\includegraphics[width=0.48\textwidth]{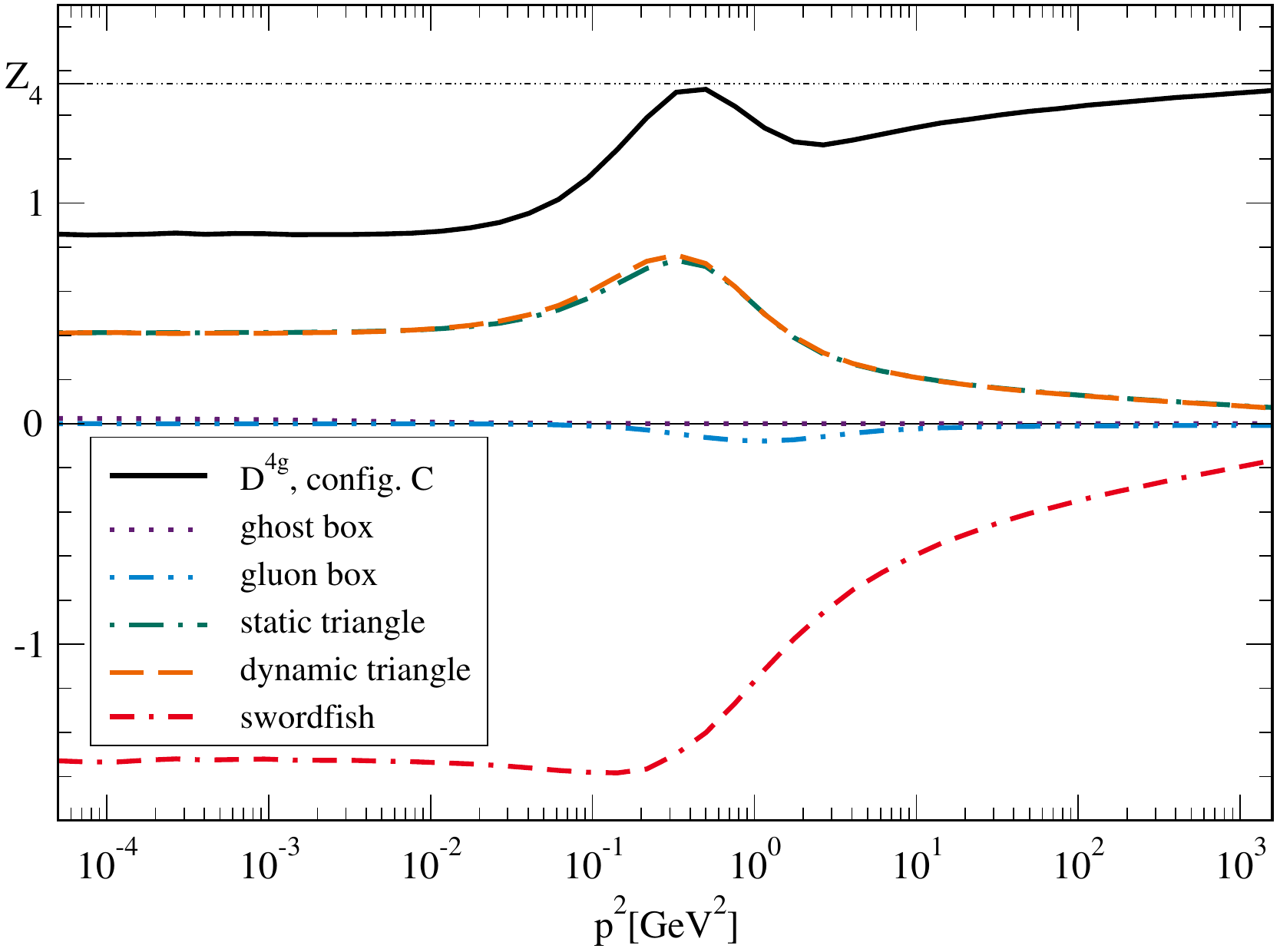}
  \includegraphics[width=0.48\textwidth]{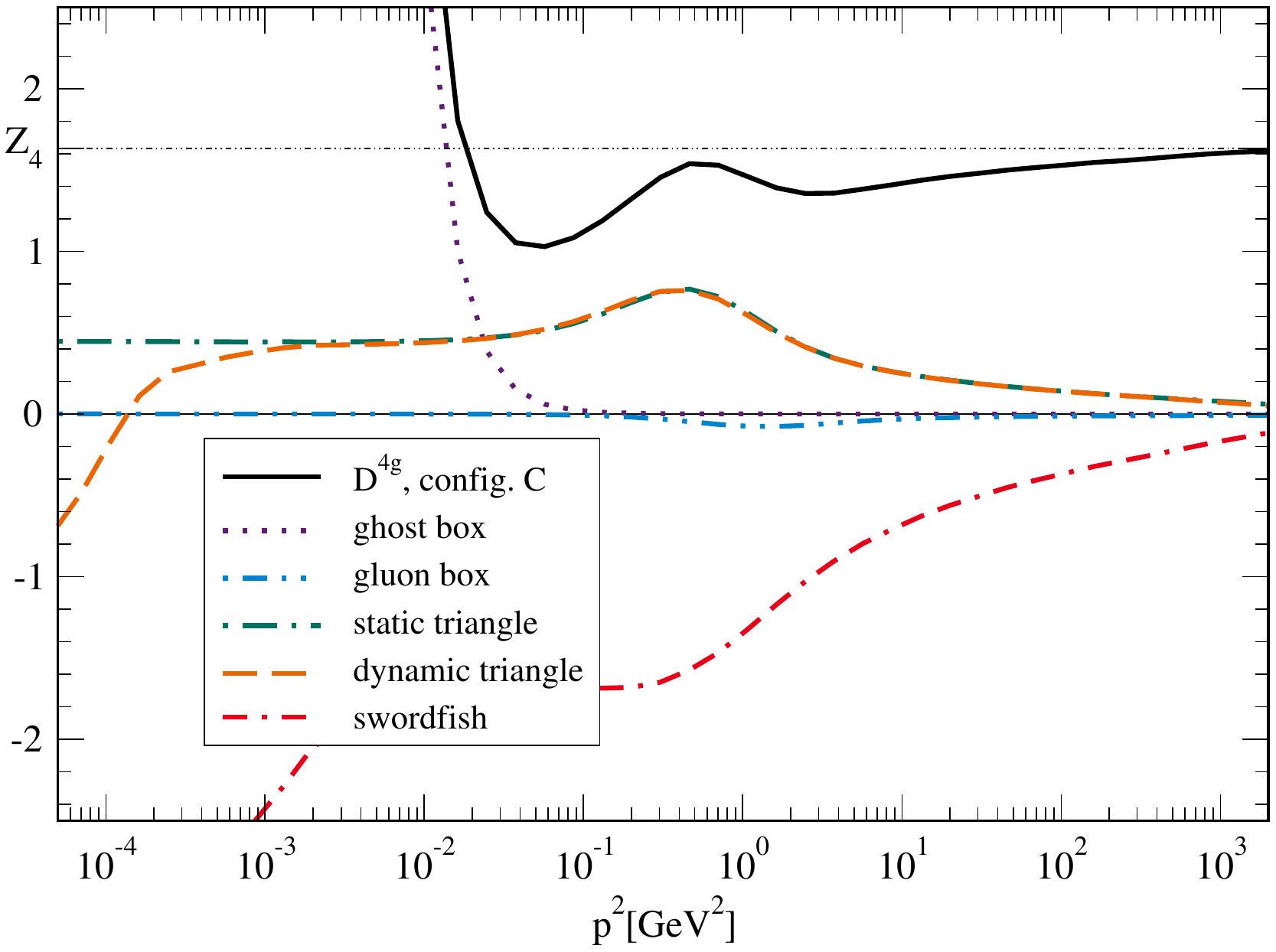}
  \caption[Contributions of individual diagrams]{Contributions of individual diagrams to the tree-level dressing function. 
  Decoupling (left) and scaling solution (right).}
  \label{fig:ContributionOfDiagrams}
\end{figure*}

The contributions of the individual diagrams to the dressing function are plotted in \fref{fig:ContributionOfDiagrams} for different configurations.
As can be seen, the scaling and the decoupling solution are very similar above \(\SI{1}{GeV}\).
Below \(\SI{1}{GeV}\), the ghost box starts to dominate the scaling solution due to the \gls{ir} divergence, \(D^\text{4g}\propto \left(p^2\right)^{-4\kappa}\).
For configuration A, for which the ghost box vanishes, the dynamic diagrams become large in the \gls{ir} but remain finite.
For all other configurations, all other diagrams of the scaling solution are insignificant in the \gls{ir} due to the dominance of the ghost box.
For the decoupling solution, since for this tensor there is no \gls{ir} divergent contribution from the ghost box, all diagrams contribute with a finite value. The mid-momentum regime is dominated by the gluonic diagrams.
Interestingly, both triangle diagrams, dynamic and static, are very similar.\footnote{Note that our separation of triangles is not the same as in Ref.~\cite{Binosi:2014kka}. There the separation into \emph{triangle 1} and \emph{triangle 2} was motivated by the simplifications occurring due to the chosen momentum configuration, which led to two classes of integrals.}
The two diagrams differ only by the dressings of the three- and four-gluon vertices, see \fref{fig:4g-DSE}.

So far, the gluonic four-point interactions had to be modelled whenever they were not neglected. In Ref.~\cite{Mader:2013ru}, e.g., a specific model for the four-gluon vertex was used to incorporate the sunset diagram of the gluon propagator \gls{dse}. In recent calculations of the three-gluon vertex \cite{Blum:2014gna,Eichmann:2014xya}, the four-gluon vertex also had to be modelled. In both studies it was found that the four-gluon vertex must be of a certain strength so that the three-gluon vertex \gls{dse} converges within the applied truncation scheme. The model employed in Ref.~\cite{Blum:2014gna} is given by
\begin{equation}
	\label{eq:FourGluonVertexModelDec}
	D_\text{model}^{4\text{g, dec}}(p,\,q,\,r,\,s) = \left( a \tanh\left(b/\bar{p}^2\right) + 1\right) D^{4\text{g}}_\text{\acrshort{rg}}(p,\,q,\,r,\,s)
\end{equation}
with $D^{4\text{g}}_\text{\acrshort{rg}}(p,\,q,\,r,\,s)$ defined in \eref{eq:RGI_four-gluon}.
It is interesting to see that such a simple form can indeed describe the four-gluon vertex tree-level dressing quite well, as we tested by fitting configurations A, B, and C. Given that the angle dependence is predominantly weak, the fit for configuration $C$, shown in \fref{fig:Comparison}, can serve as a good first approximation for the four-gluon vertex in other calculations. The values for the parameters are $a=1.15$ and $b=\SI{0.63}{GeV^2}$. However, we emphasize that this function only gives a qualitative representation of the four-gluon vertex. In particular, it describes only the tree-level tensor.

\subsection{Other tensors}
\label{sec:other_tensors}

Given the complexity of the four-gluon vertex DSE with its many tensor structures, the calculation of all dressing functions constitutes a further challenge which will not be entered here fully. However, as a first step we can take our solution approximated by the tree-level tensor and calculate other dressing functions. For the three-gluon vertex all transverse dressing functions were calculated in \cite{Eichmann:2014xya} with the result that the tree-level tensor yields indeed by far the most important contribution with the other three tensors at least an order of magnitude smaller.  When it comes to the four-gluon vertex, the situation is similar, but with an additional twist: While we find that the other tensor structures we probed are suppressed as compared to the tree-level one over a wide momentum regime, they possess a logarithmic IR divergence.

To assess the size of other dressing functions, we use the results for the tree-level dressing from the full (decoupling) calculation and calculate several other projections. This is no longer a self-consistent solution but should give us at least an idea about the magnitude of such contributions. We consider two classes of other tensors: One that also contains only the metric and no momenta and thus contains the tree-level tensor, and one whose tensors are constructed from momenta only.

The two tensors chosen from the first class are based on Refs.~\cite{Driesen:1998xc,Kellermann:2008iw} but restricted to their transverse parts. They are constructed from a subset of Bose symmetric tensors by orthogonalization. The four-gluon vertex is then written as
\begin{align}
 \Gamma^{abcd}_{\mu\nu\rho\sigma}(p,q,r,s)=\sum_{i=1}^3 V^{abcd}_{i,\mu\nu\rho\sigma}(p,q,r,s) D^{\text{4g},V_i}(p,q,r,s).
\end{align}
The basis tensors $V^{abcd}_{i,\mu\nu\rho\sigma}(p,q,r,s)$ are given in \eref{eq:NormTensors}. $V^{abcd}_{1,\mu\nu\rho\sigma}(p,q,r,s)$ corresponds to the transversely projected tree-level tensor.

Due to the orthogonality of the tensors $V^{abcd}_{i,\mu\nu\rho\sigma}(p,q,r,s)$, the corresponding dressing functions can be extracted from the four-gluon vertex DSE by replacing the tree-level tensor in the projector (\ref{eq:projected_one-loop}) by the corresponding $V^{abcd}_{i,\mu\nu\rho\sigma}(p,q,r,s)$. However, in practice we found it easier to project with the non-orthogonalized tensors given in \eref{eq:Vis_non-ortho}, because they are shorter, and calculate the dressings $D^{\text{4g},V_i}(p,q,r,s)$ from the results.

\begin{figure*}[tb]
  \centering
  \includegraphics[width=0.48\textwidth]{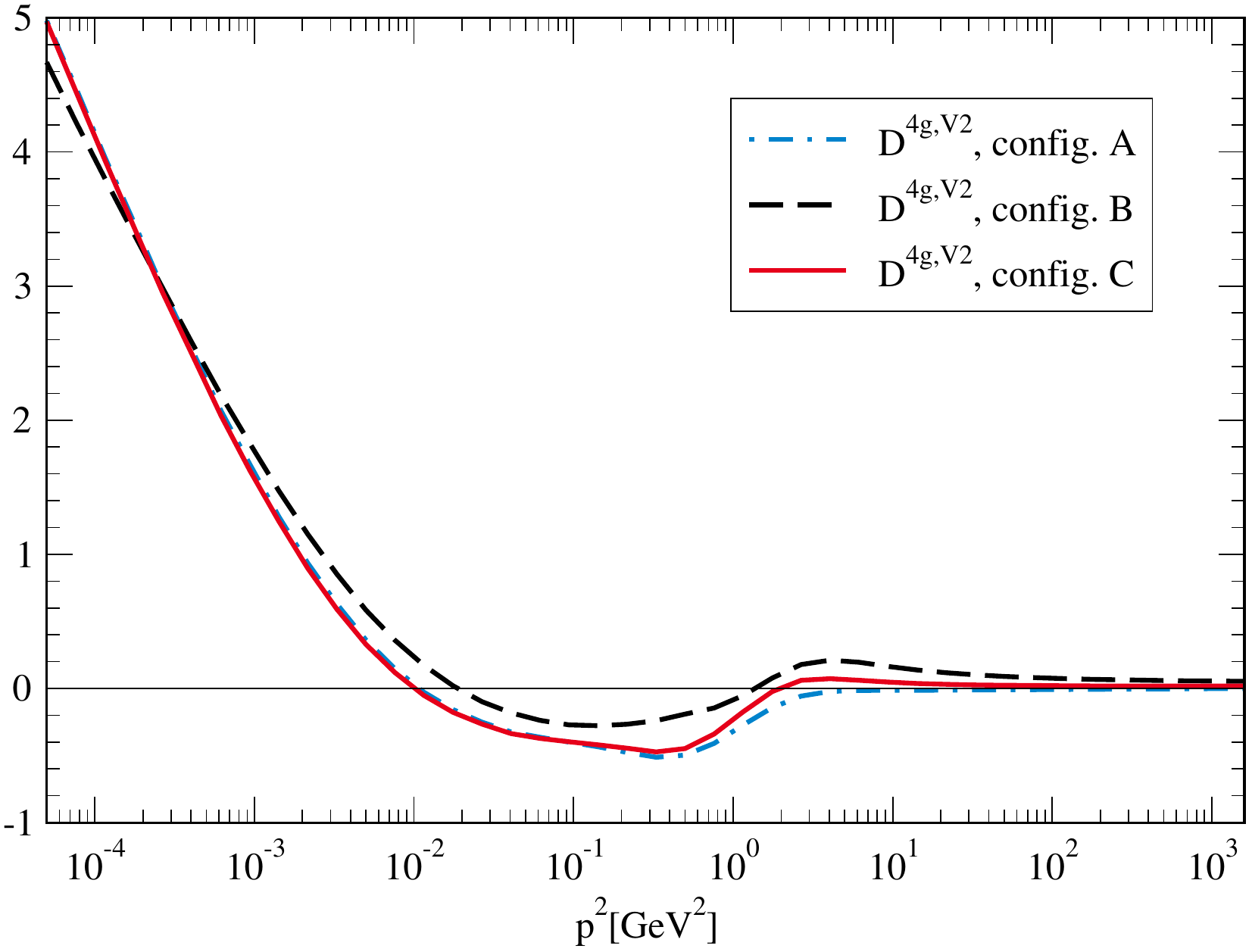}
  \includegraphics[width=0.48\textwidth]{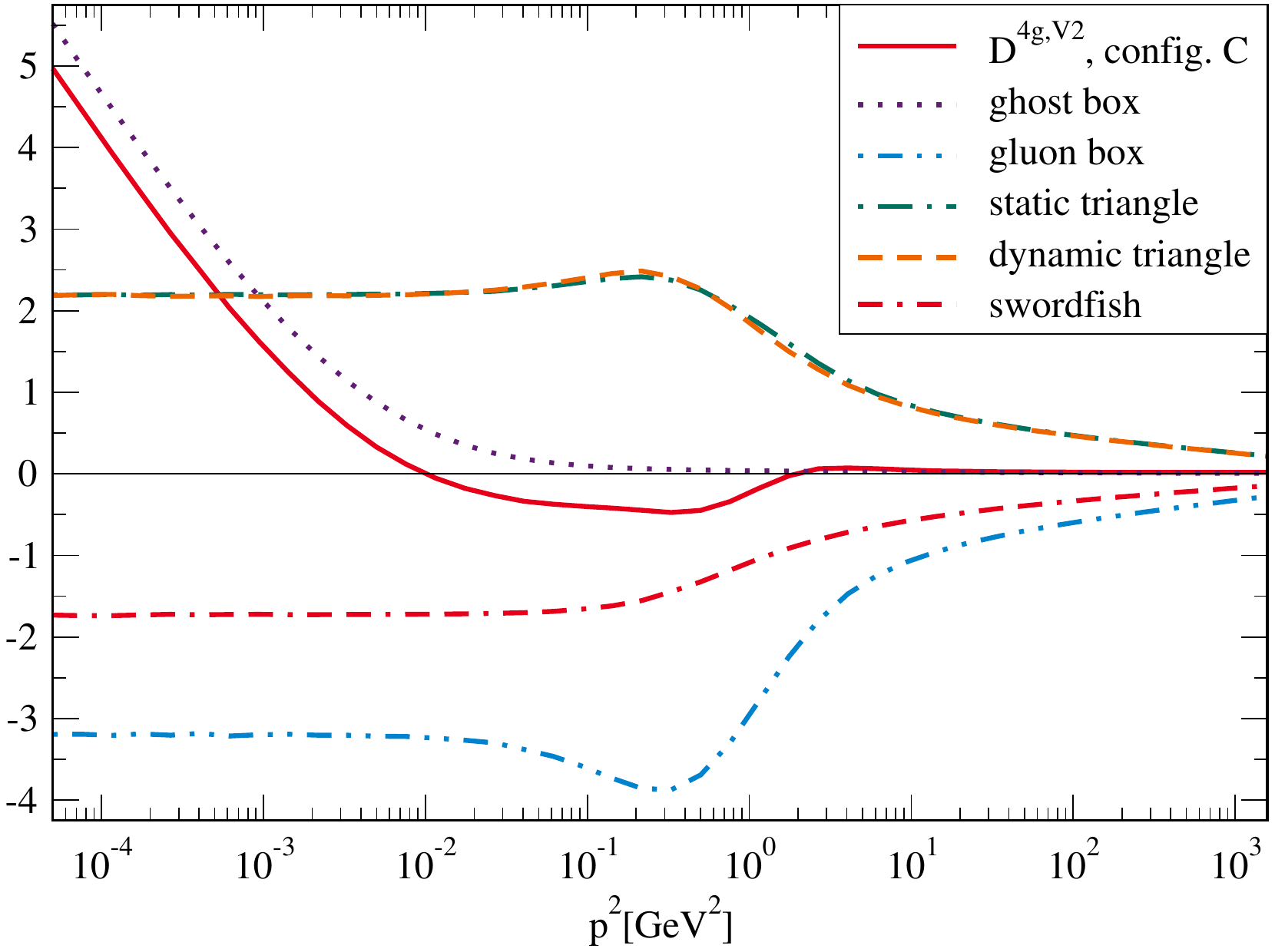}
  \caption{Results for $D^{\text{4g},V_2}$ (left) and individual contributions to configuration C (right).}
  \label{fig:4g_V2}
\end{figure*}

\begin{figure*}[tb]
  \centering
  \includegraphics[width=0.48\textwidth]{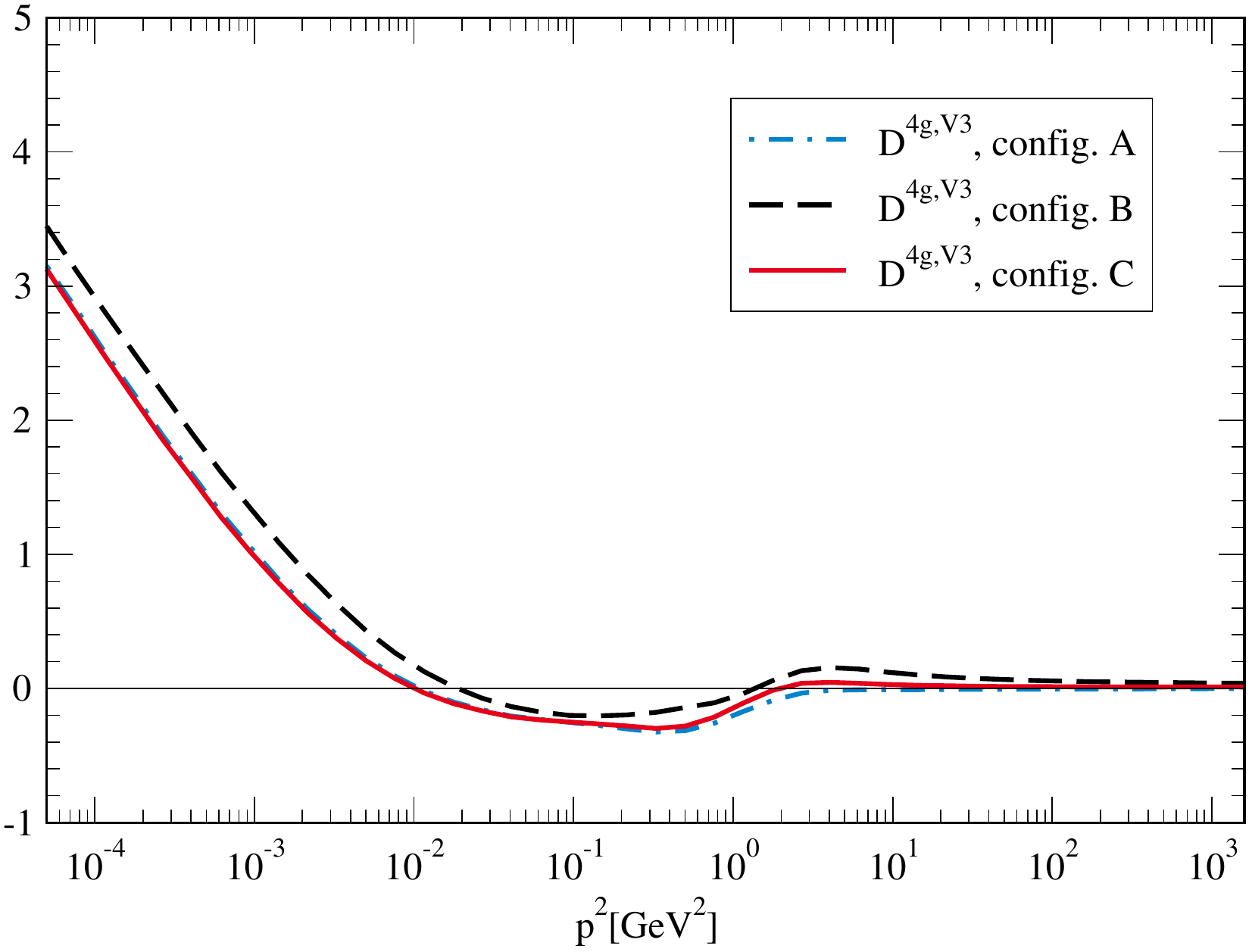}
  \includegraphics[width=0.48\textwidth]{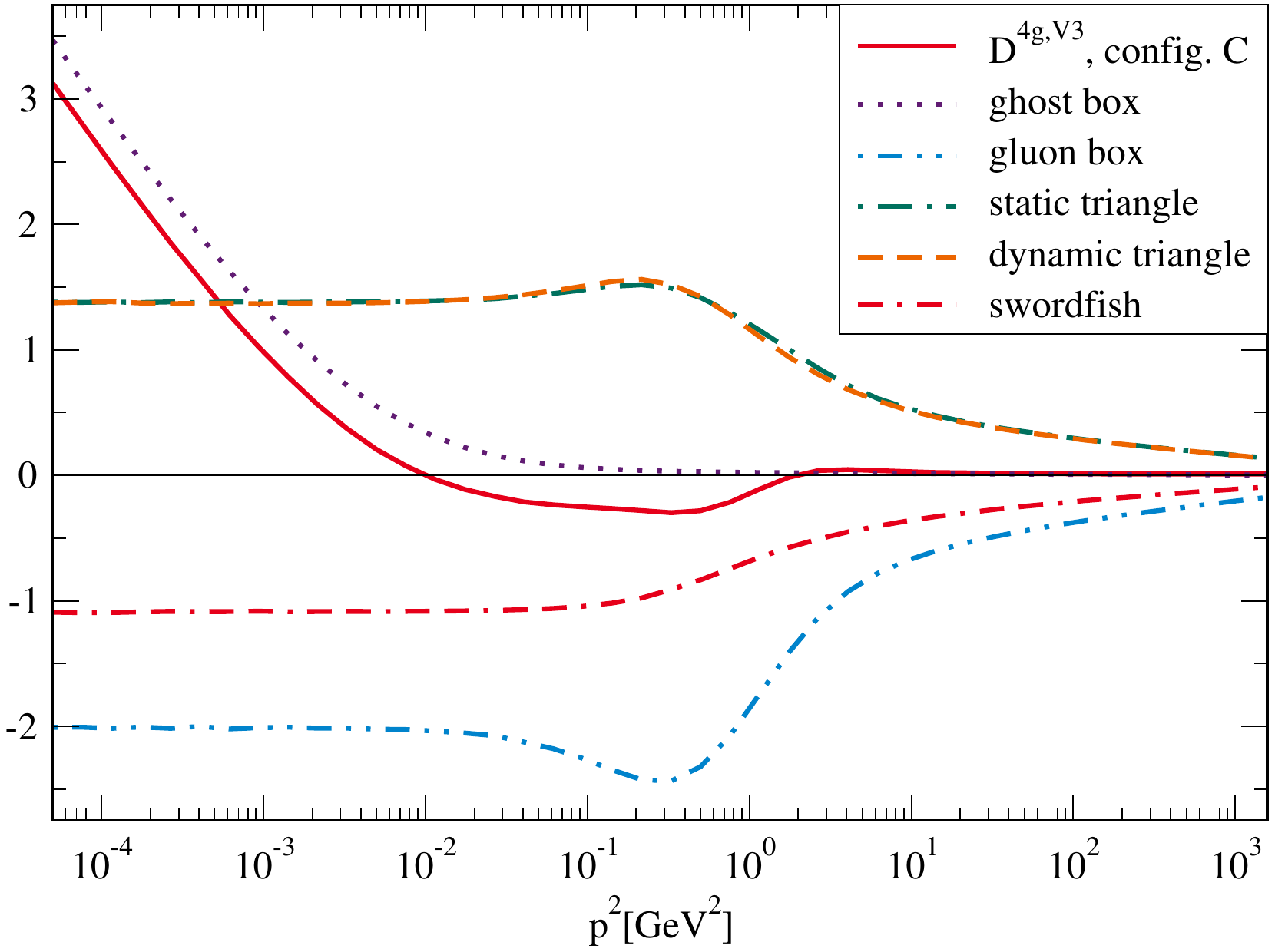}
    \caption{Results for $D^{\text{4g},V_3}$ (left) and individual contributions to configuration C (right).}
  \label{fig:4g_V3}
\end{figure*}

The two additional dressings $D^{\text{4g},V_2}(p,q,r,s)$ and $D^{\text{4g},V_3}(p,q,r,s)$ are shown in Figs.~\ref{fig:4g_V2} and \ref{fig:4g_V3}. Again we find no large dependence on the chosen configuration. Most strikingly the magnitude of the two dressings is very small compared to the tree-level dressing, but it becomes larger in the IR where they diverge logarithmically. This divergence, which is due to the ghost box, see Figs.~\ref{fig:4g_V2} and \ref{fig:4g_V3}, was also seen in Ref.~\cite{Binosi:2014kka} for the related dressing $D^{\text{4g},G}$ that belongs to the tensor $G^{abcd}_{\mu\nu\rho\sigma}$ given by
\begin{align}
\begin{split}
 G^{abcd}_{\mu\nu\rho\sigma}&=(\de^{ab}\de^{cd}+\de^{ac}\de^{bd}+\de^{ad}\de^{bc})\\&\quad
 \times(\de_{\mu\nu}\de_{\rho\sigma}+\de_{\mu\rho}\de_{\nu\sigma}+\de_{\mu\sigma}\de_{\nu\rho}).
 \end{split}
\end{align}
$G$ can be written as a linear combination of the tensors $\widetilde{V}_2$ and $\widetilde{V}_3$, see \eref{eq:Vis_non-ortho}. Since the results for $D^{\text{4g},V_2}$ and $D^{\text{4g},V_3}$ are very similar, $D^{\text{4g},G}$ resembles the two as well. In particular we do not find an enhancement in the mid-momentum regime as it was found in Ref.~\cite{Binosi:2014kka}. As we checked explicitly, the source of this enhancement lies in the truncation scheme employed in Ref.~\cite{Binosi:2014kka}, where all renormalization constants in front of the loop diagrams were set to $1$. One can see in Figs.~\ref{fig:4g_V2} and \ref{fig:4g_V3} that the contributions of the individual diagrams are by no means small and the resulting dressing is only small in the mid-momentum regime because of delicate cancelations. Since all diagrams contain either $Z_1$ or $Z_4$, which have very different values, see Table~\ref{tab:paras}, the renormalization constants naturally play an important role in this and should be taken into account properly.

\begin{figure*}[tb]
  \centering
  \includegraphics[width=0.48\textwidth]{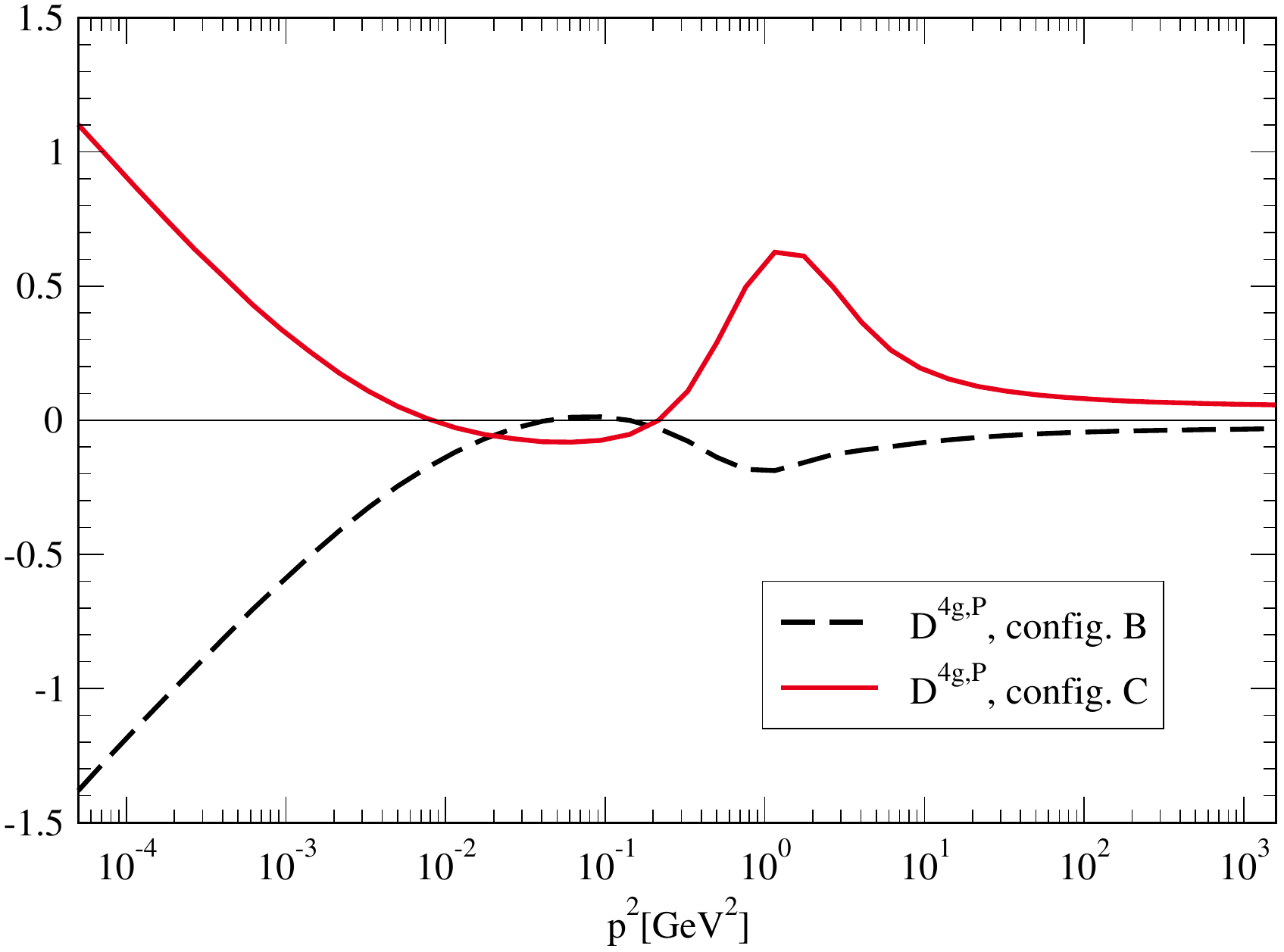}
  \includegraphics[width=0.48\textwidth]{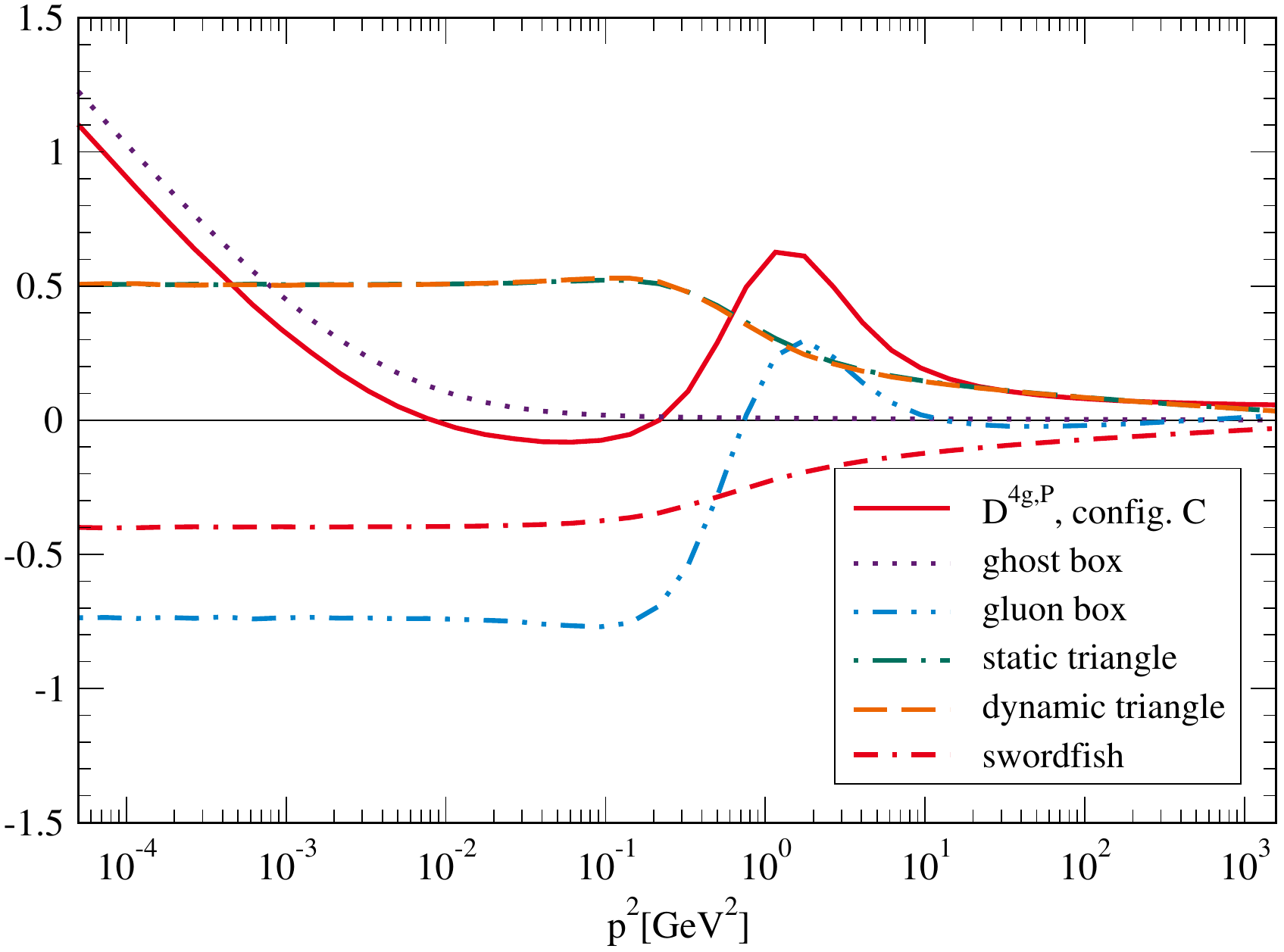}
    \caption{Results for $D^{\text{4g},P}$ (left) and individual contributions to configuration C (right).}
  \label{fig:4g_P}
\end{figure*}

As a second example we consider a tensor constructed from momenta only. This case serves to investigate if tensors from another class as before have a different behavior. As a Bose symmetric representative of such a class we choose the tensor $P^{abcd}_{\mu\nu\rho\sigma}(p,q,r,s)$ given in \eref{eq:NormTensors}. It is constructed from
\begin{align}
\begin{split}
 \widetilde{P}_{\mu\nu\rho\sigma}^{abcd}&=(\de^{ab}\de^{cd}+\de^{ac}\de^{bd}+\de^{ad}\de^{bc})\\&\quad
 \times\frac{s_\mu r_\nu q_\rho p_\sigma + r_\mu s_\nu p_\rho q_\sigma + q_\mu p_\nu s_\rho r_\sigma}{\sqrt{p^2\,q^2\,r^2\,p^2}}
\end{split}
\end{align}
by transverse projection and normalization to the tree level; see Appendix~\ref{sec:extended_tensor_basis}. As it turns out the behavior of the corresponding dressing is indeed different from that of $D^{\text{4g},V_2}$ and $D^{\text{4g},V_3}$. In particular the angle dependence is more pronounced as can be seen in \fref{fig:4g_P}. Besides the height of the bump in the mid-momentum regime also the sign depends on the configuration.
We present no results for configuration $A$, since the tensor $\widetilde{P}$ is then purely longitudinal.

\begin{figure}[tb]
  \centering
  \includegraphics[width=0.48\textwidth]{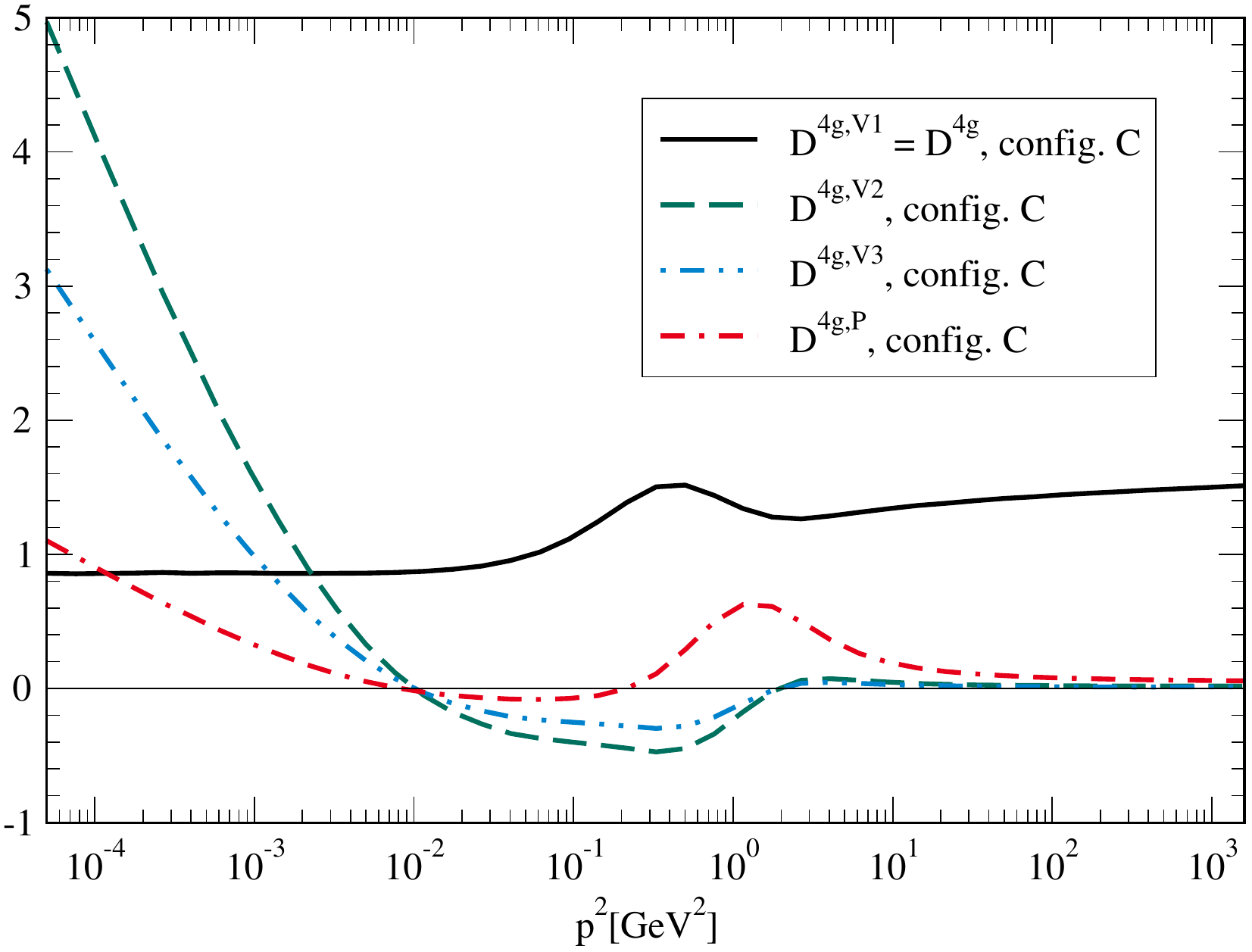}
  \caption[Comparison of dressing functions]{Comparison of the dressing functions $D^{\text{4g}}$, $D^{\text{4g},V_2}$, $D^{\text{4g},V_2}$ and $D^{\text{4g},P}$ for configuration C. }
  \label{fig:DressFuncComp}
\end{figure}

Finally we compare the contributions of different dressing functions in \fref{fig:DressFuncComp}. Clearly, the tree-level dressing function is dominant. However, this is due to the tree-level diagram itself, which contributes with the constant value $Z_4$, see \fref{fig:ContributionOfDiagrams}. The loop diagrams only account for comparatively small changes around the value of the renormalization constant $Z_4$. Taking this into account, the contributions of the loop diagrams are similar for all considered dressings. One exception is that only the non-tree-level dressing functions diverge logarithmically in the IR. However, this divergence sets in at very low momenta at around $\SI{100}{MeV}$ and the dominant dressing over a wide range of momenta is that of the tree-level tensor.

\subsection{Running coupling}
\label{sec:runningCoupling}

\begin{figure*}[tb]
  \centering
  \includegraphics[width=0.48\textwidth]{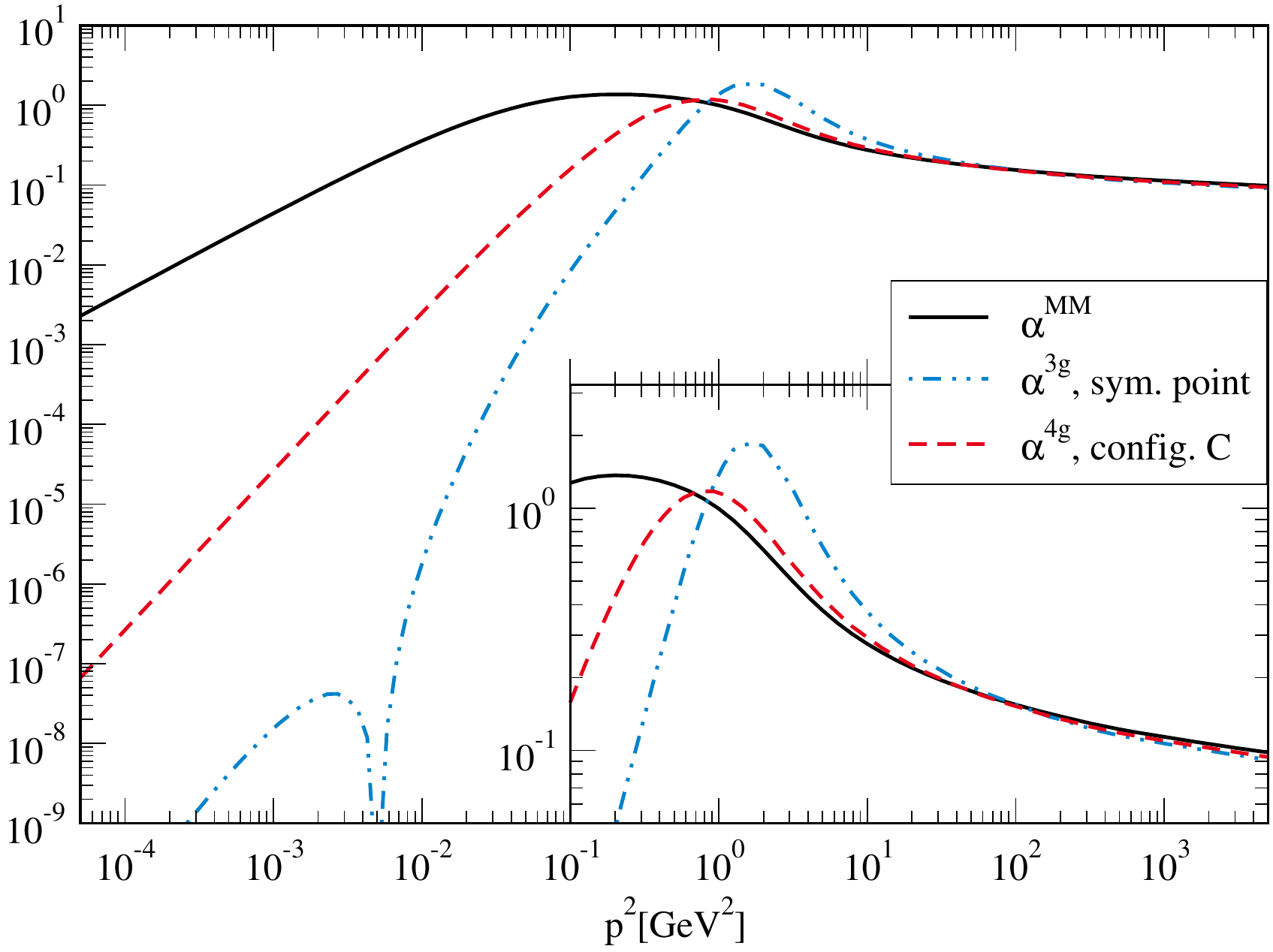}
  \includegraphics[width=0.48\textwidth]{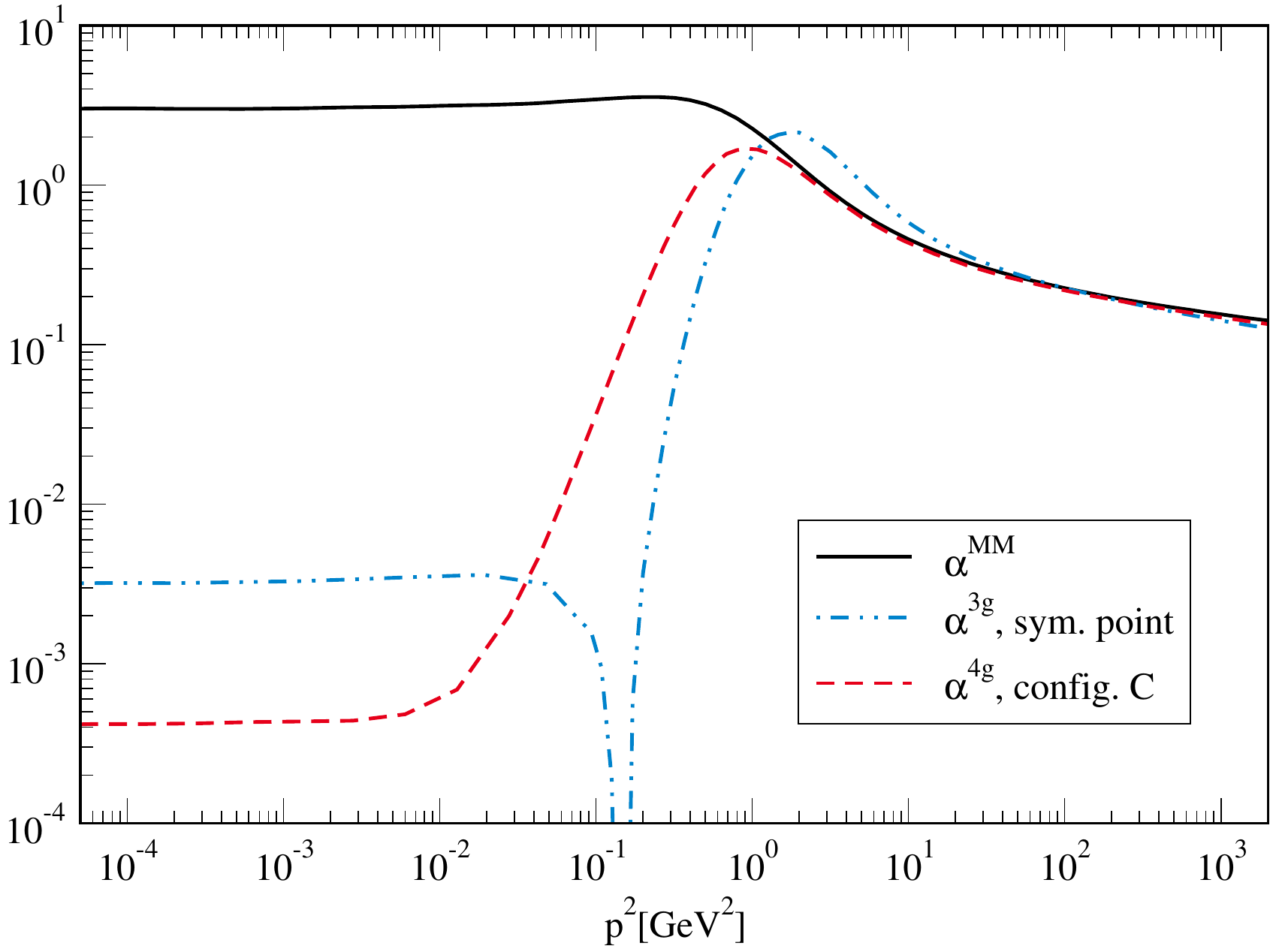}
    \caption{Comparison of the running couplings from the ghost--gluon, the three-gluon and the four-gluon vertices. Left: decoupling solution. Right: scaling solution.}
  \label{fig:RunningCoupling}
\end{figure*}

From the four-gluon vertex a renormalization group invariant running coupling can be defined, as it can be from any other vertex \cite{Alkofer:2004it,vonSmekal:1997vx}.
Up to now the couplings derived from the ghost--gluon, the three-gluon and the four-gluon vertices were calculated; see, e.g., \cite{Kellermann:2008iw,Fischer:2008uz,Huber:2012kd,Blum:2014gna,Eichmann:2014xya} for results from \glspl{dse}. They are given by \cite{Alkofer:2004it}
\begin{subequations}
	\label{eq:RunningCouplings_other}
	\begin{align}
		\label{eq:RunningCoupling_gh-gl}
		\alpha^{\text{MM}}(p^2) & = \alpha(\mu^2) \, G^2(p^2) Z(p^2),\\
		\label{eq:RunningCoupling_3Gl}
		\alpha^{\text{3g}}(p^2) & = \alpha(\mu^2) \, \frac{\left[ D^{\text{3g}}(p^2) \right]^2 Z^3(p^2)}{\left[ D^{\text{3g}}(\mu^2) \right]^2 Z^3(\mu^2)},\\
		 \label{eq:RunningCoupling_4Gl}
		\alpha^{\text{4g}}(p^2) & = \vphantom{\Big]^2}\alpha(\mu^2) \, \frac{D^{\text{4g}}(p^2) \, Z^2(p^2)}{D^{\text{4g}}(\mu^2) \, Z^2(\mu^2)}.
	\end{align}
\end{subequations}
where \(\alpha(\mu^2)=g^2/4\pi\) is the running coupling at the renormalization scale $\mu$. For the arguments of the three- and four-gluon vertices a generic scale $p^2$ was given. Which kinematic configuration is chosen is in principle free. For the three-gluon vertex we choose the symmetric point and for the four-gluon vertex configuration $C$. Note that the denominators of the three- and four-gluon vertex couplings are not unity, because we work here in the \textit{MiniMOM} scheme where $G^2(\mu^2) Z(\mu^2)=1$. Choosing, e.g., $\left[ D^{\text{3g}}(\mu^2) \right]^2 Z^3(\mu^2)=1$ would correspond to a different scheme. Hence we have to take this factor into account.

In \fref{fig:RunningCoupling} we show the different running couplings. For large momenta they all agree as they should with only small deviations. At low momenta the four-gluon vertex running coupling vanishes like $p^4$.
In the scaling case all couplings exhibit an IR fixed point \cite{Alkofer:2004it}. For the four-gluon vertex it is determined by the ghost box as it was also found in Ref.~\cite{Kellermann:2008iw} without transverse projection. However, if only the transverse part is considered, the contribution for configuration $A$, which was used in \cite{Kellermann:2008iw}, vanishes, as discussed in Sect.~\ref{sec:ghost_box}. Thus this configuration is not suited for calculating the running coupling within this truncation. Since the existence of the \gls{ir} fixed point should not depend on the angular configuration, we believe that this is a shortcoming of the truncation and not a general feature. Indeed there are two (scaling) IR leading diagram types, namely swordfish-like and triangle diagrams with internal ghost lines, which we neglected. They may yield a contribution that does not vanish in the \gls{ir} and provides a non-vanishing value for the running coupling. We did not follow this further, as it would require ghost--gluon four- and five-point functions. For configuration $C$ we extract a value of $\alpha^{\text{4g}}(0)=\num{0.00042}$. This is in accordance with the findings of Ref.~\cite{Kellermann:2008iw} that the fixed point value of the four-gluon vertex is much lower than that of the ghost--gluon vertex, which is $\alpha^{\text{MM}}(0)\approx 2.97$ \cite{Lerche:2002ep}. For the symmetric point of the three-gluon vertex we extract $\alpha^{\text{3g}}(0)=\num{0.0032}$. Our value for $\alpha^{\text{3g}}(0)$ deviates from the value found in Ref.~\cite{Eichmann:2014xya}, $\alpha^{\text{3g}}(0)\simeq\num{0.0016}$, but this is due to the different ghost propagator input.

\section{Summary and conclusions}
\label{sec:summary}

In the past the two- and three-point functions of Yang--Mills theory were intensively scrutinized with functional equations. However, within the employed truncations these calculations still relied on model input for higher Green functions. Using lattice data, these models could be tuned such that neglected contributions could be effectively taken into account which led to good agreement with lattice results. Thus, as far as two-and three-point functions are concerned, we have quantitatively reliable input for other calculations such as those performed here.

Employing the by now common truncation to the UV leading diagrams, which also contain IR leading diagrams, the DSE of the four-gluon vertex does not rely on any model. This is, within this truncation scheme, a unique feature among the primitively divergent Green functions. However, given the tensorial complexity of the vertex, we solved the DSE self-consistently only for the tree-level tensor. A posteriori, we confirmed for some additional dressing functions that their magnitude is much smaller than that of the tree-level tensor over a wide momentum regime. In the IR, on the other hand, the tree-level dressing is in the decoupling case constant, whereas other dressings can diverge logarithmically. These divergences set in at about $\SI{100}{MeV}$. Since a four-gluon vertex is within a functional equation for \gls{proper} functions contracted with at least two gluon dressing functions, which are IR suppressed, these divergences most likely do not have a strong effect on other Green functions as long as no other IR divergent expressions appear in the integrand.

For the dominant tree-level structure we found that the box diagrams are almost negligible and the triangle and swordfish diagrams yield the largest contributions. However, for other tensors this is no longer the case. Especially the gluon box can have a sizable impact and the ghost box leads to the IR divergence mentioned before.

It is important to note that the fact that other dressings functions are so small in the mid-momentum regime is in no way trivial, as it comes from cancelations between all the diagrams taken into account. If it turns out that also dressing functions beyond those we investigated here follow this pattern, the four-gluon vertex can be well approximated by one tensor only. This would alleviate its use in future studies, like its effect in the gluon propagator or three-gluon vertex DSEs, considerably. As a first approximation we provide a fit that describes the dressing qualitatively well.

An important aspect of our calculations was that we used only the transverse subspace, since it is sufficient for the Landau gauge. By studying the ghost box diagram explicitly also without transverse projection we found that there can be a considerable influence of this restriction which most likely also exists for other diagrams. Thus in future studies this restriction should always be taken into account.

From the four-gluon vertex a running coupling can be extracted. Qualitatively it behaves as expected, viz., it agrees in the perturbative regime with the couplings from the ghost--gluon and three-gluon vertices and then turns toward zero in the IR for decoupling and toward an IR fixed point for scaling. We confirmed that the value of this fixed point is very small compared to that of the \textit{MiniMOM} coupling.

The four-gluon vertex was the last primitively divergent Green function of Landau gauge Yang--Mills theory for which a self-consistent solution was lacking. Its calculation constitutes an important step toward a fully self-contained description of Yang--Mills Green functions from functional equations and will enable the study of its impact on the propagators and the three-gluon vertex.

\section*{Acknowledgements}

We thank Reinhard Alkofer, Adrian~L.~Blum, Gernot Eichmann, and Christian S.~Fischer for useful discussions. This work was supported by the Helmholtz International Center for FAIR within the LOEWE program of the State of Hesse and NAWI Graz.

\appendix

\section{Color calculations}
\label{sec:app_color}

To calculate color traces, we employ the following two well-known identities:
\begin{subequations}
	\label{eq:ColourIdentities}
	\begin{align}
		f^{am'n'} f^{bm'n'} & = N_\text{c} \delta^{ab},\\
		f^{aa'b'}f^{bb'c'}f^{cc'a'} & = \frac{N_\text{c}}{2}f^{abc}.
	\end{align}
\end{subequations}
If the trace contains six structure constants, it is sometimes necessary to insert the Jacobi identity,
\begin{equation}
  \label{eq:jacobi_identity}
  C^{abcd}_4 + C^{abcd}_5 + C^{abcd}_6 = 0,
\end{equation}
before Eqs.~\eqref{eq:ColourIdentities} can be applied. The rank-4 color tensors $C_i$ are given in Eqs.~\eqref{eq:color_tensors_basis} and \eqref{eq:C7}.

We show now that \(C_7\), defined in \eref{eq:C7}, is not linearly independent of the tensor set given in \eref{eq:color_tensors_basis} for $N_c<4$.
To start with, consider the set of tensors \(C_1, \ldots, C_5\) and \(C_7\).
The scalar product \(\langle C_i,\,C_j \rangle\) of two tensors is given by the trace over the color indices.
Thus, the metric tensor of a six-dimensional color space defined by \(C_1, \ldots, C_5\) and \(C_7\) is given by
\begin{equation}
  \label{eq:metric_colour_tensor}
  g^\text{color}_6 =
    \left(
    \begin{matrix}
      \langle \widetilde{C}_1,\,\widetilde{C}_1\rangle  & \ldots & \langle \widetilde{C}_1,\,\widetilde{C}_5\rangle  & \langle \widetilde{C}_1,\,\widetilde{C}_7\rangle  \\
      \vdots      & \ddots & \vdots      & \vdots      \\
      \langle \widetilde{C}_5,\,\widetilde{C}_1\rangle  & \ldots & \langle \widetilde{C}_5,\,\widetilde{C}_5\rangle  & \langle \widetilde{C}_5,\,\widetilde{C}_7\rangle  \\
      \langle \widetilde{C}_7,\,\widetilde{C}_1\rangle  & \ldots & \langle \widetilde{C}_7,\,\widetilde{C}_5\rangle  & \langle \widetilde{C}_7,\,\widetilde{C}_7\rangle 
    \end{matrix}
    \right),
\end{equation}
where $\widetilde{C}_i$ is defined as the normalized $C_i$, viz., $\widetilde{C}_i=C_i/\sqrt{\langle C_i,\, C_i\rangle}$.
Calculating the determinant yields
\begin{equation}
  \det \left(g^\text{color}_6\right) = \frac{1}{4}\frac{\left(N_\text{c}^2-9\right)\cdot \left(N_\text{c}^2-4\right)^3}{\left(N_\text{c}^2-1\right)\cdot \left(N_\text{c}^2 + 12\right)}.
\end{equation}
Hence, if \(N_\text{c}=3\) or \(N_\text{c}=2\), \(\det(g^\text{color}_6)=0\).
If the determinant of the metric tensor is zero, the tensors (i.e.,  \(C_1 \ldots C_5\) and \(C_7\)) are linearly dependent.
One can similarly show that the determinant of the metric tensor of the color space spanned by the tensors \(C_1 \ldots C_5\) is not zero, \(\det(g^\text{color}_5) \neq 0\,\).
Thus, \(C_7\) can be expressed in terms of \(C_1 \ldots C_5\):
\begin{equation}
  \label{eq:C7_undetermined}
  C^{abcd}_7 = \sum_{i=1}^{5} a_i C^{abcd}_i.
\end{equation}
Multiplying \eref{eq:C7_undetermined} by the basis tensors \eref{eq:color_tensors_basis} gives five equations.
Solving this linear system of equations yields the \(a_i\,'s\) and thus \eref{eq:C7_SU3identity} in the case of $SU(3)$ and \eref{eq:C7_SU2identity} in the case of $SU(2)$.

\section{Technical details of the calculation}
\label{sec:tech_details}

The four-gluon vertex was calculated with the \textit{CrasyDSE} framework \cite{Huber:2011xc}.
In a first step, we calculate all static diagrams, the diagrams that do not depend on the four-gluon vertex (the diagrams in the second line in Fig.~\ref{fig:4g-DSE}).
Calculating on cores of Intel Xeon E5-2670 processors, the first step takes typically $\num{30000}$ core hours.
For the actual iteration process we only need to calculate the \acrlong{sf} and the \acrlong{dt} diagrams, where one iteration step typically takes $\num{10000}$ core hours.
Fortunately, the four-gluon vertex DSE converges relatively fast within \(6\) to \(8\) iteration steps.
We perform the integration as detailed below. 

\begin{figure*}[tb]
  \includegraphics[width=\textwidth]{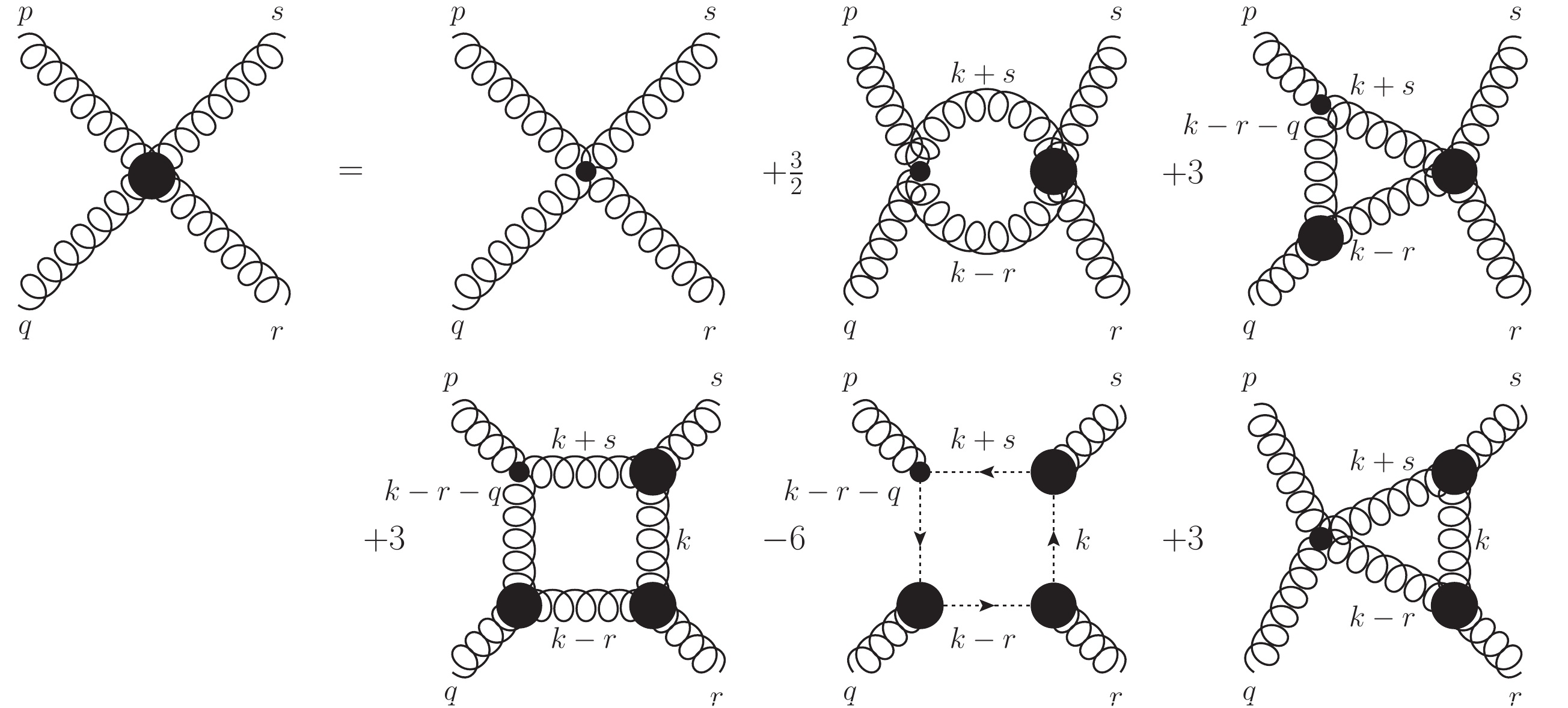}
  \caption{The momentum routing used in the truncated four-gluon vertex DSE.  The factors of \(3\) for each diagram represent the fact that we do not need to calculate the different permutations of the diagrams if we symmetrize the DSE with respect to all external legs.
  }
  \label{fig:4g-DSE_Routing}
\end{figure*}

The integration itself is done by a standard Gau\ss-Legendre quadrature. Unfortunately we cannot integrate out any variables analytically and have to perform all four integrations numerically.
We use spherical coordinates given by
\begin{align}
	\label{eq:LoopMomentum}
	k=K\cdot\colvec{4}
	{\cos(\theta_k)\phantom{\sin(\psi_k)\cos(\phi_k)}}
	{\sin(\theta_k)\cos(\psi_k)\phantom{\cos(\phi_k)}}
	{\sin(\theta_k)\sin(\psi_k)\cos(\phi_k)}
	{\sin(\theta_k)\sin(\psi_k)\sin(\phi_k)}.
\end{align}
The integral measure is then (with \(K=\abs{k}\))
\begin{align}
	\label{eq:integral}
  \begin{split}
	&\int \text{d} k^4 =\frac{1}{2} 
		\int_{\epsilon^2}^{\Lambda^2} \text{d} K^2 \; K^2
		\int_0^\pi \text{d} \theta_k \; \sin^2(\theta_k)\\&\quad\times
		\int_0^\pi \text{d} \psi_k \; \sin(\psi_k)
		\int_0^{2\pi} \text{d} \phi_k.
  \end{split}
\end{align}
Exploiting that all scalar products of \(k\) and \(s,\,r\) or \(q\), and thus the kernels, depend on \(\cos(\phi_k)\) only, the \(\phi_k\) integration can be simplified:
\(\int_0^{2\pi} \text{d} \phi_k \dots = 2 \int_0^{\pi} \text{d} \phi_k \dots\;.\)
We split each integration into several integration regions in order to not integrate over the singularities arising from the internal propagators.
We choose the momentum routing such that the loop momenta of the gluon box and the ghost box are given by
\begin{subequations}
	\label{eq:LoopMomenta}
	\begin{align}
		k_1 & = k ,\\
		k_2 & = k + s ,\\
		k_3 & = k - r ,\\
		k_4 & = k - r - q.
	\end{align}
\end{subequations}
The routing is illustrated in \fref{fig:4g-DSE_Routing}. 
The integrand diverges if \(k_1^2=0\), \(k_2^2=0\), \(k_3^2=0\) or \(k_4^2=0\).
Due to the choice of coordinates, this leads to four integration regions in \(K^2\), three in \(\theta_k\), two in \(\psi_k\) and one in \(\phi_k\).
The momentum routing of the swordfish and the triangle diagrams can be chosen such that only loop momenta given in \eref{eq:LoopMomenta} appear.
This allows us to use the same quadrature for all diagrams.
For the static diagrams we use \(\num{30}\) (\(12\)) integration nodes for every momentum (angular) integration region (for the dynamic diagrams we use up to \(15\) integration nodes per angular integration region).
Thus, for each of the \(15^3 \times 7^3 = \num{1157625}\) external grid points, we need to evaluate \(\num{1244160}\) internal grid points.
The most complicated kernel is that of the \acrlong{glb}. To evaluate the \acrlong{glb} kernel once, we need about \(\num{50000}\) multiplications even though we optimized the kernel with a specialized {\it Mathematica} \cite{Wolfram:2004} algorithm.

In the evaluation of the kernels values for the variables of the four-gluon vertex may appear that are outside of the grid. In this case an appropriate extrapolation must be performed.
For the angles (\(\theta_r,\,\theta_q,\,\psi_q\)), it can happen that they are (slightly) higher than the highest angular grid point, since we did set the highest grid point slightly below $\pi$ to avoid the appearance of a (well-defined) $0/0$.
In that case we simply approximate the dressing function by the boundary value.

The squared momenta, on the other hand, can be below as well as above the range defined.
In the former case, the \gls{ir} extrapolation, we again use the boundary value.
This is reasonable if the \gls{ir} behavior of the dressing function is insignificant for the iteration, or the dressing function is constant in the \gls{ir}.
The latter is approximately the case in the decoupling solution (see, e.g., \fref{fig:ContributionOfDiagrams}),
the former in the scaling solution:
The ghost box is the leading diagram of the scaling solution in the \gls{ir}.
Since the ghost box does not depend on the four-gluon vertex, it does also not depend on the extrapolation. Thus the IR region is not affected by the extrapolation of the four-gluon vertex.

For the \gls{uv} extrapolation we use a Bose symmetric model in analogy to that used for the three-gluon vertex \cite{Blum:2014gna}.
For that purpose we define the averaged momentum \(\bar{p}\) by \(\bar{p}^2 = (p^2+q^2+r^2+s^2)/2\) and denote the highest averaged momentum that lies within the grid by \(\bar{p}_0^2=(p_0^2+q_0^2+r_0^2+s_0^2)/2\,\).
Using the exponents from \eref{eq:Exponents}, we approximate the dressing function in the \gls{uv} by
\begin{align}
	\label{eq:UVextrapolation}
  \begin{split}
	& D_\text{\acrshort{uv}}^\text{4g}(p,\,q,\,r,\,s)\\&\quad =
		D^\text{4g}(p_0,\,q_0,\,r_0,\,s_0) \,
		\left(\frac{G(\bar{p}^2)}{G(\bar{p}_0^2)}\right)^{\alpha_{4\text{g}}} \,
		\left(\frac{Z(\bar{p}^2)}{Z(\bar{p}_0^2)}\right)^{\beta_{4\text{g}}}.
    \end{split}
\end{align}
We find that the highest calculated \gls{uv} points show the expected \gls{uv} behavior given by \eref{eq:UVextrapolation}.
In addition, \eref{eq:UVextrapolation} connects the extrapolated momentum region with the non-extrapolated region smoothly.
Therefore, the \gls{uv} extrapolation by \eref{eq:UVextrapolation} is a reasonable extrapolation. However, in some isolated cases it can happen that the results deviate. One such example is configuration $A$ with $p^2$ equal to the highest point on the momentum grid. For specific permutations the angle configuration corresponds to one of the configurations with a large angle dependence; see \fref{fig:3DPlots}. Thus the extrapolation gives a value that is off. However, such configurations are rare and we expect them to have a negligible influence on the total calculation.

An extrapolation is also necessary for some points in the symmetrization process. However, the UV extrapolation function is valid for the dressing function $D^\text{4g}(p,q,r,s)$ and not $L(p,q,r,s)$ given in \eref{eq:projected_one-loop}. Thus we work with
\begin{align}
\label{eq:D4g-3L}
 D^\text{4g}(p,q,r,s)=Z_4+3L(p,q,r,s)
\end{align}
which should approximate the dressing function well enough to allow a good extrapolation. The final dressing function is obtained from averaging over all $12$ $D^\text{4g}$.

\section{Extended tensor basis}
\label{sec:extended_tensor_basis}

The tensors employed for the investigation of alternative dressings in Sect.~\ref{sec:other_tensors} are discussed here. We consider two classes that differ in the structure of the Lorentz part. The tree-level belongs to the first class, which contains only tensors constructed from the metric. The second class consists of tensors constructed of momenta only.

The tensors of the first class were also used in Refs.~\cite{Driesen:1998xc,Kellermann:2008iw}, but for our purposes some modifications were necessary, which we summarize here.
Starting from the color tensors given in \eref{eq:color_tensors_basis} and the Lorentz tensors
\begin{align}
 L^1_{\mu\nu\rho\sigma}=\de_{\mu\nu}\de_{\rho\sigma}, \quad L^2_{\mu\nu\rho\sigma}=\de_{\mu\rho}\de_{\nu\sigma}, \quad L^3_{\mu\nu\rho\sigma}=\de_{\mu\sigma}\de_{\nu\rho},
\end{align}
a three-dimensional Bose symmetric subspace is given by	
\begin{subequations}
\label{eq:Vis_non-ortho}
\begin{align}
 &\widetilde{V}_{1,\mu\nu\rho\sigma}^{abcd}=\Gamma^{(0),abcd}_{\mu\nu\rho\sigma},\\
 &\widetilde{V}_{2,\mu\nu\rho\sigma}^{abcd}=\de^{ab}\de^{cd}\de_{\mu\nu}\de_{\rho\sigma}+\de^{ac}\de^{bd}\de_{\mu\rho}\de_{\nu\sigma}+\de^{ad}\de^{bc}\de_{\mu\sigma}\de_{\nu\rho},\\
 &\widetilde{V}_{3,\mu\nu\rho\sigma}^{abcd}=\nnnl
 &\quad(\de^{ac}\de^{bd}+\de^{ad}\de^{bc})\de_{\mu\nu}\de_{\rho\sigma}+(\de^{ab}\de^{cd}+\de^{ad}\de^{bc})\de_{\mu\rho}\de_{\nu\sigma}\nnnl
 &\quad\quad+(\de^{ab}\de^{cd}+\de^{ac}\de^{bd})\de_{\mu\sigma}\de_{\nu\rho}.
\end{align}
\end{subequations}
Orthonormalizing this basis leads to the expressions given in Eq.~(10) of Ref.~\cite{Kellermann:2008iw}. However, we found that the expression given for $V_3$ contains several errors.

Here we are only interested in the transverse part of the four-gluon vertex and thus construct a basis from the transversely projected tensors. Combining this with a Gram--Schmidt orthogonalization we obtain the following basis, where all indices and arguments are suppressed and $T$ denotes the transverse projection of all four Lorentz indices:
 \begin{subequations}
 \label{eq:Volis}
  \begin{align}
   \overline{V}_1&=T\widetilde{V}_1,\\
   \overline{V}_2&=T\left(\widetilde{V}_2-\overline{V}_1 \frac{ \overline{V}_1 \cdot\wt{V}_2}{\overline{V}_1  \cdot\overline{V}_1}\right),\\
   \overline{V}_3&=T\left(\widetilde{V}_3-\overline{V}_1 \frac{ \overline{V}_1 \cdot\wt{V}_3}{\overline{V}_1  \cdot\overline{V}_1}-\overline{V}_2\frac{ \overline{V}_2 \cdot\wt{V}_3}{\overline{V}_2\cdot \overline{V}_2}\right).
  \end{align}
 \end{subequations}
 
As tensor of the second class we consider
\begin{align}
 \label{eq:Ptilde}
 \widetilde{P}_{\mu\nu\rho\sigma}^{abcd}&=(\de^{ab}\de^{cd}+\de^{ac}\de^{bd}+\de^{ad}\de^{bc})\nnnl
 &\times\frac{s_\mu r_\nu q_\rho p_\sigma + r_\mu s_\nu p_\rho q_\sigma + q_\mu p_\nu s_\rho r_\sigma}{\sqrt{p^2\,q^2\,r^2\,p^2}},
\end{align}
which is also Bose symmetric. The tensor $\widetilde{P}$ is orthogonal to $\widetilde{V}_1$ due to the color part but not to $\widetilde{V}_2$ and $\widetilde{V}_3$. For estimating the magnitude of the corresponding dressing functions orthogonality to the tree-level tensor suffices.
Hence, we simply use the transversely projected tensor (suppressing indices again)
 \begin{equation}
  \label{eq:Pol}
  \overline{P}=T\widetilde{P}.
 \end{equation}
 
For meaningful comparisons the tensors need to be normalized.
Thus, we introduce normalization factors:
\begin{equation}
 \label{eq:NormTensors}
 V_i=\mathcal{N}_i \overline{V}_i\,,\quad
 P=\mathcal{N}_P \overline{P}
\end{equation}
To simplify notation, we denote the norm of a tensor by
\begin{equation*}
 \norm{X}=\sqrt{X_{\mu\nu\rho\sigma}^{abcd} X_{\mu\nu\rho\sigma}^{abcd}}\,.
\end{equation*}
A straightforward choice would be to normalize the tensors to $1$, viz., $\norm{V_i}=1$ and $\norm{P}=1$.
However,  then $V_1$ does not coincide with the transversely projected tree-level tensor.
Thus we use $\mathcal{N}_1=1$.
In order to make the dressing functions comparable, we demand
\begin{equation}
 \label{eq:NormConditions}
 \norm{V_1}=\norm{V_2}=\norm{V_3}=\norm{P},
\end{equation}
which leads to
\begin{equation}
 \label{eq:NormConstants}
 \mathcal{N}_2=-\,\frac{\norm{\overline{V}_1}}{\norm{\overline{V}_2}},\;\;
 \mathcal{N}_3=-\,\frac{\norm{\overline{V}_1}}{\norm{\overline{V}_3}},\;\;\text{and}\;\;
 \mathcal{N}_P=\frac{\norm{\overline{V}_1}}{\norm{\overline{P}}}.
\end{equation}
The signs of the tensors are in principle arbitrary and we chose them such that the corresponding dressing functions have the same sign as the tree-level.
Note that via the normalization factor each tensor gets a factor of $g^2$. 
In general, the normalization factors depend on the momenta. However, for a fixed momentum configuration with one momentum scale, as used in our plots, the norms are constant for purely dimensional reasons. Thus, the normalization factors are also constant then.

Finally we mention the influence of this choice of normalization on \fref{fig:DressFuncComp}. To obtain tensors normalized to $1$, we need to divide the tensors $V_1$, $V_2$, $V_3$ and $P$ by $\norm{\overline{V}_1}$.
Since the $\norm{\overline{V}_1}$ is constant for a specific configuration, the relative values of the dressing functions in \fref{fig:DressFuncComp} would have been identical if we had chosen to work with tensors normalized to $1$. To be precise, a plot for tensors normalized to \(1\) can be obtained by multiplying all values of the dressing functions by $\norm{\overline{V}_1}=\frac{7}{2}g^2 N_\text{c} \sqrt{3(N_\text{c}^2-1)}\approx 51.44\, g^2$ for configuration $C$.

\bibliographystyle{utphys_mod}
\bibliography{literature_4g_vertex}

\providecommand{\href}[2]{#2}\begingroup\raggedright\begin{thebibliography}{10}

\bibitem{Alkofer:2000wg}
R.~Alkofer and L.~von Smekal, {\em Phys. Rept.} {\bfseries 353} (2001) 281,
\href{http://arxiv.org/abs/hep-ph/0007355}{{\ttfamily arXiv:hep-ph/0007355}}.

\bibitem{Fischer:2006ub}
C.~S. Fischer, \href{http://dx.doi.org/10.1088/0954-3899/32/8/R02}{{\em J.
  Phys.} {\bfseries G32} (2006) R253--R291},
\href{http://arxiv.org/abs/hep-ph/0605173}{{\ttfamily arXiv:hep-ph/0605173}}.

\bibitem{Fischer:2009wc}
C.~S. Fischer, \href{http://dx.doi.org/10.1103/PhysRevLett.103.052003}{{\em
  Phys. Rev. Lett.} {\bfseries 103} (2009) 052003},
\href{http://arxiv.org/abs/0904.2700}{{\ttfamily arXiv:0904.2700 [hep-ph]}}.

\bibitem{Braun:2009gm}
J.~Braun, L.~M. Haas, F.~Marhauser, and J.~M. Pawlowski,
  \href{http://dx.doi.org/10.1103/PhysRevLett.106.022002}{{\em Phys. Rev.
  Lett.} {\bfseries 106} (2011) 022002},
\href{http://arxiv.org/abs/0908.0008}{{\ttfamily arXiv:0908.0008 [hep-ph]}}.

\bibitem{Braun:2010cy}
J.~Braun, A.~Eichhorn, H.~Gies, and J.~M. Pawlowski,
  \href{http://dx.doi.org/10.1140/epjc/s10052-010-1485-1}{{\em Eur.Phys.J.}
  {\bfseries C70} (2010) 689--702},
\href{http://arxiv.org/abs/1007.2619}{{\ttfamily arXiv:1007.2619 [hep-ph]}}.

\bibitem{Bashir:2012fs}
A.~Bashir, L.~Chang, I.~Cloet, B.~El-Bennich, Y.-X. Liu, C.~Roberts, and
  P.~Tandy, \href{http://dx.doi.org/10.1088/0253-6102/58/1/16}{{\em
  Commun.Theor.Phys.} {\bfseries 58} (2012) 79--134},
\href{http://arxiv.org/abs/1201.3366}{{\ttfamily arXiv:1201.3366 [nucl-th]}}.

\bibitem{Fischer:2012vc}
C.~S. Fischer and J.~Luecker,
  \href{http://dx.doi.org/10.1016/j.physletb.2012.11.054}{{\em Phys.Lett.}
  {\bfseries B718} (2013) 1036--1043},
\href{http://arxiv.org/abs/1206.5191}{{\ttfamily arXiv:1206.5191 [hep-ph]}}.

\bibitem{Hopfer:2012qr}
M.~Hopfer, M.~Mitter, B.-J. Schaefer, and R.~Alkofer,
  \href{http://dx.doi.org/10.5506/APhysPolBSupp.6.353}{{\em Acta
  Phys.Polon.Supp.} {\bfseries 6} (2013) 353--358},
\href{http://arxiv.org/abs/1211.0166}{{\ttfamily arXiv:1211.0166 [hep-ph]}}.

\bibitem{Eichmann:2013afa}
G.~Eichmann,
\href{http://dx.doi.org/10.1088/1742-6596/426/1/012014}{{\em J.Phys.Conf.Ser.}
  {\bfseries 426} (2013) 012014}.

\bibitem{Fischer:2013eca}
C.~S. Fischer, L.~Fister, J.~Luecker, and J.~M. Pawlowski,
  \href{http://dx.doi.org/10.1016/j.physletb.2014.03.057}{{\em Phys.Lett.}
  {\bfseries B732} (2014) 273--277},
\href{http://arxiv.org/abs/1306.6022}{{\ttfamily arXiv:1306.6022 [hep-ph]}}.

\bibitem{Tripolt:2013jra}
R.-A. Tripolt, N.~Strodthoff, L.~von Smekal, and J.~Wambach,
  \href{http://dx.doi.org/10.1103/PhysRevD.89.034010}{{\em Phys.Rev.}
  {\bfseries D89} (2014) 034010},
\href{http://arxiv.org/abs/1311.0630}{{\ttfamily arXiv:1311.0630 [hep-ph]}}.

\bibitem{Braun:2007bx}
J.~Braun, H.~Gies, and J.~M. Pawlowski,
  \href{http://dx.doi.org/10.1016/j.physletb.2010.01.009}{{\em Phys. Lett.}
  {\bfseries B684} (2010) 262--267},
\href{http://arxiv.org/abs/0708.2413}{{\ttfamily arXiv:0708.2413 [hep-th]}}.

\bibitem{Marhauser:2008fz}
F.~Marhauser and J.~M. Pawlowski,
\href{http://arxiv.org/abs/0812.1144}{{\ttfamily arXiv:0812.1144 [hep-ph]}}.

\bibitem{Fister:2013bh}
L.~Fister and J.~M. Pawlowski,
  \href{http://dx.doi.org/10.1103/PhysRevD.88.045010}{{\em Phys.Rev.}
  {\bfseries D88} (2013) 045010},
\href{http://arxiv.org/abs/1301.4163}{{\ttfamily arXiv:1301.4163 [hep-ph]}}.

\bibitem{Cucchieri:2007md}
A.~Cucchieri and T.~Mendes, {\em PoS} {\bfseries LAT2007} (2007) 297,
\href{http://arxiv.org/abs/0710.0412}{{\ttfamily arXiv:0710.0412 [hep-lat]}}.

\bibitem{Cucchieri:2008fc}
A.~Cucchieri and T.~Mendes,
  \href{http://dx.doi.org/10.1103/PhysRevD.78.094503}{{\em Phys. Rev.}
  {\bfseries D78} (2008) 094503},
\href{http://arxiv.org/abs/0804.2371}{{\ttfamily arXiv:0804.2371 [hep-lat]}}.

\bibitem{Sternbeck:2007ug}
A.~Sternbeck, L.~von Smekal, D.~Leinweber, and A.~Williams, {\em PoS}
  {\bfseries LAT2007} (2007) 340,
\href{http://arxiv.org/abs/0710.1982}{{\ttfamily arXiv:0710.1982 [hep-lat]}}.

\bibitem{Bogolubsky:2009dc}
I.~L. Bogolubsky, E.~M. Ilgenfritz, M.~M\"uller-Preussker, and A.~Sternbeck,
  \href{http://dx.doi.org/10.1016/j.physletb.2009.04.076}{{\em Phys. Lett.}
  {\bfseries B676} (2009) 69--73},
\href{http://arxiv.org/abs/0901.0736}{{\ttfamily arXiv:0901.0736 [hep-lat]}}.

\bibitem{Oliveira:2012eh}
O.~Oliveira and P.~J. Silva,
  \href{http://dx.doi.org/10.1103/PhysRevD.86.114513}{{\em Phys.Rev.}
  {\bfseries D86} (2012) 114513},
\href{http://arxiv.org/abs/1207.3029}{{\ttfamily arXiv:1207.3029 [hep-lat]}}.

\bibitem{Sternbeck:2012mf}
A.~Sternbeck and M.~M\"uller-Preussker,
  \href{http://dx.doi.org/10.1016/j.physletb.2013.08.017}{{\em Phys.Lett.}
  {\bfseries B726} (2013) 396--403},
\href{http://arxiv.org/abs/1211.3057}{{\ttfamily arXiv:1211.3057 [hep-lat]}}.

\bibitem{vonSmekal:1997is}
L.~von Smekal, R.~Alkofer, and A.~Hauck,
  \href{http://dx.doi.org/10.1103/PhysRevLett.79.3591}{{\em Phys. Rev. Lett.}
  {\bfseries 79} (1997) 3591--3594},
\href{http://arxiv.org/abs/hep-ph/9705242}{{\ttfamily arXiv:hep-ph/9705242}}.

\bibitem{vonSmekal:1997vx}
L.~von Smekal, A.~Hauck, and R.~Alkofer,
  \href{http://dx.doi.org/10.1006/aphy.1998.5806}{{\em Ann. Phys.} {\bfseries
  267} (1998) 1},
\href{http://arxiv.org/abs/hep-ph/9707327}{{\ttfamily arXiv:hep-ph/9707327}}.

\bibitem{Atkinson:1997tu}
D.~Atkinson and J.~C.~R. Bloch,
  \href{http://dx.doi.org/10.1103/PhysRevD.58.094036}{{\em Phys. Rev.}
  {\bfseries D58} (1998) 094036},
\href{http://arxiv.org/abs/hep-ph/9712459}{{\ttfamily arXiv:hep-ph/9712459}}.

\bibitem{Zwanziger:2001kw}
D.~Zwanziger, \href{http://dx.doi.org/10.1103/PhysRevD.65.094039}{{\em
  Phys.Rev.} {\bfseries D65} (2002) 094039},
\href{http://arxiv.org/abs/hep-th/0109224}{{\ttfamily arXiv:hep-th/0109224
  [hep-th]}}.

\bibitem{Lerche:2002ep}
C.~Lerche and L.~von Smekal,
  \href{http://dx.doi.org/10.1103/PhysRevD.65.125006}{{\em Phys.Rev.}
  {\bfseries D65} (2002) 125006},
\href{http://arxiv.org/abs/hep-ph/0202194}{{\ttfamily arXiv:hep-ph/0202194}}.

\bibitem{Zwanziger:2002ia}
D.~Zwanziger, \href{http://dx.doi.org/10.1103/PhysRevD.67.105001}{{\em
  Phys.Rev.} {\bfseries D67} (2003) 105001},
\href{http://arxiv.org/abs/hep-th/0206053}{{\ttfamily arXiv:hep-th/0206053
  [hep-th]}}.

\bibitem{Fischer:2002hn}
C.~S. Fischer and R.~Alkofer,
  \href{http://dx.doi.org/10.1016/S0370-2693(02)01809-9}{{\em Phys. Lett.}
  {\bfseries B536} (2002) 177--184},
\href{http://arxiv.org/abs/hep-ph/0202202}{{\ttfamily arXiv:hep-ph/0202202}}.

\bibitem{Zwanziger:2003cf}
D.~Zwanziger, \href{http://dx.doi.org/10.1103/PhysRevD.69.016002}{{\em Phys.
  Rev.} {\bfseries D69} (2004) 016002},
\href{http://arxiv.org/abs/hep-ph/0303028}{{\ttfamily arXiv:hep-ph/0303028}}.

\bibitem{Boucaud:2008ji}
P.~Boucaud {\em et~al.},
  \href{http://dx.doi.org/10.1088/1126-6708/2008/06/012}{{\em JHEP} {\bfseries
  06} (2008) 012},
\href{http://arxiv.org/abs/0801.2721}{{\ttfamily arXiv:0801.2721 [hep-ph]}}.

\bibitem{Aguilar:2008xm}
A.~Aguilar, D.~Binosi, and J.~Papavassiliou,
  \href{http://dx.doi.org/10.1103/PhysRevD.78.025010}{{\em Phys.Rev.}
  {\bfseries D78} (2008) 025010},
  \href{http://arxiv.org/abs/0802.1870}{{\ttfamily arXiv:0802.1870 [hep-ph]}}.

\bibitem{Alkofer:2008jy}
R.~Alkofer, M.~Q. Huber, and K.~Schwenzer,
  \href{http://dx.doi.org/http://link.aps.org/doi/10.1103/PhysRevD.81.105010}{{\em
  Phys. Rev.} {\bfseries D81} (2010) 105010},
\href{http://arxiv.org/abs/0801.2762}{{\ttfamily arXiv:0801.2762 [hep-th]}}.

\bibitem{Fischer:2008uz}
C.~S. Fischer, A.~Maas, and J.~M. Pawlowski,
  \href{http://dx.doi.org/10.1016/j.aop.2009.07.009}{{\em Ann.Phys.} {\bfseries
  324} (2009) 2408--2437},
\href{http://arxiv.org/abs/0810.1987}{{\ttfamily arXiv:0810.1987 [hep-ph]}}.

\bibitem{Fischer:2009tn}
C.~S. Fischer and J.~M. Pawlowski,
  \href{http://dx.doi.org/10.1103/PhysRevD.80.025023}{{\em Phys. Rev.}
  {\bfseries D80} (2009) 025023},
\href{http://arxiv.org/abs/0903.2193}{{\ttfamily arXiv:0903.2193 [hep-th]}}.

\bibitem{Huber:2009tx}
M.~Q. Huber, R.~Alkofer, and S.~P. Sorella,
  \href{http://dx.doi.org/10.1103/PhysRevD.81.065003}{{\em Phys. Rev.}
  {\bfseries D81} (2010) 065003},
\href{http://arxiv.org/abs/0910.5604}{{\ttfamily arXiv:0910.5604 [hep-th]}}.

\bibitem{Pennington:2011xs}
M.~Pennington and D.~Wilson,
  \href{http://dx.doi.org/10.1103/PhysRevD.84.094028}{{\em Phys.Rev.}
  {\bfseries D84} (2011) 119901},
\href{http://arxiv.org/abs/1109.2117}{{\ttfamily arXiv:1109.2117 [hep-ph]}}.

\bibitem{LlanesEstrada:2012my}
F.~J. Llanes-Estrada and R.~Williams,
  \href{http://dx.doi.org/10.1103/PhysRevD.86.065034}{{\em Phys.Rev.}
  {\bfseries D86} (2012) 065034},
\href{http://arxiv.org/abs/1207.5950}{{\ttfamily arXiv:1207.5950 [hep-th]}}.

\bibitem{Strauss:2012dg}
S.~Strauss, C.~S. Fischer, and C.~Kellermann,
  \href{http://dx.doi.org/10.1103/PhysRevLett.109.252001}{{\em Phys.Rev.Lett.}
  {\bfseries 109} (2012) 252001},
\href{http://arxiv.org/abs/1208.6239}{{\ttfamily arXiv:1208.6239 [hep-ph]}}.

\bibitem{Pawlowski:2003hq}
J.~M. Pawlowski, D.~F. Litim, S.~Nedelko, and L.~von Smekal,
  \href{http://dx.doi.org/10.1103/PhysRevLett.93.152002}{{\em Phys. Rev. Lett.}
  {\bfseries 93} (2004) 152002},
\href{http://arxiv.org/abs/hep-th/0312324}{{\ttfamily arXiv:hep-th/0312324}}.

\bibitem{Quandt:2013wna}
M.~Quandt, H.~Reinhardt, and J.~Heffner,
  \href{http://dx.doi.org/10.1103/PhysRevD.89.065037}{{\em Phys.Rev.}
  {\bfseries D89} (2014) 065037},
\href{http://arxiv.org/abs/1310.5950}{{\ttfamily arXiv:1310.5950 [hep-th]}}.

\bibitem{Tissier:2010ts}
M.~Tissier and N.~Wschebor,
  \href{http://dx.doi.org/10.1103/PhysRevD.82.101701}{{\em Phys.Rev.}
  {\bfseries D82} (2010) 101701},
\href{http://arxiv.org/abs/1004.1607}{{\ttfamily arXiv:1004.1607 [hep-ph]}}.

\bibitem{Gribov:1977wm}
V.~Gribov,
\href{http://dx.doi.org/10.1016/0550-3213(78)90175-X}{{\em Nucl.Phys.}
  {\bfseries B139} (1978) 1}.

\bibitem{Zwanziger:1992qr}
D.~Zwanziger,
\href{http://dx.doi.org/10.1016/0550-3213(93)90506-K}{{\em Nucl.Phys.}
  {\bfseries B399} (1993) 477--513}.

\bibitem{Zwanziger:1993dh}
D.~Zwanziger,
\href{http://dx.doi.org/10.1016/0550-3213(94)90396-4}{{\em Nucl. Phys.}
  {\bfseries B412} (1994) 657--730}.

\bibitem{Dudal:2008sp}
D.~Dudal, J.~A. Gracey, S.~P. Sorella, N.~Vandersickel, and H.~Verschelde,
  \href{http://dx.doi.org/10.1103/PhysRevD.78.065047}{{\em Phys. Rev.}
  {\bfseries D78} (2008) 065047},
\href{http://arxiv.org/abs/0806.4348}{{\ttfamily arXiv:0806.4348 [hep-th]}}.

\bibitem{Dudal:2007cw}
D.~Dudal, S.~P. Sorella, N.~Vandersickel, and H.~Verschelde,
  \href{http://dx.doi.org/10.1103/PhysRevD.77.071501}{{\em Phys. Rev.}
  {\bfseries D77} (2008) 071501},
\href{http://arxiv.org/abs/0711.4496}{{\ttfamily arXiv:0711.4496 [hep-th]}}.

\bibitem{Dudal:2011gd}
D.~Dudal, S.~Sorella, and N.~Vandersickel,
  \href{http://dx.doi.org/10.1103/PhysRevD.84.065039}{{\em Phys.Rev.}
  {\bfseries D84} (2011) 065039},
\href{http://arxiv.org/abs/1105.3371}{{\ttfamily arXiv:1105.3371 [hep-th]}}.

\bibitem{Schleifenbaum:2004id}
W.~Schleifenbaum, A.~Maas, J.~Wambach, and R.~Alkofer,
  \href{http://dx.doi.org/10.1103/PhysRevD.72.014017}{{\em Phys.Rev.}
  {\bfseries D72} (2005) 014017},
\href{http://arxiv.org/abs/hep-ph/0411052}{{\ttfamily arXiv:hep-ph/0411052
  [hep-ph]}}.

\bibitem{Cucchieri:2008qm}
A.~Cucchieri, A.~Maas, and T.~Mendes,
  \href{http://dx.doi.org/10.1103/PhysRevD.77.094510}{{\em Phys. Rev.}
  {\bfseries D77} (2008) 094510},
\href{http://arxiv.org/abs/0803.1798}{{\ttfamily arXiv:0803.1798 [hep-lat]}}.

\bibitem{Alkofer:2008dt}
R.~Alkofer, M.~Q. Huber, and K.~Schwenzer,
  \href{http://dx.doi.org/10.1140/epjc/s10052-009-1066-3}{{\em Eur. Phys. J.}
  {\bfseries C62} (2009) 761--781},
\href{http://arxiv.org/abs/0812.4045}{{\ttfamily arXiv:0812.4045 [hep-ph]}}.

\bibitem{Ilgenfritz:2006he}
E.~M. Ilgenfritz, M.~M{\"u}ller-Preussker, A.~Sternbeck, A.~Schiller, and I.~L.
  Bogolubsky, {\em Braz. J. Phys.} {\bfseries 37} (2007) 193,
\href{http://arxiv.org/abs/hep-lat/0609043}{{\ttfamily arXiv:hep-lat/0609043}}.

\bibitem{Boucaud:2011eh}
P.~Boucaud, D.~Dudal, J.~Leroy, O.~Pene, and J.~Rodriguez-Quintero,
  \href{http://dx.doi.org/10.1007/JHEP12(2011)018}{{\em JHEP} {\bfseries 1112}
  (2011) 018},
\href{http://arxiv.org/abs/1109.3803}{{\ttfamily arXiv:1109.3803 [hep-ph]}}.

\bibitem{Fister:2011uw}
L.~Fister and J.~M. Pawlowski,
\href{http://arxiv.org/abs/1112.5440}{{\ttfamily arXiv:1112.5440 [hep-ph]}}.

\bibitem{Huber:2012kd}
M.~Q. Huber and L.~von Smekal,
  \href{http://dx.doi.org/10.1007/JHEP04(2013)149}{{\em JHEP} {\bfseries 1304}
  (2013) 149},
\href{http://arxiv.org/abs/1211.6092}{{\ttfamily arXiv:1211.6092 [hep-th]}}.

\bibitem{Pelaez:2013cpa}
M.~Pelaez, M.~Tissier, and N.~Wschebor,
  \href{http://dx.doi.org/10.1103/PhysRevD.88.125003}{{\em Phys.Rev.}
  {\bfseries D88} (2013) 125003},
\href{http://arxiv.org/abs/1310.2594}{{\ttfamily arXiv:1310.2594 [hep-th]}}.

\bibitem{Aguilar:2013xqa}
A.~Aguilar, D.~Ib\'a\~nez, and J.~Papavassiliou,
  \href{http://dx.doi.org/10.1103/PhysRevD.87.114020}{{\em Phys.Rev.}
  {\bfseries D87} (2013) 114020},
\href{http://arxiv.org/abs/1303.3609}{{\ttfamily arXiv:1303.3609 [hep-ph]}}.

\bibitem{Aguilar:2013vaa}
A.~Aguilar, D.~Binosi, D.~Ib\'a\~nez, and J.~Papavassiliou,
  \href{http://dx.doi.org/10.1103/PhysRevD.89.085008}{{\em Phys.Rev.}
  {\bfseries D89} (2014) 085008},
\href{http://arxiv.org/abs/1312.1212}{{\ttfamily arXiv:1312.1212 [hep-ph]}}.

\bibitem{Blum:2014gna}
A.~Blum, M.~Q. Huber, M.~Mitter, and L.~von Smekal,
  \href{http://dx.doi.org/10.1103/PhysRevD.89.061703}{{\em Phys. Rev. D}
  {\bfseries 89} (2014) 061703(R)},
\href{http://arxiv.org/abs/1401.0713}{{\ttfamily arXiv:1401.0713 [hep-ph]}}.

\bibitem{Eichmann:2014xya}
G.~Eichmann, R.~Williams, R.~Alkofer, and M.~Vujinovic,
  \href{http://dx.doi.org/10.1103/PhysRevD.89.105014}{{\em Phys.Rev.}
  {\bfseries D89} (2014) 105014},
\href{http://arxiv.org/abs/1402.1365}{{\ttfamily arXiv:1402.1365 [hep-ph]}}.

\bibitem{Bloch:2003yu}
J.~C. Bloch, \href{http://dx.doi.org/10.1007/s00601-003-0013-3}{{\em Few Body
  Syst.} {\bfseries 33} (2003) 111--152},
\href{http://arxiv.org/abs/hep-ph/0303125}{{\ttfamily arXiv:hep-ph/0303125}}.

\bibitem{Mader:2013ru}
V.~Mader and R.~Alkofer, {\em PoS} {\bfseries ConfinementX} (2012) 063,
\href{http://arxiv.org/abs/1301.7498}{{\ttfamily arXiv:1301.7498 [hep-th]}}.

\bibitem{Meyers:2014iwa}
J.~Meyers and E.~S. Swanson,
\href{http://arxiv.org/abs/1403.4350}{{\ttfamily arXiv:1403.4350 [hep-ph]}}.

\bibitem{Pascual:1980yu}
P.~Pascual and R.~Tarrach,
\href{http://dx.doi.org/10.1016/0550-3213(80)90193-5}{{\em Nucl.Phys.}
  {\bfseries B174} (1980) 123}.

\bibitem{Gracey:2014ola}
J.~Gracey,
\href{http://arxiv.org/abs/1406.1618}{{\ttfamily arXiv:1406.1618 [hep-ph]}}.

\bibitem{Driesen:1998xc}
L.~Driesen and M.~Stingl, \href{http://dx.doi.org/10.1007/s100500050247}{{\em
  Eur.Phys.J.} {\bfseries A4} (1999) 401--419},
\href{http://arxiv.org/abs/hep-th/9808155}{{\ttfamily arXiv:hep-th/9808155
  [hep-th]}}.

\bibitem{Kellermann:2008iw}
C.~Kellermann and C.~S. Fischer,
  \href{http://dx.doi.org/10.1103/PhysRevD.78.025015}{{\em Phys. Rev.}
  {\bfseries D78} (2008) 025015},
\href{http://arxiv.org/abs/0801.2697}{{\ttfamily arXiv:0801.2697 [hep-ph]}}.

\bibitem{Binosi:2014kka}
D.~Binosi, D.~Ibañez, and J.~Papavassiliou,
  \href{http://dx.doi.org/10.1007/JHEP09(2014)059}{{\em JHEP} {\bfseries 1409}
  (2014) 059},
\href{http://arxiv.org/abs/1407.3677}{{\ttfamily arXiv:1407.3677 [hep-ph]}}.

\bibitem{Aguilar:2010gm}
A.~Aguilar, D.~Binosi, and J.~Papavassiliou,
  \href{http://dx.doi.org/10.1007/JHEP07(2010)002}{{\em JHEP} {\bfseries 1007}
  (2010) 002},
\href{http://arxiv.org/abs/1004.1105}{{\ttfamily arXiv:1004.1105 [hep-ph]}}.

\bibitem{Rivers:1988pi}
R.~J. Rivers, {\em Path Integrals Methods in Quantum Field Theory}.
\newblock Cambridge University Press, Cambridge, 1988.

\bibitem{Roberts:1994dr}
C.~D. Roberts and A.~G. Williams, {\em Prog. Part. Nucl. Phys.} {\bfseries 33}
  (1994) 477--575,
\href{http://arxiv.org/abs/hep-ph/9403224}{{\ttfamily hep-ph/9403224}}.

\bibitem{Huber:2011xc}
M.~Q. Huber and M.~Mitter,
  \href{http://dx.doi.org/10.1016/j.cpc.2012.05.019}{{\em Comput.Phys.Commun.}
  {\bfseries 183} (2012) 2441--2457},
\href{http://arxiv.org/abs/1112.5622}{{\ttfamily arXiv:1112.5622 [hep-th]}}.

\bibitem{Alkofer:2008nt}
R.~Alkofer, M.~Q. Huber, and K.~Schwenzer,
  \href{http://dx.doi.org/10.1016/j.cpc.2008.12.009}{{\em Comput. Phys.
  Commun.} {\bfseries 180} (2009) 965--976},
\href{http://arxiv.org/abs/0808.2939}{{\ttfamily arXiv:0808.2939 [hep-th]}}.

\bibitem{Huber:2011qr}
M.~Q. Huber and J.~Braun,
  \href{http://dx.doi.org/10.1016/j.cpc.2012.01.014}{{\em Comput.Phys.Commun.}
  {\bfseries 183} (2012) 1290--1320},
\href{http://arxiv.org/abs/1102.5307}{{\ttfamily arXiv:1102.5307 [hep-th]}}.

\bibitem{Binosi:2003yf}
D.~Binosi and L.~Theussl,
  \href{http://dx.doi.org/10.1016/j.cpc.2004.05.001}{{\em Comput.Phys.Commun.}
  {\bfseries 161} (2004) 76--86},
\href{http://arxiv.org/abs/hep-ph/0309015}{{\ttfamily arXiv:hep-ph/0309015}}.

\bibitem{Alkofer:2004it}
R.~Alkofer, C.~S. Fischer, and F.~J. Llanes-Estrada,
  \href{http://dx.doi.org/10.1016/j.physletb.2008.11.068,
  10.1016/j.physletb.2005.02.043}{{\em Phys.Lett.} {\bfseries B611} (2005)
  279--288},
\href{http://arxiv.org/abs/hep-th/0412330}{{\ttfamily arXiv:hep-th/0412330
  [hep-th]}}.

\bibitem{Huber:2012zj}
M.~Q. Huber, A.~Maas, and L.~von Smekal,
  \href{http://dx.doi.org/10.1007/JHEP11(2012)035}{{\em JHEP} {\bfseries 1211}
  (2012) 035},
\href{http://arxiv.org/abs/1207.0222}{{\ttfamily arXiv:1207.0222 [hep-th]}}.

\bibitem{Huber:2007kc}
M.~Q. Huber, R.~Alkofer, C.~S. Fischer, and K.~Schwenzer,
  \href{http://dx.doi.org/10.1016/j.physletb.2007.10.073}{{\em Phys. Lett.}
  {\bfseries B659} (2008) 434--440},
\href{http://arxiv.org/abs/0705.3809}{{\ttfamily arXiv:0705.3809 [hep-ph]}}.

\bibitem{Carimalo:1992ia}
C.~Carimalo,
\href{http://dx.doi.org/10.1063/1.530334}{{\em J.Math.Phys.} {\bfseries 34}
  (1993) 4930--4963}.

\bibitem{Eichmann:2011vu}
G.~Eichmann, \href{http://dx.doi.org/10.1103/PhysRevD.84.014014}{{\em
  Phys.Rev.} {\bfseries D84} (2011) 014014},
\href{http://arxiv.org/abs/1104.4505}{{\ttfamily arXiv:1104.4505 [hep-ph]}}.

\bibitem{Eichmann:2014pc}
G.~Eichmann, private communication, 2014.

\bibitem{vonSmekal:2009ae}
L.~von Smekal, K.~Maltman, and A.~Sternbeck,
  \href{http://dx.doi.org/10.1016/j.physletb.2009.10.030}{{\em Phys.Lett.}
  {\bfseries B681} (2009) 336--342},
\href{http://arxiv.org/abs/0903.1696}{{\ttfamily arXiv:0903.1696 [hep-ph]}}.

\bibitem{Huber:2014tva}
M.~Q. Huber and L.~von Smekal,
  \href{http://dx.doi.org/10.1007/JHEP06(2014)015}{{\em JHEP} {\bfseries 1406}
  (2014) 015},
\href{http://arxiv.org/abs/1404.3642}{{\ttfamily arXiv:1404.3642 [hep-ph]}}.

\bibitem{Sternbeck:2006rd}
A.~Sternbeck, \href{http://arxiv.org/abs/hep-lat/0609016}{{\ttfamily
  arXiv:hep-lat/0609016}}, PhD thesis, Humboldt-Universit\"at zu Berlin,
2006.

\bibitem{Blum:2014mt}
A.~L. Blum, master thesis, Technische Universit\"at Darmstadt, 2014.

\bibitem{Wolfram:2004}
S.~Wolfram, {\em The Mathematica Book}.
\newblock Wolfram Media and Cambridge University Press, 2004.

\end{thebibliography}\endgroup

\end{document}